%% file: aph.tex
\documentclass[iop]{emulateapj-rtx4}
\usepackage{natbib}
\usepackage{comment}
\usepackage{lscape}
\usepackage{amsmath}
\usepackage{hyperref,breakurl}
\usepackage{epsfig}

\shorttitle{GBT 33\,GHz OBSERVATIONS OF GALAXY NUCLEI AND EXTRA-NUCLEAR STAR-FORMING REGIONS}
\shortauthors{MURPHY ET AL.}
\slugcomment{Submitted to ApJ July 16, 2012; Accepted October 10, 2012}

\begin{document}
\title{The Star Formation in Radio Survey: GBT 33\,GHz Observations of Nearby Galaxy Nuclei and Extranuclear Star-Forming Regions}

\author{E.J.\,Murphy\altaffilmark{1}, J.\,Bremseth\altaffilmark{2}, B.S.\,Mason\altaffilmark{3}, J.J.\,Condon\altaffilmark{3}, E.\,Schinnerer\altaffilmark{4}, G.\,Aniano\altaffilmark{5}, L.\,Armus \altaffilmark{5}, G.\,Helou\altaffilmark{6}, J.L.\,Turner\altaffilmark{7}, T.H.\,Jarrett\altaffilmark{8}}
\altaffiltext{1}{Observatories of the Carnegie Institution for Science, 813 Santa Barbara Street, Pasadena, CA 91101, USA; emurphy@obs.carnegiescience.edu}
\altaffiltext{2}{Department of Physics and Astronomy, Pomona College, Claremont, CA 91711, USA}
\altaffiltext{3}{National Radio Astronomy Observatory, 520 Edgemont Road, Charlottesville, VA 22903, USA}
\altaffiltext{4}{Max Planck Institut f\"{u}r Astronomie, K\"{o}nigstuhl 17, Heidelberg D-69117, Germany}
\altaffiltext{5}{Department of Astrophysical Sciences, Princeton University, Princeton, NJ 08544, USA}
\altaffiltext{5}{ {\it Spitzer Science Center,} California Institute of Technology, MC 314-6, Pasadena CA, USA}
\altaffiltext{6}{California Institute of Technology, MC 100-22, Pasadena, CA 91125, USA}
\altaffiltext{7}{Department of Physics and Astronomy, UCLA, Los Angeles, CA 90095, USA}
\altaffiltext{8}{Infrared Processing and Analysis Center, California Institute of Technology, MC 100-22, Pasadena, CA 91125, USA}


\begin{abstract}  
We present 33\,GHz photometry of 103 galaxy nuclei and extranuclear star-forming complexes taken with the Green Bank Telescope (GBT) as part of the Star Formation in Radio Survey (SFRS).  
Among the sources without evidence for an AGN, 
 and also having lower frequency radio data, we find a median thermal fraction at 33\,GHz of $\approx$76\% with a dispersion of $\approx$24\%.  
For all sources resolved on scales $\la$0.5\,kpc, the thermal fraction is even larger, being $\ga$90\%.  
This suggests that the rest-frame 33\,GHz emission provides a sensitive measure of the ionizing photon rate from young star-forming regions, thus making it a robust star formation rate indicator.  
Taking the 33\,GHz star formation rates as a reference, we investigate other empirical calibrations relying on different combinations of warm 24\,$\mu$m dust, total infrared (IR; $8-1000\,\mu$m), H$\alpha$ line, and far-UV continuum emission.  
The recipes derived here generally agree with others found in the literature, albeit with a large dispersion that most likely stems from a combination of effects.  
Comparing the 33\,GHz to total IR flux ratios as a function of the radio spectral index, measured between 1.7 and 33\,GHz, we find that the ratio increases as the radio spectral index flattens which does not appear to be a distance effect.  
Consequently, the ratio of non-thermal to total IR emission appears relatively constant, suggesting only moderate variations in the cosmic-ray electron injection spectrum and ratio of synchrotron to total cooling processes among star-forming complexes.  
Assuming that this trend solely arises from an increase in the thermal fraction sets a maximum on the scatter of the non-thermal spectral indices among the star-forming regions of $\sigma_{\alpha^{\rm NT}} \la 0.13$.  

\end{abstract}

\section{Introduction}
Radio emission from galaxies is powered by a combination of distinct physical processes.  
And although it is energetically weak with respect to a galaxy's bolometric luminosity, it provides critical information on the massive star formation activity, as well as access to the relativistic [magnetic field $+$ cosmic rays (CRs)] component in the interstellar medium (ISM) of galaxies.  

Stars more massive than $\sim8\,M_{\sun}$ end their lives as core-collapse supernovae, whose remnants are thought to be the primary accelerators of CR electrons \citep[e.g.,][]{kk95} giving rise to the diffuse synchrotron emission observed from star-forming galaxies \citep{jc92}.  
These same massive stars are also responsible for the creation of H{\sc ii} regions that produce radio free-free emission, whose strength is directly proportional to the production rate of ionizing (Lyman continuum) photons. 

Microwave frequencies, nominally spanning $\sim$$1-100$\,GHz and observable from the ground, are particularly useful in probing such processes.  
The non-thermal emission component typically has a steep spectrum ($S_{\nu} \propto \nu^{-\alpha}$, where $\alpha \sim 0.8$), while the thermal (free-free) component is relatively flat ($\alpha\sim0.1$).  
Accordingly, for globally integrated measurements of star-forming galaxies, lower frequencies (e.g., 1.4\,GHz) are generally dominated by non-thermal emission, while the observed thermal fraction of the emission increases with frequency, eventually being dominated by free-free emission once beyond $\sim$30\,GHz \citep{cy90}.  
For typical H{\sc ii} regions, the thermal fraction at 33\,GHz can be considerably higher, being $\sim$80\% \citep{ejm11b}.   
Thus, observations at such frequencies, which are largely unbiased by dust, provide an excellent diagnostic for the current star formation rate (SFR) of galaxies.  

It is worth noting that the presence of an ``anomalous" emission component in excess of free-free emission between $\sim$10 and 90\,GHz, generally attributed to electric dipole rotational emission from ultrasmall ($a \la 10^{-6}$\,cm) grains \citep[e.g.,][]{wce57,dl98a,dl98b, plsd11} or magnetic dipole emission from thermal fluctuations in the magnetization of interstellar dust grains \citep[][]{dl99}, may complicate this picture.
For a single outer-disk star-forming region in NGC\,6946, \citet{ejm10} reported an excess of 33\,GHz emission relative to what is expected given existing lower frequency radio data. 
This result has been interpreted as the first detection of so-called ``anomalous" dust emission outside of the Milky Way. 
Given that the excess was only detected for a single region, this emission component may be negligible for globally integrated measurements.

Due to the faintness of galaxies at high (i.e., $\ga$15\,GHz) microwave frequencies, existing work has been restricted to the brightest objects, and small sample sizes.  
For example, past studies demonstrating the link between high-frequency free-free emission and massive star formation include investigations of Galactic star-forming regions \citep[e.g.,][]{pgm67a}, nearby dwarf irregular galaxies \citep[e.g.,][]{kg86}, galaxy nuclei \citep[e.g.,][]{th83,th94}, nearby starbursts \citep[e.g.,][]{kwm88,th85}, and super star clusters within nearby blue compact dwarfs \citep[e.g.,][]{thb98,kj99}.  
And while these studies focus on the free-free emission from galaxies, each was conducted at frequencies $\la$30\,GHz.  
With recent improvements to the backends of existing radio telescopes, such as the Caltech Continuum Backend (CCB) on the Green Bank Telescope (GBT) and the Wideband Interferometric Digital ARchitecture (WIDAR) correlator on the Karl G. Jansky Very Large Array (VLA), the availability of increased bandwidth is making it possible to conduct investigations for large samples of objects at frequencies $\sim$30\,GHz.  

Here we present such an investigation using GBT 33\,GHz photometry for 103 galaxy nuclei and extranuclear star-forming regions in a sample of 46 nearby galaxies included in the Star Formation in Radio Survey (SFRS; see $\S$\ref{sec-sample}).  
These galaxies, which are included in the {\it Spitzer} Infrared Nearby Galaxies Survey \citep[SINGS;][]{rck03} and Key Insights on Nearby Galaxies: a Far-Infrared Survey with {\it Herschel} \citep[KINGFISH;][]{kf11} legacy programs, are well studied and have a wealth of ancillary data available.  
We are currently in the process of collecting complementary interferometric observations of these same targets using the VLA, which will be presented in a forthcoming paper.  
This paper is organized as follows:   
In $\S$2 we describe our sample selection and the data used in the present study.  
In $\S$3 we describe our analysis procedures.  
Our results are presented in $\S$4 and discussed in $\S$5.  
Finally, in $\S$6, we summarize our main conclusions.

\section{Sample and Data}
In this section we describe the sample selection.  
We additionally present the GBT observations and provide a description of the ancillary data utilized for the present study.  

\subsection{Sample Selection}
\label{sec-sample}
The Star Formation in Radio Survey (SFRS) sample comprises nuclear and extranuclear star-forming regions in 56 nearby galaxies ($d < 30$\,Mpc) observed as part of the SINGS \citep{rck03} and KINGFISH \citep{kf11} legacy programs.  
Each of these nuclear and extranuclear star-forming complexes have $\sim1\arcmin \times 0\farcm5$ sized mid-infrared (i.e., low resolution from $5-14\,\mu$m and high resolution from $10-37\,\mu$m) spectral mappings carried out by the IRS instrument on board {\it Spitzer}, and $47\arcsec\times47\arcsec$ sized {\it Herschel}/PACS far-infrared spectral mappings for a combination of the principal atomic ISM cooling lines of [OI]63\,$\mu$m, [OIII]88\,$\mu$m, [NII]122,205\,$\mu$m, and [CII]158\,$\mu$m.  
Two galaxies that are exceptions include NGC\,5194 and NGC\,2403; these galaxies were part of the SINGS sample, but are not formally included in KINGFISH.  
They were observed with {\it Herschel} as part of the Very Nearby Galaxy Survey (VNGS; PI: C. Wilson).  
Similarly, there are additional KINGFISH galaxies that were not part of SINGS, but have existing {\it Spitzer} data:  NGC\,5457 (M\,101), IC\,342, NGC\,3077, and NGC\,2146.  

SINGS and KINGFISH galaxies were chosen to cover the full range of integrated properties and ISM conditions found in the local Universe, spanning the full range in morphological types, a factor of $\sim10^{5}$ in infrared (IR: $8-1000\,\mu$m) luminosity, a factor of $\sim10^{3}$ in $L_{\rm IR}/L_{\rm opt}$, and a large range in star formation rate ($\lesssim10^{-3} - 10\,M_{\odot}\,{\rm yr}^{-1}$).  
Similarly, spectroscopically targeted extranuclear sources included in SINGS and KINGFISH were selected to cover the full range of physical conditions and spectral characteristics found in (bright) infrared sources in nearby galaxies, requiring optical and infrared selections.  
Optically selected extranuclear regions were chosen to span a large range in physical properties including extinction-corrected production rate of ionizing photons [$Q(H^{0}) \sim10^{49}-10^{52}$ s$^{-1}$], metallicity ($\sim 0.1-3\,Z_{\odot}$), visual extinction ($A_{V} \la 4$\,mag), radiation field intensity (100-fold range), ionizing stellar temperature ($T_{\rm eff} \sim 3.5 - 5.5\times10^{4}$\,K), and local H$_{2}$/H{\sc i} ratios ($\lesssim 0.1 - \gtrsim10$).  
A sub-sample of infrared-selected extranuclear targets were chosen to span a range in $f_{\nu}(8\,\mu{\rm m})/f_{\nu}(24\,\mu{\rm m})$ and $f_{\rm H\alpha}/f_{\nu}(8\,\mu{\rm m})$ ratios.  

The total sample over the entire sky consists of 118 star-forming complexes (56 nuclei and 62 extranuclear regions), 103 of which (46 nuclei and 57 extranuclear regions; see Tables \ref{tbl-1} and \ref{tbl-2}, respectively) are observable with the GBT (i.e., having $\delta > -30\degr$).  
Galaxy morphologies, adopted distances, and optically-defined nuclear types are given in Table \ref{tbl-1}.  
In Figure \ref{fig-gbtobs}, we show the location of each target region on {\it Spitzer} 24\,$\mu$m images; 
the corresponding circle diameters are 25\arcsec, which match the FWHM of our lowest resolution data (i.e., the beam of the GBT 33\,GHz radio data) for the present multiwavelength study.  

\input{tbl-1.tex}
\input{tbl-2.tex}

\subsection{GBT Observations and Data Reduction}
\label{sec-gbtobs}
Observations in the Ka band ($26-40$\,GHz) were taken using the Caltech Continuum Backend (CCB) on the 100\,m Robert C. Byrd Green Bank Telescope \citep[GBT;][]{jp04} over a two year period spanning 2009 March through 2011 January.  
The CCB simultaneously measures the entire Ka bandwidth over four equally spaced frequency channels (i.e., 27.75, 31.25, 34.75, and 38.25\,GHz) and synchronously reads out and demodulates the beam-switched signal to remove atmospheric fluctuation and/or gain variations. 
The FWHM of the GBT beam across the full Ka band was typically $\approx$25\arcsec among our sets of observations (see $\S$\ref{sec-gbtphot}).  
Given the range of distances to the sample galaxies, this projects to linear scales of $0.37 - 3.7$\,kpc.  
Reference beams used for sky subtraction are measured by nodding 1\farcm3 away from the source, and their positions are identified on top of {\it Spitzer} 24\,$\mu$m images in Figure \ref{fig-gbtobs}.  

Observations were made using an ``On-the-Fly Nod" variant of a symmetric nodding procedure \citep[i.e., double-differencing technique;][]{tr89} with data being collected continuously through an entire observation, including slews between beams.      
This procedure alternately places the source of interest in each of the two beams of the Ka-band receiver in an A/B/B/A pattern.  
Such a sequence is able to cancel means and gradients in the atmospheric or receiver noise with time.  
Each nodding cycle lasts $\approx$70\,s: an on-source dwell time of 10\,s for each of the 4 phases, along with 10\,s spent on the initial position acquisition, and 10\,s for the 2 slews between beams.  
Thus, $\approx$40\,s is spent on source per nodding cycle.  
A detailed description on the performance of the CCB receiver, the data reduction pipeline, and error estimates is given in \citet{bm09}.
In addition to any systematic errors, we assign a calibration error of 10\% to the flux density measurement at each frequency channel.  

Our observing strategy was constructed to make the most efficient use of the telescope.  
Thus, given the large range in brightness among our targeted regions, we varied the time spent on source based on an estimate of the expected 33\,GHz flux density using the {\it Spitzer} 24\,$\mu$m maps.  
The number of nods spent on each source was chosen to reach a S/N ratio of $\approx$10, while never spending $\la$5\,min (i.e., $\approx$4 nods) nor $\ga$70\,min (i.e., $\approx$60 nods) of total integration time on a given source.  
Naturally, not all nods were deemed usable given weather conditions and technical difficulties, and sources were revisited if possible.  
We list the number of nods used and taken, along with the corresponding time spent on source, for the nuclear and extranuclear regions in Appendix Tables \ref{tbl-A1} and \ref{tbl-A2}, respectively.  

\input{tbl-3.tex}
\input{tbl-4.tex}

\subsection{Ancillary UV, Optical, Infrared, and Radio Data}
{\it Galaxy Evolution Explorer (GALEX)} far-UV (FUV; 1528\,\AA) and near-UV (NUV; 2271\,\AA) data were taken from the {\it GALEX} archive (GR6) and will be included in the {\it GALEX} Large Galaxy Atlas (M. Seibert et al. 2012, in preparation).   
We also made use of star masks generated by the {\it GALEX} team to identify and remove foreground stars for the analysis.  
The angular resolutions of the FUV and NUV images are 4\farcs25 and 5\farcs25, respectively, while the calibration uncertainty for these data is $\approx$15\% in both bands.  

The H$\alpha$ imaging used in the analysis is taken from \citet{akl11}, where details about the data quality and preparation (e.g., correction for [NII] emission) can be found.  
Like the UV data, H$\alpha$ images were corrected for foreground stars.   
The typical resolution of the H$\alpha$ images is $\approx$2\arcsec, and the calibration uncertainty among these maps is taken to be $\approx$20\%.  

Archival {\it Spitzer} 24\,$\mu$m data were largely taken from the SINGS and Local Volume Legacy (LVL) legacy programs, and have a calibration uncertainty of $\approx$5\%.    
Details on the associated observation strategies and data reduction steps can be found in \citet{dd07} and \citet{dd09}, respectively.  
Two galaxies, IC\,342 and NGC\,2146, were not a part of SINGS or LVL; their 24\,$\mu$m imaging comes from \citet{ce08}.  

We additionally made use of {\it Herschel} 70, 100, 160, and 250\,$\mu$m data from KINGFISH \citep[see][]{kf11}.  
Observations with the PACS instrument \citep{ap10} were carried out at 70, 100, and 160\,$\mu$m in the Scan-Map mode and reduced by the Scanamorphos data reduction pipeline, version 12.5 \citep{hr12}.  
The goal of the Scanamorphos reduction is to preserve low-surface-brightness emission.  
The 250\,$\mu$m observations were made using the SPIRE instrument \citep{mg10} and reduced using the HIPE version spire-5.0.1894.
For NGC\,2403 and NGC\,5194, we make use of {\it Herschel} 70, 160, and 250\,$\mu$m imaging obtained as part of the VNGS program.  
Information on the observational strategy and data processing can be found in \citet{gjb11} for NGC\,2403 and E. Mentuch (2012, in preparation) for NGC\,5194.  

Ancillary radio data at 1.365 and 1.697\,GHz for 24 of the sample galaxies are available from the Westerbork Synthesis Radio Telescope SINGS survey \citep{rb07}.  
The intrinsic FWHM of the radio beams is approximately 11\arcsec~east-west by $11/\sin{\delta}\arcsec$ north-south at 1.5\,GHz and scales as wavelength, where $\delta$ is the source declination.  
For 5 galaxies (i.e., NGC\,628,  NGC\,3627, NGC\,4254, NGC\,4321, NGC\,4569) the north-south axis of the WSRT-SINGS beam at 1.7\,GHz is larger than median FWHM of the 33\,GHz GBT beam (including the 1\,$\sigma$ scatter), ranging between 36\farcs5 and 44\arcsec.  
Similarly, for 7 galaxies (i.e., NGC\,628,  NGC\,3627, NGC\,4254, NGC\,4321, NGC\,4569, and NGC\,4725, and NGC\,4826) the north-south axis of the WSRT-SINGS beam at 1.4\,GHz is larger than the median $+1\sigma$ scatter on the 33\,GHz GBT beam, ranging between 29\arcsec~and 55\farcs5.  
Since this will affect the accuracy of the matched photometry with the GBT 33\,GHz data, we do not include these sources in the radio spectral index analysis.  
The flux density calibration of the radio maps is better than $\approx$5\%.

\section{Photometry and Analysis}
\label{sec-phot}

In the following section we describe our procedure to produce resolution-matched photometry among all data sets.  
Special considerations when dealing with the GBT single-beam photometry are also described in detail. 

We also note that the beam of the GBT averages over significantly large physical areas given the distances of the sample galaxies (e.g., as large as $\sim$3.7\,kpc with a median linear scale of $\approx$0.9\,kpc among all nuclei and extranuclear sources targeted), and likely contains a number of non-coeval H{\sc ii} regions, photodissociation regions, and diffuse emission.  
This is similar to the physical scale of $\approx$0.8\,kpc investigated by \citet{ejm11b}, who found that conducting their analysis with and without local background subtractions did not qualitatively affect their main conclusions.  
We therefore do not attempt to subtract local background estimates, which would add more uncertainty to the present analysis.  

\subsection{GBT Observations: Additional Considerations \label{sec-gbtphot}}
Among all sets of observations, we measure a median beam width over all channels of $25\farcs07$ with a dispersion of $3\farcs1$.  
However, given that we are targeting resolved sources, and the beam size varies over the full Ka band, we apply a correction to the flux densities at each frequency channel as if their beam were at the nominal 25\arcsec.  
We use the scaling factors given in \citet{ejm10}, since the distance to NGC\,6946 is close to the median distance among the entire SFRS sample.  
These scale factors were derived by computing the photometry for NGC\,6946 on the 8.5 GHz map using the beam sizes for each frequency channel averaged over that entire observing session, and are 0.89, 0.95, 1.00, and 1.04 at 27.75, 31.25, 34.75, and 38.25 GHz, respectively.  
While the exact values of these corrections factors will change depending on source geometry/morphology, 
the final 33\,GHz photometry, averaged over the entire band, should be robust.    
The combination of resolved sources and the changing beam size across the Ka band will complicate the interpretation of in-band radio spectral indices, and is therefore not presented.  

While the unblocked aperture and active surface of the GBT mitigate the significance of sidelobes, we inspected calibration scans from each observing session to determine if there were instances of days exhibiting sidelobes with large amplitudes.  
The added power may cause an overestimation of the 33 GHz flux density. 
We find that sidelobes, when measurable, have an amplitude that is $\la$2\% of the beam peak, on average, over the entire 2 year observing campaign.  
At this level, our results should not be significantly affected by the presence of sidelobes given that \citet{ejm10} found negligible differences in their ancillary radio, submillimeter, and infrared photometry using a model beam having sidelobes at 5\% of the beam peak.  

The averaged Ka-band flux densities, weighted by the errors from each channel over the full band, are given in Tables \ref{tbl-1} and \ref{tbl-2} along with uncertainties;
the corresponding effective frequency 
is given for each source in Appendix Tables \ref{tbl-A1} and \ref{tbl-A2}, and is $\approx$33\,GHz for each position 
(i.e., a median of 32.92\,GHz with a dispersion of 0.79\,GHz, corresponding to a $\approx$2\% scatter in flux densities assuming a spectral index near 33\,GHz of $\sim$0.5). 
For sources not detected at the 3\,$\sigma$ level, we list the 3\,$\sigma$ upper limit.  
Flux densities of the individual channels, {\it before applying any corrections}, are given with 1\,$\sigma$ errors in the Appendix Tables \ref{tbl-A1} and \ref{tbl-A2} for the galaxy nuclei and extranuclear regions, respectively.  
We do not give upper limits for the individual channels, but rather list the actual measured values and estimated errors.  

Since the reference beam throw is only 1\farcm3, real signal in our reference positions is a concern; 
having reference positions on the galaxy may result in an underestimation of the 33 GHz flux density (see Figure \ref{fig-gbtobs}). 
For NGC\,6946, we note that the positions presented here correct those originally given in \citet{ejm10}, however this slight correction does not impact their results.  
To quantitatively assess the severity of having real signal in the reference beam for our measurement, we estimate over-subtractions using the 24\,$\mu$m data by comparing the flux densities measured at the reference positions compared to the on-source position.  
To correct for these losses, we add back to the 33 GHz flux densities the average flux densities measured among the reference positions using modeled 33 GHz maps by scaling the 24$\mu$m data using Equation 2 of \citet{ejm06b}. 
The minimum flux density of the reference beam positions for each target was then taken as the local sky and subtracted. 

These corrections are typically small, having a median value of $\approx$2\% among all sources.  
We conservatively apply corrections to only those 22 sources estimated to be missing $>$15\% of the actual flux density based on the 24\,$\mu$m photometry alone.    
The median correction to the 33\,GHz flux density for these sources is $\approx$24\%.  
The uncorrected 33\,GHz flux densities are given in the final columns of Tables \ref{tbl-1} and \ref{tbl-2}.   
We note that for the case of NGC\,6946, using 8.5 GHz-derived 33\,GHz maps resulted in corrections that agreed to those based on the 24\,$\mu$m-derived values to within a few percent \citep{ejm10}.  

\subsection{Ancillary Data}
\label{sec-ancphot}
The photometry was carried out on the ancillary data sets after matching their resolution, cropping each image to a common field-of-view, and re-gridding to a common pixel scale. 
To accurately match the photometry from these images to the GBT measurements, maps were convolved to the resolution of the GBT beam in the Ka band (i.e., 25\arcsec) following the image registration method of \citet{ga11}.  
Using the combination of all infrared data, total infrared (IR; $8-1000\,\mu$m) emission and uncertainty maps were constructed using the models of \citet{dl07} as described in G. Aniano (2012, submitted).  

For each data set, the flux density at the position of each GBT pointing is the surface brightness multiplied by the beam solid angle.  
Uncertainties on the photometry are estimated as a combination of the calibration and beam-size uncertainties; 
we assign a 17\% uncertainty to account for the dispersion in the beam areas among our sets of GBT observations as given in \S\ref{sec-gbtphot}.  

In the case of the UV and H$\alpha$ photometry, we correct each region for Milky Way extinction using \citet{ds98} assuming $A_{V}/E(B-V)=3.1$ and the modeled extinction curves of \citet{wd01,bd03}.  
The multi-wavelength photometry is given in Tables \ref{tbl-3} and \ref{tbl-4}, along with the corresponding 1-$\sigma$ errors, for the nuclei and extranuclear regions, respectively.

\subsection{Star Formation Rate Calibrations}
\label{sec-sfr}
For the non-AGN sources in the sample, we can estimate star formation rates using the new GBT 33\,GHz data.  
In this section we present the calibrations given in \citet{ejm11b}, where more details about their derivations can be found.      
These calibrations, which update those found in \citet{rck98} and have been adopted by \citet{ke12}, were calculated using Starburst99 \citep[][]{cl99} for a common IMF so that each diagnostic can be compared fairly.  
A summary of these relations can be found in the Appendix Table \ref{tbl-A3}.   

We choose a Kroupa \citep{pk01} IMF, having a slope of $-1.3$ for stellar masses between $0.1-0.5~M_{\sun}$ and $-2.3$ for stellar masses ranging between $0.5-100\,M_{\sun}$.  
Assuming solar metallicity and a continuous, constant star formation rate over $\sim$100\,Myr, Starburst99 stellar population models yield the following relation between the star formation rate and production rate of ionizing photons, $Q(H^{0})$:   
\begin{equation}
\label{eq-Nly}
\left(\frac{\rm SFR}{M_{\sun}\,{\rm yr^{-1}}}\right) = 7.29\times10^{-54}\left[\frac{Q(H^{0})}{\rm s^{-1}}\right].  
\end{equation}
Since the ionizing flux comes from very massive stars with lifetimes $\la$10~Myr, we note that the coefficient in Equation \ref{eq-Nly} is nearly independent of starburst age under the assumption of continuous star formation so long as it is $\ga$10~Myr.  
Accordingly, such measurements sample the current (i.e., $\sim$10~Myr) star formation activity.  
The integrated UV spectrum is also dominated by young stars, however, it is sensitive to a significantly longer timescale of recent \citep[$\sim$$10-100$~Myr;][]{rck98,dc05,ss07} star formation activity.  
We convolve the output Starburst99 spectrum with the {\it GALEX} FUV transmission curve to obtain the following conversion between star formation rate and FUV luminosity, 
\begin{equation}
\label{eq-sfrfuv}
\left(\frac{\rm SFR_{FUV}}{M_{\sun}\,{\rm yr^{-1}}}\right) = 4.42\times10^{-44}\left(\frac{L_{\rm FUV}}{\rm erg~s^{-1}}\right).
\end{equation}
Similarly, we can write such an expression for the {\it GALEX} NUV band such that, 
\begin{equation}
\label{eq-sfrnuv}
\left(\frac{\rm SFR_{NUV}}{M_{\sun}\,{\rm yr^{-1}}}\right) = 7.15\times10^{-44}\left(\frac{L_{\rm NUV}}{\rm erg~s^{-1}}\right).
\end{equation}

The ionizing photon rate can of course be expressed as a (extinction corrected) H~recombination line flux, such that for Case B recombination, and assuming an electron temperature $T_{\rm e} = 10^{4}$~K, the H$\alpha$ recombination line strength is related to the star formation rate by 
\begin{equation}
\label{eq-sfrha}
\left(\frac{\rm SFR_{H\alpha}}{M_{\sun}~{\rm yr^{-1}}}\right) = 5.37\times10^{-42}\left(\frac{L_{{\rm H}\alpha}}{\rm erg~s^{-1}}\right).  
\end{equation}
The above equation indicates that the star formation rate is directly proportional to the H$\alpha$ line luminosity, assuming that a constant fraction (i.e., $\approx$45\%) of the ionized H~atoms will emit an H$\alpha$ photon as they recombine, and that the extinction correction is accurate.
However, if a significant fraction of ionizing photons are absorbed by dust, the above equation will underestimate the star formation rate, and the $=$ sign should be replaced by $\geq$.  

Similarly, at high radio frequencies, where $\tau \ll 1$, the ionizing photon rate is directly proportional to the thermal spectral luminosity, $L_{\nu}^{\rm T}$, varying only weakly with electron temperature $T_{\rm e}$ \citep{rr68}, such that 
\begin{equation}
\label{eq-Nlyrad}
\begin{split}
\left[\frac{Q(H^{0})}{s^{-1}}\right] &= 6.3\times10^{25} \\ 
&\left(\frac{T_{\rm e}}{10^{4}\,{\rm K}}\right)^{-0.45} \left(\frac{\nu}{\rm GHz}\right)^{0.1} \left(\frac{L_{\nu}^{\rm T}}{\rm erg~s^{-1}~Hz^{-1}}\right).  
\end{split}
\end{equation}
By combining Equations \ref{eq-Nly} and \ref{eq-Nlyrad}, one can derive a relation between the star formation rate and  thermal radio emission: 
\begin{equation}
\label{eq-sfrt}
\begin{split}
\left(\frac{\rm SFR_{\nu}^{T}}{M_{\sun}\,{\rm yr^{-1}}}\right) &= 4.6\times10^{-28}\\
&\left(\frac{T_{\rm e}}{10^{4}\,{\rm K}}\right)^{-0.45} \left(\frac{\nu}{\rm GHz}\right)^{0.1} \left(\frac{L_{\nu}^{\rm T}}{\rm erg~s^{-1}~Hz^{-1}}\right).  
\end{split}
\end{equation}
As with the H~recombination line fluxes, $Q(H^{0})$, and consequently the star formation rate, may in fact be underestimated by the free-free emission if  a significant fraction of ionizing photons are absorbed by dust; in this case the $=$ sign in the above equation should be replaced by $\geq$.   
However, it is worth noting that unlike free-free emission, which arises directly from the ionized gas, optical/NIR H~recombination line fluxes may also suffer extinction internal to the H{\sc ii} region itself, resulting in an even larger deficit. 
For example, the extinction {\it internal} to H{\sc ii} regions in the starbursting dwarf galaxy NGC \,5253 is measured to be quite large \citep[$A_{V} = 16-18$\,mag;][]{jt03}.

At lower radio frequencies, which are typically dominated by non-thermal synchrotron emission, calibrations between the supernova rate, and thus the star formation rate, have been developed.  
From the output of Starburst99, which assumed a supernova cut-off mass of 8\,$M_{\sun}$, we find that the total core-collapse supernova rate, $\dot{N}_{\rm SN}$, is related to the star formation rate by,   
\begin{equation}
\label{eq-sfrqsnr}
\left(\frac{\rm SFR}{M_{\sun}\,{\rm yr^{-1}}}\right) =  86.3 \left(\frac{\dot{N}_{\rm SN}}{\rm~yr^{-1}}\right).
\end{equation}
Work comparing the non-thermal spectral luminosity with the supernova rate in the Galaxy has yielded an empirical calibration such that
\begin{equation}
\label{eq-lntqsnr}
\left(\frac{L_{\nu}^{\rm NT}}{\rm erg~s^{-1}~Hz^{-1}}\right) = 1.3\times10^{30} \left(\frac{\dot{N}_{\rm SN}}{\rm yr^{-1}}\right) \left(\frac{\nu}{\rm GHz}\right)^{-\alpha^{\rm NT}}, 
\end{equation}
where $\alpha^{\rm NT}$ is the non-thermal radio spectral index \citep{gtam82,cy90}.
By combining Equations \ref{eq-sfrqsnr} and \ref{eq-lntqsnr}, we can express the star formation rate as a function of the non-thermal radio emission where 
\begin{equation}
\label{eq-sfrnt}
\left(\frac{\rm SFR_{\nu}^{NT}}{M_{\sun}\,{\rm yr^{-1}}}\right) =6.64\times10^{-29}  \left(\frac{\nu}{\rm GHz}\right)^{\alpha^{\rm NT}}\left(\frac{L_{\nu}^{\rm NT}}{\rm erg~s^{-1}~Hz^{-1}}\right).
\end{equation}
Since it will take $\sim$30\,Myr for 8\,$M_{\sun}$ stars to go supernova, and the radiating lifetimes of CR electrons can be on the order of tens of Myr in normal galaxies, the non-thermal emission is sensitive to slightly longer timescales than free-free or H~recombination line emission.  

The observed radio continuum emission comprises both free-free and synchrotron emission.  
Therefore, we can combine Equations \ref{eq-sfrt} and \ref{eq-sfrnt} to construct a single expression for the star formation rate from the total radio continuum emission at a given frequency such that: 
\begin{equation}
\label{eq-sfrrad}
\begin{split}
\left(\frac{\rm SFR_{\nu}}{M_{\sun}\,{\rm yr^{-1}}}\right) &= 10^{-27}  
\left[2.18 \left(\frac{T_{\rm e}}{10^{4}\,{\rm K}}\right)^{0.45} \left(\frac{\nu}{\rm GHz}\right)^{-0.1}\right. + \\
&\left.15.1 \left(\frac{\nu}{\rm GHz}\right)^{-\alpha^{\rm NT}}\right]^{-1} \left(\frac{L_{\nu}}{\rm erg~s^{-1}~Hz^{-1}}\right).  
\end{split}
\end{equation}
This equation essentially weights the observed radio continuum luminosity based on the expected thermal fraction at a given frequency.  
As pointed out above, the thermal and non-thermal emission timescales are mismatched, with free-free emission being sensitive to massive stars with ages $\la$10\,Myr and the non-thermal emission being most sensitive to the lowest mass (i.e., $\ga$8\,$M_{\sun}$) supernova progenitors having lifetimes of $\la$30\,Myr.   
However, Equation \ref{eq-sfrrad} should hold under the assumption of continuous star formation on timescales $\ga$30\,Myr.  

In our analysis we have assumed an electron temperature of $T_{\rm e} = 10^{4}$\,K when calculating star formation rates using Equations \ref{eq-sfrt} and \ref{eq-sfrrad}.  
However, variations in the actual electron temperatures will inject scatter into our observed trends.  
For instance, assuming a value of $T_{\rm e}$ as low as 5000\,K will result in star formation rates that are 37 and 21\% larger than assuming $T_{\rm e} = 10^{4}$\,K using equations \ref{eq-sfrt} and \ref{eq-sfrrad}, respectively.  
Thus, we expect any scatter introduced by the assumption of a constant electron temperature to be smaller than these values.  

\setcounter{figure}{1}
\begin{figure}[tc]
\begin{center}
\epsscale{1.1}
\plotone{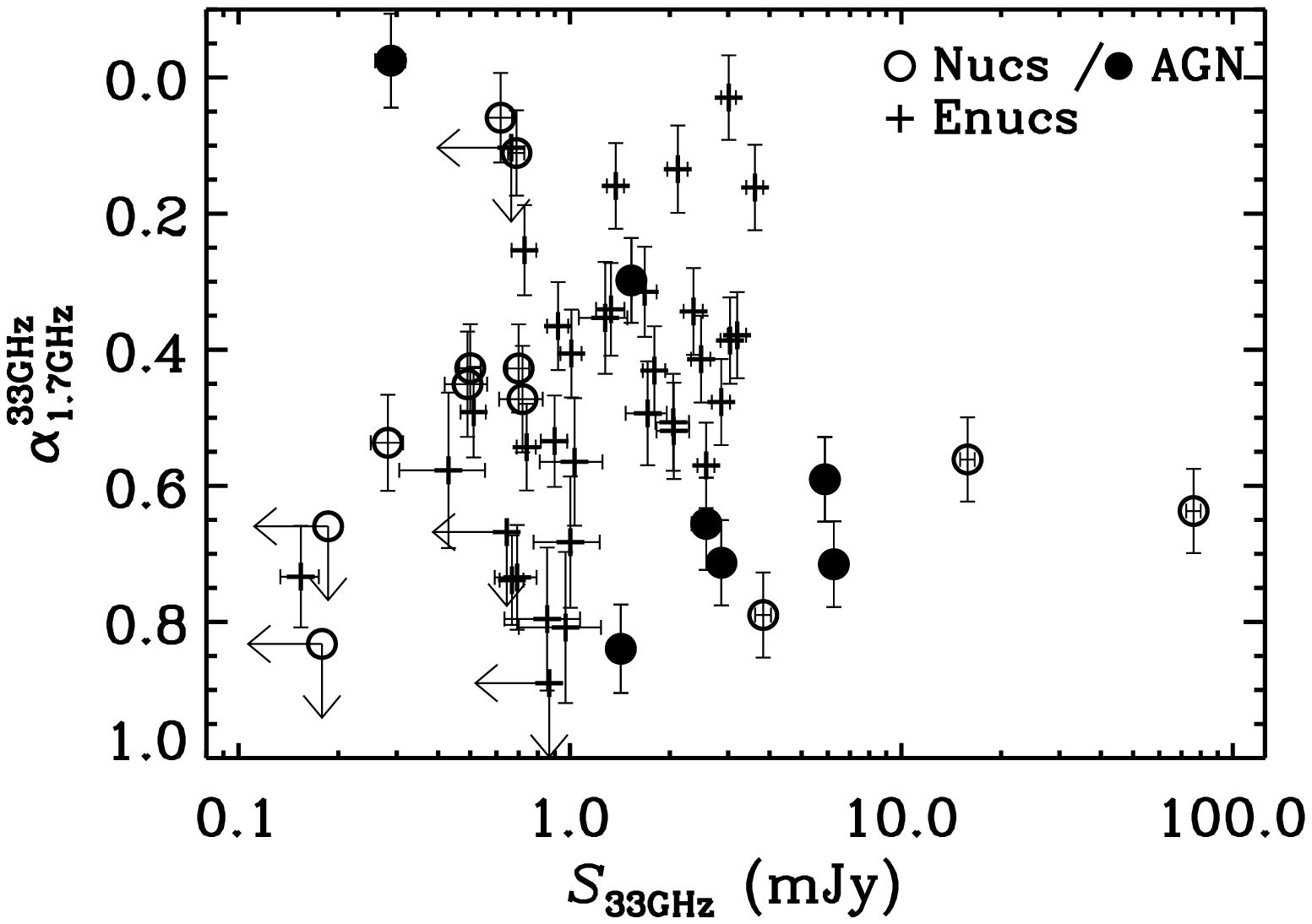}
\end{center}
\caption{
Radio spectral indices measured between 1.7 and 33\,GHz plotted against 33\,GHz flux densities for all 53 galaxy nuclei (circles) and extranuclear star-forming regions (crosses) having 1.7\,GHz measurements.    
Nuclei identified as AGN in Table \ref{tbl-1} are plotted using filled symbols.  
Upper limits for the 33\,GHz flux densities, and corresponding spectral index measurements, are shown as arrows.  
Sources for which the major axis of the 1.7\,GHz data was larger than the FWHM of the GBT beam at 33\,GHz are not included due to imprecise matched photometry.  
}
\label{fig-spx-s33}
\end{figure}

\begin{figure}[tc]
\begin{center}
\epsscale{1.1}
\plotone{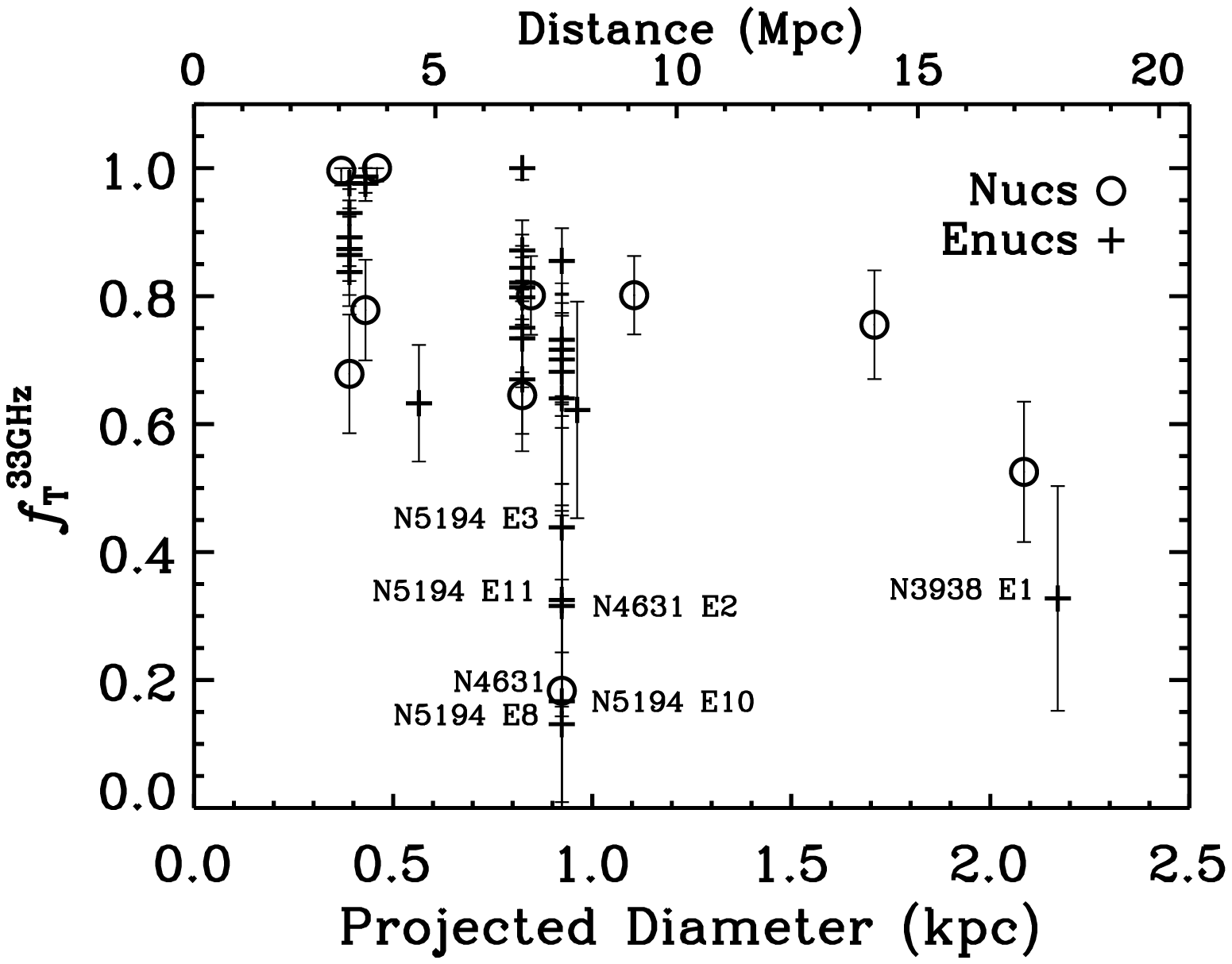}
\end{center}
\caption{
Estimates for the thermal fraction at 33\,GHz plotted against the projected diameter of the GBT beam 
for all (i.e., 41) non-AGN sources with 33\,GHz detections and corresponding 1.7\,GHz data.  
Nuclei and extranuclear regions are shown by circles and crosses, respectively.  
Only a weak trend is found with distance, suggesting that the 33\,GHz star formation rates are not significantly affected by averaging over larger physical areas, which should increase the amount of non-thermal emission at 33\,GHz, for the more distant sources.  
The seven sources having 33\,GHz thermal fractions $<$50\% are identified, five of which required a correction for badly placed off-nod positions.  
These are the same seven sources having star formation rates discrepant by more than a factor of $>$1.5 in Figure \ref{fig-sfr33-comp} (see \S\ref{sec-spx}).  
}
\label{fig-tfrac-dist}
\end{figure}

\begin{figure}[tc]
\begin{center}
\epsscale{1.1}
\plotone{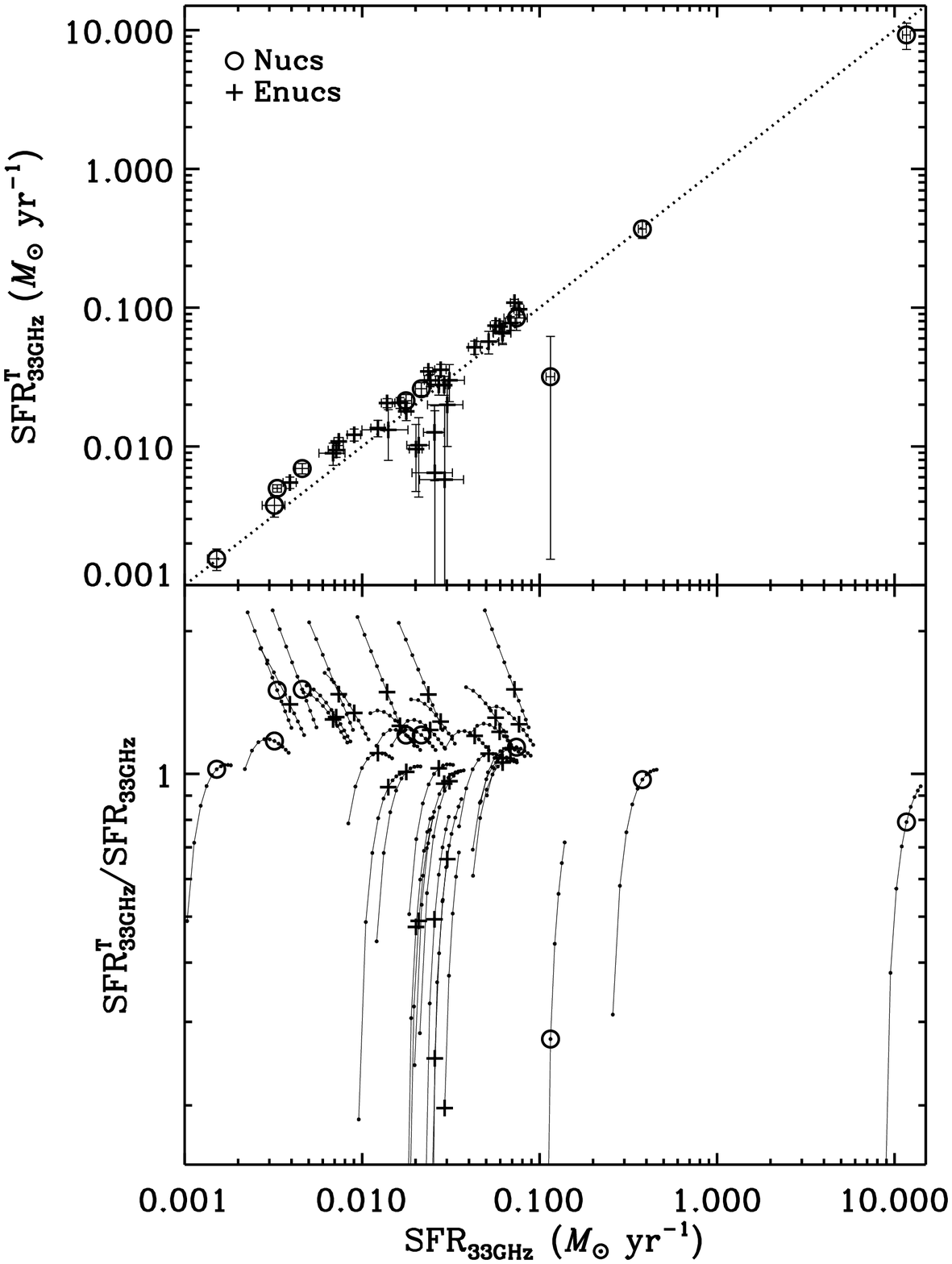}
\end{center}
\caption{
{\it Top:} The 33\,GHz star formation rates calculated in two ways for non-AGN sources having lower frequency radio data, allowing for an estimate for the 33\,GHz thermal fraction.  
Star formation rates calculated using the thermal emission at 33\,GHz and Equation \ref{eq-sfrt} are given along the ordinate while star formation rates calculated using the total 33\,GHz spectral luminosity along with Equation \ref{eq-sfrrad}, assuming a non-thermal radio spectral index of $\alpha^{\rm NT} = -0.85$ are given along the abscissa.  
Also shown is a one-to-one (dotted) line.  
The seven sources having star formation rates discrepant by more than a factor of $>$1.5 are the same seven sources identified in Figure \ref{fig-tfrac-dist} (see \S\ref{sec-spx}).  
{\it Bottom:} The ratio of the two 33\,GHz star formation rate estimates, illustrating the effect of assuming different values for the non-thermal radio spectral index, ranging between $0.6 - 1.05$ (left to right) in increments of 0.05.  
The open circles assume $\alpha^{\rm NT} = -0.85$, as in the above comparison.  
}
\label{fig-sfr33-comp}
\end{figure}

\begin{figure}[tc]
\begin{center}
\epsscale{1.1}
\plotone{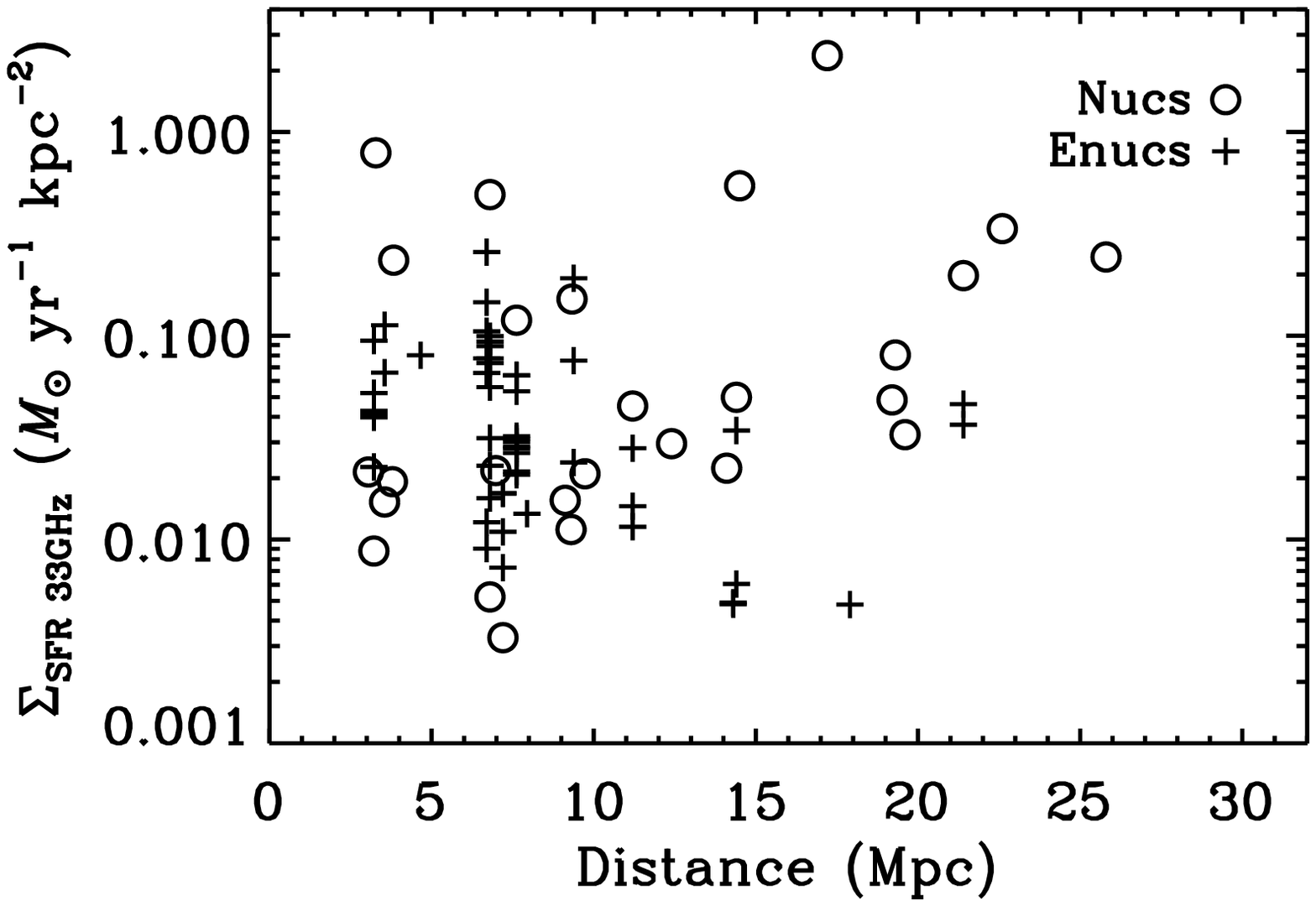}
\end{center}
\caption{
The 33\,GHz star formation rates normalized by the projected area of the GBT beam (i.e, spanning $\approx0.15-11\,{\rm kpc}^{2}$) plotted against distance for all (i.e., 81) non-AGN detected sources in the sample.  
Nuclei and extranuclear regions are shown by circles and crosses, respectively.  
Only a weak trend is found with distance, suggesting that the 33\,GHz star formation rates are not significantly affected by averaging over larger physical areas, which should increase the amount of non-thermal emission at 33\,GHz, for the more distant sources.  
Error bars are not shown since they are typically smaller than the plotting symbols.  
}
\label{fig-sfrd33-dist}
\end{figure}

\begin{figure}[tc]
\begin{center}
\epsscale{1.1}
\plotone{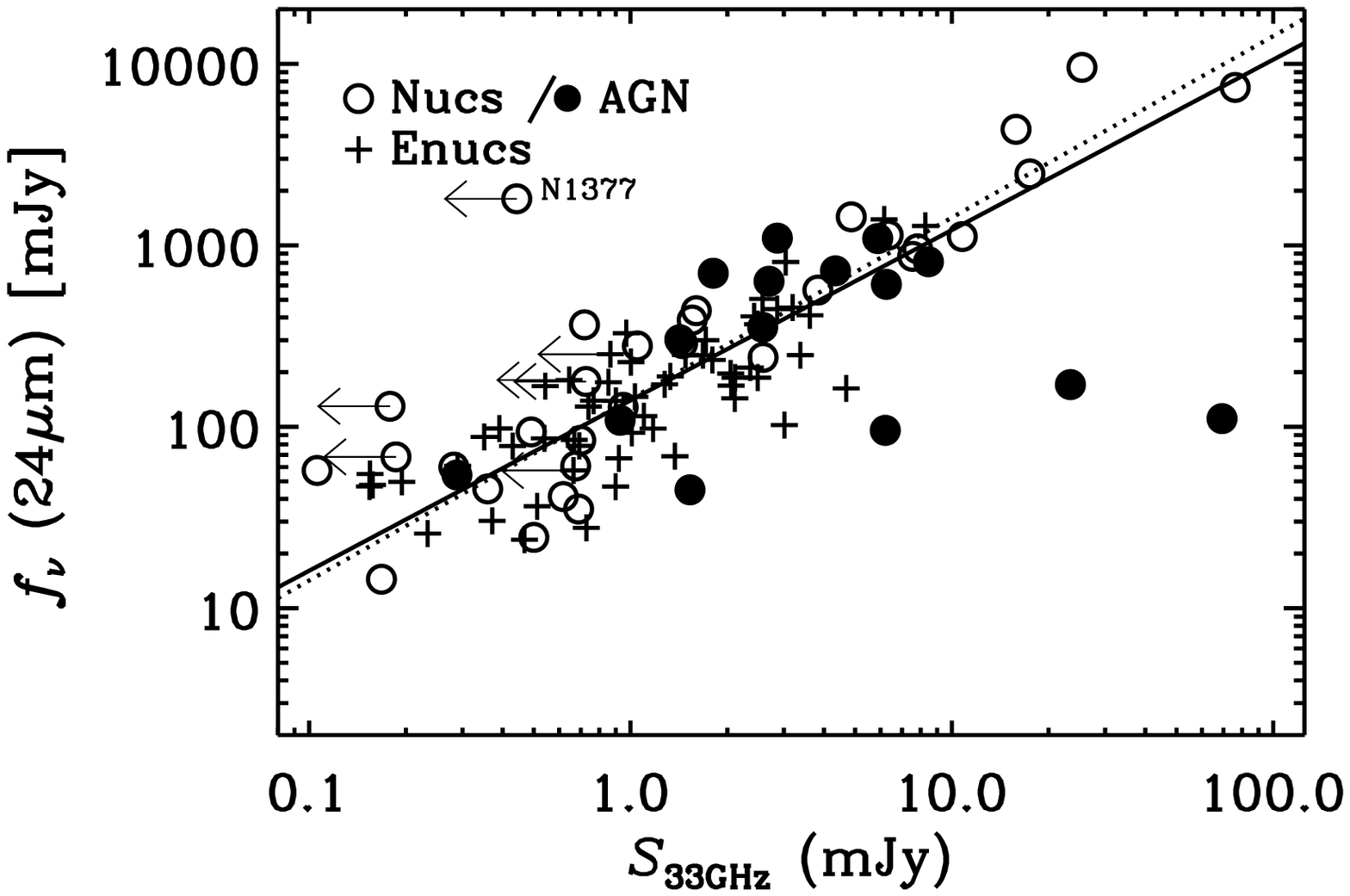}
\end{center}
\caption{
The 24\,$\mu$m flux densities plotted against corresponding (corrected) 33\,GHz flux densities for the entire sample of 103 galaxy nuclei (circles) and extranuclear star-forming regions (crosses).  
Nuclei identified as AGN in Table \ref{tbl-1} are indicated by filled symbols.  
Upper limits for the 33\,GHz flux densities are shown as arrows.  
Error bars are not shown since they are typically smaller than the plotting symbols.  
Over-plotted is an ordinary least squares fit to the detected extranuclear regions and non-AGN nuclei (solid line), along with a one-to-one line  scaled by the median 24\,$\mu$m to 33\,GHz flux density ratio $<f_{\nu}(24\,\mu{\rm m}) / S_{\rm 33\,GHz}> = 142$ (dotted line).  
A number of AGN are clearly radio loud at 33\,GHz relative to the main correlation between the 24\,$\mu$m and 33\,GHz flux densities.  
The galaxy furthest above the fit to the main trend, which is detected at 24\,$\mu$m but not at 33\,GHz, is NGC\,1377.  
This galaxy is identified as being possibly a nascent starburst \citep[e.g.,][]{hr03,hr06}. 
}
\label{fig-f24-s33}
\end{figure}

\begin{figure}[tc]
\begin{center}
\epsscale{1.1}
\plotone{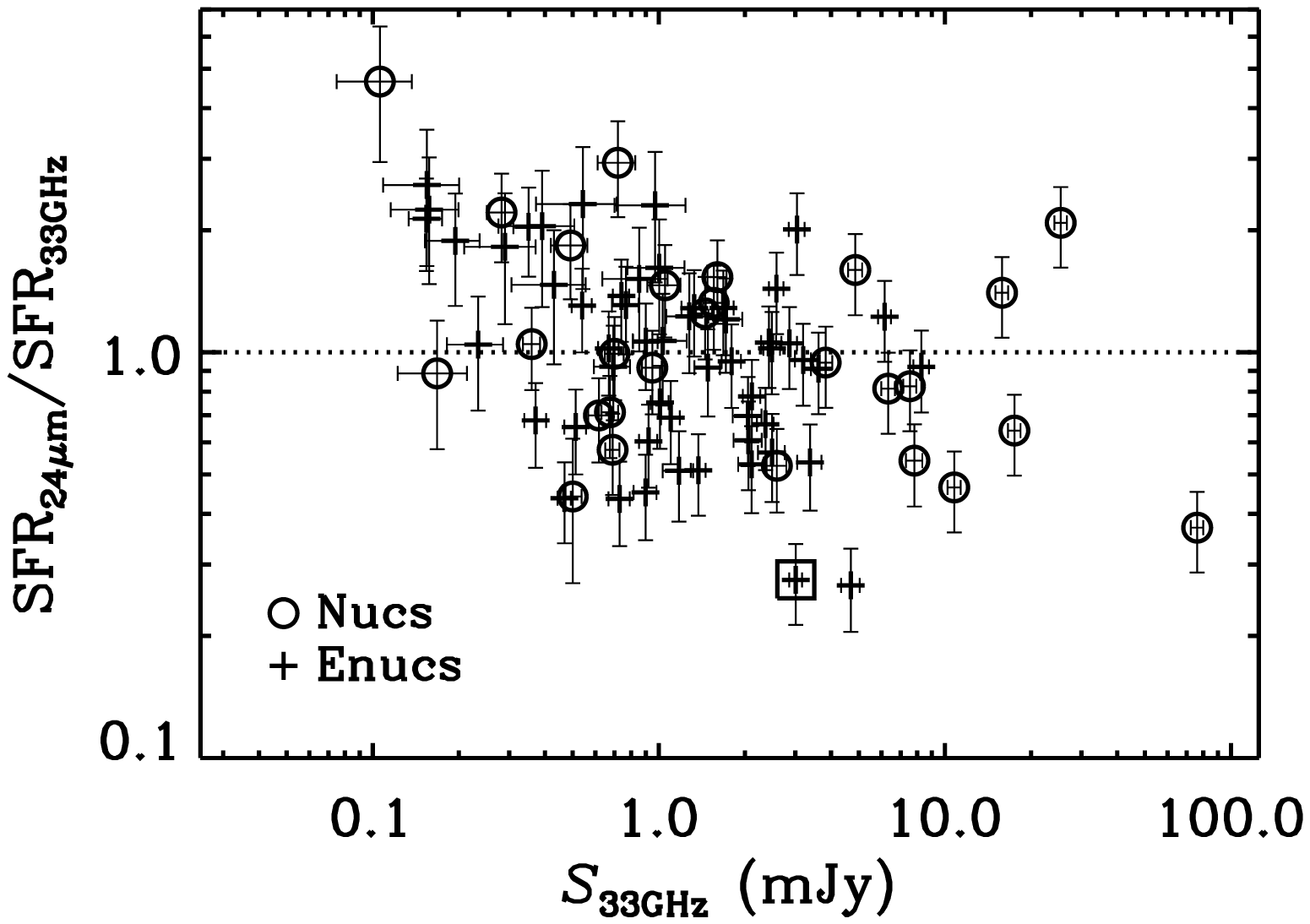}
\end{center}
\caption{
The ratio of 24\,$\mu$m to 33\,GHz star formation rates plotted against the 33\,GHz star formation rates for all (i.e., 81) non-AGN detected sources in the sample.  
Nuclei and extranuclear regions are shown by circles and crosses, respectively.  
The scatter about unity is $\approx$0.25\,dex.  
Extranuclear region 4 in NGC\,6946 is identified using a square, as the 33\,GHz flux density of this sources is complicated by the presence of anomalous dust emission \citep{ejm10,as10}.  
}
\label{fig-sfr24-sfr33}
\end{figure}

\section{Results}
In the following section we estimate the fraction of thermal emission at 33\,GHz and compare a number of star formation rate diagnostics with those derived using the GBT 33\,GHz data.  
The assumption is made that the 33\,GHz data provide the most reliable estimate as they are highly sensitive to free-free emission arising from the ionized gas in H{\sc ii} regions.  
While we adopt the calibrations given above \citep[i.e.,][]{ejm11b}, we derive our own empirical relations and compare with others found in the literature.

\subsection{Radio Spectral Indices and Estimates for the Thermal Fraction at 33\,GHz}
\label{sec-spx}
In Figure \ref{fig-spx-s33} the radio spectral indices measured between 1.7 and 33\,GHz are shown for all 53 sources having data at 1.7\,GHz with a resolution that is equal to, or higher than, the GBT data.   
AGN are identified among the galaxy nuclei, typically having radio spectral indices that are steeper than the rest of the sources.  
Using the spectral index information over this rather long lever-arm, we can estimate the thermal emission fraction at 33\,GHz if we assume a fixed non-thermal spectral index.  
We assume a constant non-thermal radio spectral index of $\alpha^{\rm NT} = 0.85$ given that this was the average non-thermal spectral index found among the 10 star-forming regions studied in NGC\,6946 by \citet{ejm11b}.  
Furthermore, this is very close to the value found by \citet[][i.e., $ \alpha^{\rm NT} = 0.83$ with a scatter of $ \sigma_{\alpha^{\rm NT}} = 0.13$]{nkw97} 
for a sample of 74 nearby galaxies.  
We do not estimate the thermal fractions for galaxy nuclei containing AGN since there is likely to be more variation in their intrinsic non-thermal slope.  
Following \citet{kwb84}, and assuming that the free-free emission is optically thin, we estimate the thermal fraction such that
\begin{equation}
\label{eq-tfrac}
f_{\rm T}^{\nu_{1}} = \frac{\left(\frac{\nu_{2}}{\nu_{1}}\right)^{-\alpha} - \left(\frac{\nu_{2}}{\nu_{1}}\right)^{-\alpha^{\rm NT}}}
	{\left(\frac{\nu_{2}}{\nu_{1}}\right)^{-0.1} - \left(\frac{\nu_{2}}{\nu_{1}}\right)^{-\alpha^{\rm NT}}}, 
\end{equation}
where, for our specific case, $\nu_{1} = 33$\,GHz, $\nu_{2} = 1.7$\,GHz, $\alpha$ is the observed radio spectral index between $\nu_{1}$ and $\nu_{2}$, and we fix the thermal radio spectral index to $0.1$.  
We find a median 33\,GHz thermal fraction of $\approx$76\% among all sources, with a lower and upper quartile of 64 and 87\%, respectively   
(see Figure \ref{fig-tfrac-dist}).  
While this method has been shown to overestimate thermal fractions for star-forming regions in a high-resolution radio study of M\,33 \citep{fat07b}, those authors assumed a much steeper non-thermal index (i.e., $\alpha^{\rm NT} = 1.0$) than adopted here.  
Non-thermal spectral indices have been found to be significantly flatter than 1.0 around giant H{\sc ii} regions based on comparisons between multifrequency radio and de-reddened H$\alpha$ observations in NGC\,6946 (F. Tabatabaei et al. 2012, in preparation).  
Assuming $\alpha^{\rm NT} = 1.0$ would increase our thermal fraction estimates to $<f_{\rm T}^{\rm 33\,GHz}> \approx 85$ with a dispersion of 15\%.  

In Figure \ref{fig-sfr33-comp} we plot star formation rates for all non-AGN sources having low frequency radio data, allowing us to estimate the 33\,GHz thermal fraction, in two ways.  
We first compute 33\,GHz star formation rates using only the thermal fraction at 33\,GHz and Equation \ref{eq-sfrt}.  
In the top panel of \ref{fig-sfr33-comp}, these are plotted against 33\,GHz star formation rates using the total 33\,GHz spectral luminosity with Equation \ref{eq-sfrrad}, again assuming a non-thermal spectral index of $\alpha^{\rm NT} = 0.85$.  
There are 6 sources (i.e, NGC\,3938~Enuc.\,1, NGC\,4631, NGC\,4631~Enuc.\,2, NGC\,5194~Enuc.\,8, NGC\,5194~Enuc.\,10, and NGC\,5194~Enuc.\,11) with star formation rate estimates that are discrepant by more than a factor of $\sim$2.  
This discrepancy may arise from our assumption for the non-thermal spectral index being too flat; 
these are the only sources shown in Figure \ref{fig-spx-s33} for which $\alpha^{\rm 33GHz}_{\rm 1.7GHz} > 0.70$, and are also the only sources in Figure \ref{fig-tfrac-dist} having thermal fractions $<$40\%.  
For the two observations towards NGC\,4631, we might expect a steeper spectral index as the galaxy is observed to be edge-on, and has a very prominent non-thermal radio component.  
Observing through the disk of this source could lead to a significant amount of diffuse non-thermal emission in the beam, steepening the observed spectrum.  
However, for the remaining four sources, NGC\,3938~Enuc.\,1, NGC\,5194~Enuc.\,10, and NGC\,5194~Enuc.\,11, it is worth noting that the 33\,GHz photometry required a correction for badly placed off-nod positions; 
underestimating the corrections will result in artificially steep spectral index estimates.  
In the top panel of Figure \ref{fig-sfr33-comp} we also plot a one-to-one line, showing that these two estimates are generally consistent with one another.  
The dispersion about an ordinary least squares fit line is less than $\approx$0.05\,dex (i.e., 12\%).  

In the bottom panel of Figure \ref{fig-sfr33-comp}, we plot the ratio of these two star-formation rate estimates, illustrating the effect for different assumptions of $\alpha^{\rm NT}$.  
We allow $\alpha^{\rm NT}$ to range between $0.6-1.05$ (left to right) in increments of 0.05.  
Except for the six discrepant sources identified above, the two estimates appear generally consistent within errors when varying the assumption for the non-thermal radio spectral index.  
Given the extremely good agreement between these two estimates, which may not be terribly surprising given that the 33\,GHz flux densities are generally dominated by thermal emission, star formation rates estimated using Equation \ref{eq-sfrrad} are assumed as our reference value for comparisons throughout the paper.  
By using Equation \ref{eq-sfrrad}, we are able to calculate star formation rates for the entire sample (i.e., lower frequency radio data are not necessarily required).  
Before moving on, it is worth investigating how the variations in the physical size of the GBT beam may affect the star formation rate estimates given that Equation \ref{eq-sfrrad} does not take into account morphological differences between the (compact) free-free and (diffuse) non-thermal emission in galaxies.  

\subsubsection{Distance Effects}
The distances to the galaxies being investigated vary by a factor of $\sim$10.  
Consequently, the physical areas our observations average over span a factor of $\sim$100.  
By averaging over larger pieces of each galaxy, more diffuse emission is introduced into each beam relative to additional emission from  compact H{\sc ii} regions and, for the 33\,GHz observations, likely introduces more non-thermal emission.  
To investigate the consequence of this effect, we plot the 33\,GHz thermal fraction estimates as a function of the projected diameter of the region being measured in Figure \ref{fig-tfrac-dist}.  
We find large (i.e., $\approx 93^{+7}_{-10}$\,\%) thermal fractions for all sources having sizes $\la$0.5\,kpc, and a larger mix of thermal fractions for larger diameters.  
Excluding the 6 sources having thermal fractions $\la$40\%, as they may be artificially low (see above), there does appear to be a weak trend of decreasing thermal fraction with increasing projected area.   
However, it is hard to reliably quantify any trend given that there are so few sources with lower frequency radio data at the large distance end of the sample.  

To look at the effect of distance in another way, we plot the 33\,GHz star formation rates, normalized by the projected area of the beam, versus distance in Figure \ref{fig-sfrd33-dist} for all detected, non-AGN sources.  
We calculate star formation rates using Equation \ref{eq-sfrrad}, with $\alpha^{\rm NT} = 0.85$, since this expression takes both thermal and non-thermal emission components into account.  
If the star formation rate calibration is being affected by the inclusion of a substantial amount of non-thermal emission for the more distant galaxies, we would expect a trend of increasing star formation rate per unit area as the beam size increases.  
{Perhaps there is a weak trend; the median star formation rate surface density at distances below and above 15\,Mpc is $\approx$0.03 and $\approx$0.08\,$M_{\sun}\,{\rm yr^{-1}\,kpc}^{2}$, respectively.   
However, the lack of a meaningful number of sources, especially extranuclear regions, at $d > 15$\,Mpc, makes it difficult to quantify any trend.  
Thus, is it appears that larger beams do not significantly affect our 33\,GHz star formation rate estimates.  }

\begin{figure}[tc]
\begin{center}
\epsscale{1.1}
\plotone{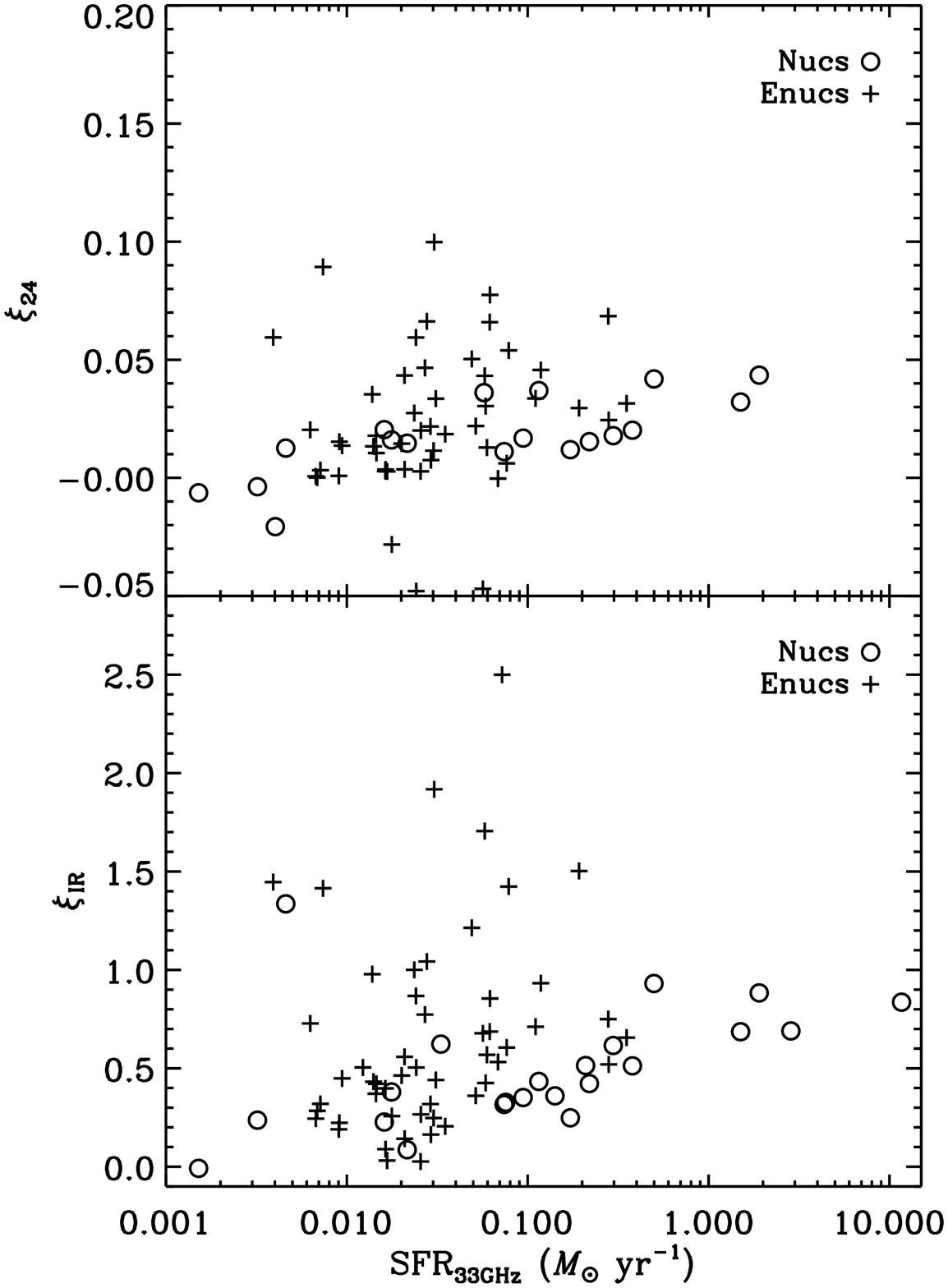}
\end{center}
\caption{
{\it Top:} The 24\,$\mu$m scaling factor for computing H$\alpha + 24\,\mu$m star formation rates plotted against the 33\,GHz star formation rates for all non-AGN detected regions.  
{\it Bottom:} The IR scaling factor for computing FUV$+$IR star formation rates plotted against the 33\,GHz star formation rates for all non-AGN detected regions.  
}
\label{fig-scl24-sclir}
\end{figure}

\subsection{Comparison with 24\,$\mu$m Warm Dust Emission}
\label{sec-sfr24}
Unlike the far-infrared continuum emission, for which a non-negligible fraction may be powered by an older stellar population \citep[e.g.,][]{st92,wg96,gjb10}, warm dust emission appears to be more tightly correlated to the current star formation activity in the disks of galaxies \citep[e.g.,][]{gxh04,ejm06a}.  
Consequently, a number of calibrations relating the warm dust emission at 24\,$\mu$m to star formation rates have been introduced in the literature \citep[e.g.,][]{hwu05, ppg06,aah06,dc07,mr07,ynz08,gr09}; a detailed comparison among each of these relations can be found in \citet{dc10}.  

In Figure \ref{fig-f24-s33} we plot 24\,$\mu$m flux densities against corresponding (corrected) 33\,GHz flux densities for the entire sample of 103 galaxy nuclei and extranuclear regions.  
Apart from the galaxy nuclei that are known to harbor AGN, a number of which are radio loud at 33\,GHz relative to the observed 24\,$\mu$m emission, there is a fairly tight correlation between the 24\,$\mu$m and 33\,GHz flux densities.  
The only other clearly outlying source is NGC\,1377, which has a large 24\,$\mu$m flux and remains undetected at 33\,GHz.  
This galaxy is thought to be experiencing a nascent starburst, for which it is within a few Myr of the onset of an intense star formation episode after being quiescent for at least $\sim$100\,Myr  \citep[e.g.,][]{hr03,hr06}.   
This starburst, while heating the dust, has yet to produce detectable free-free emission, nor observable signatures of CRs.  
Over-plotted in Figure \ref{fig-f24-s33} are both a one-to-one line, scaled by the median 24\,$\mu$m to 33\,GHz flux density ratio $<f_{\nu}(24\,\mu{\rm m}) / S_{\rm 33\,GHz}> = 142$, and an ordinary least squares fit to all non-AGN sources detected at 33\,GHz given by, 
\begin{equation}
\label{eq-f24-s33}
\left[\frac{f_{\nu}(24\,\mu{\rm m})}{\rm mJy}\right] = 140\left(\frac{S_{\rm 33\,GHz}}{\rm mJy}\right)^{0.94}, 
\end{equation}
which are consistent  within errors.  
The scatter about the ordinary least squares fit is $\sim$0.26\,dex.  

After converting the 24\,$\mu$m flux densities into luminosities, we relate these to corresponding 33\,GHz star formation rate estimates to come up with the following relation among all non-AGN detected sources, 
\begin{equation}
\label{eq-sfr24}
\left(\frac{\rm SFR_{24\,\micron}}{M_{\sun}\,{\rm yr^{-1}}}\right) = 3.1\times10^{-38} \left[\frac{\nu L_{\nu}(24\,\micron)} {\rm erg~s^{-1}} \right]^{0.88}.  
\end{equation}
The fitting coefficients are consistent with others in the literature \citep[for a detailed compilation see][]{dc10}.   
In Figure \ref{fig-sfr24-sfr33} we plot the ratio of the 24\,$\mu$m to 33\,GHz star formation rate against the 33\,GHz star formation rates finding that the scatter about this relation is $\approx$0.25\,dex.  
The square identifies the location of the anomalous dust detection in NGC\,6946 \citep{ejm10, as10}, which may also explain why other sources lie significantly below the unity line.  

We might expect different relations between warm 24\,$\mu$m dust emission per unit star formation rate among the extranuclear regions and nuclei due to additional old stellar population heating of grains by galaxy bulges.  
To investigate this, we can separate our sample, and measure the median ratio of the 24\,$\mu$m  spectral luminosity to 33\,GHz star formation rate for the nuclei and extranuclear regions.  
The average ratios, along with the associated dispersions, are given in Table \ref{tbl-5}.  
Within the scatter, which is just under a factor of $\sim$2, the median ratios are consistent with one another, as well as with the median ratio for all non-AGN detected sources.    
Comparing the ratio of the total IR luminosity to 33\,GHz star formation rate for nuclei, extranuclear regions, and all non-AGN detected sources, we again find that the median values are consistent.  
Accordingly, we can use these average ratios to write empirical relations between the 24\,$\mu$m spectral luminosity and total IR luminosity with the star formation rate that should be reliable to within a factor of $\sim$2 such that, 
\begin{equation}
\label{eq-sfr24avg}
\left(\frac{\rm SFR^{\rm Avg.}_{24\,\micron}}{M_{\sun}\,{\rm yr^{-1}}}\right) = 2.45\times10^{-43} \left[\frac{\nu L_{\nu}(24\,\micron)} {\rm erg~s^{-1}} \right],  
\end{equation}
and 
\begin{equation}
\label{eq-sfriravg}
\left(\frac{\rm SFR^{\rm Avg.}_{\rm IR}}{M_{\sun}\,{\rm yr^{-1}}}\right) = 3.15\times10^{-44} \left(\frac{L_{\rm IR} }{\rm erg~s^{-1}} \right).  
\end{equation}
We note that the the coefficient equating the total IR luminosity with star formation rate derived in \citet{ejm11b} is $\approx$23\% larger than this empirically measured coefficient, and thus consistent given the nearly factor of $\sim$2 scatter.

\input{tbl-5.tex}

\subsection{Comparison with H$\alpha$ and 24\,$\mu$m Emission}
\label{sec-sfrha24}
Given that not all of the UV/optical photons will be absorbed and re-radiated by dust, a series of new empirical calibrations based on the linear combination of observed 24\,$\mu$m (obscured star formation) and H$\alpha$ (unobscured star formation) luminosities have been developed \citep[e.g.,][]{dc07,rck07,rck09,ynz08}.  
These empirical star formation recipes usually take the form of 
\begin{equation}
\label{eq-sfrscl24}
\left(\frac{\rm SFR_{\rm mix}}{M_{\sun}\,{\rm yr^{-1}}}\right) = 5.37\times10^{-42} \left[\frac{L_{\rm H\alpha} + \xi_{\rm 24}\nu L_{\nu}(24\,\micron)} {\rm erg~s^{-1}}\right], 
\end{equation}
where the coefficient $\xi_{\rm 24}$, scaling the warm 24\,$\mu$m dust emission, has been found to vary depending on the physical scale being investigated.  
For instance, on the scale of individual H{\sc ii} regions, $\xi_{\rm 24} \approx 0.031$ \citep{dc07}, where as for whole galaxies $\xi_{\rm 24} \approx 0.020$ \citep{rck09}.  
A luminosity dependence has also been found, with $\xi_{\rm 24}$ increasing with the 24\,$\mu$m luminosity of a galaxy \citep{dc10}.  

Combining the 33\,GHz star formation rates with the available H$\alpha$ and 24\,$\mu$m photometry, we can estimate $\xi_{\rm 24}$ for the star-forming (i.e., non-AGN) regions in our sample.  
These values are plotted in the top panel of Figure \ref{fig-scl24-sclir} against the 33\,GHz star formation rates, showing a large scatter.  
The median and dispersion are $\xi_{\rm 24} = 0.018\pm0.004$ and $\sigma_{\xi_{\rm 24}} =  0.031$, respectively.  
Thus, on $\sim$kpc scales, which likely average over many non-coeval H{\sc ii} regions, the scaling coefficient appears to be similar to that for entire galaxies.  
It is also worth noting that the scatter is quite large, which most likely arises from both the large range in physical sizes being covered by the GBT beam among the entire sample, as well as the difficulty in matching the GBT and imaging photometry.  

\subsection{Comparison with Total IR and UV Emission}
\label{sec-sfruvir}
Similarly, by combining the star formation rate estimates from the total IR and (observed) UV emission, one can account for the obscured and unobscured emission contributing to the total (bolometric) star formation rate, which we define as: 
\begin{equation}
\label{eq-sfrtot}
{\rm SFR_{tot}}  =  {\rm SFR_{FUV}} + {\rm SFR_{IR}}. 
\end{equation}
This diagnostic is often used to characterize star formation rates from galaxies in both low- \citep[e.g.,][]{jip06,vb07,vb11} and high-$z$ \citep[e.g.,][]{jip04,de07,ed07a,nr10} studies.

Using the above calibration for the FUV star formation rate (i.e., Equation \ref{eq-sfrfuv}), this relation can be expresses as a linear combination of the UV and total IR emission such that, 
\begin{equation}
\label{eq-sfrsclir}
\left(\frac{\rm SFR_{\rm tot}}{M_{\sun}\,{\rm yr^{-1}}}\right) = 4.42\times10^{-44} \left(\frac{L_{\rm FUV} + \xi_{\rm IR}L_{\rm IR}} {\rm erg~s^{-1}} \right),   
\end{equation}
where the coefficient $\xi_{\rm IR}$, scaling the total IR emission, has been found be $\sim$0.46 for normal star-forming galaxies \citep{cnh11} and as large as $\sim$0.6 for starburst galaxies \citep[e.g.,][]{gm99, dc01}.  
These empirical values are significantly smaller than the value of 0.88 reported in \citet{ejm11b}, which assumed that the entire Balmer continuum was absorbed and reradiated in the infrared for their derived IR star formation rate calibration of individual star-forming complexes.    

Using our 33\,GHz star formation rates, we estimate $\xi_{\rm IR}$ for the star-forming regions in our sample and plot them in the bottom panel of Figure \ref{fig-scl24-sclir} against the 33\,GHz star formation rates.  
The median and dispersion are $\xi_{\rm IR} = 0.50\pm0.05$ and $\sigma_{\xi_{\rm IR}} = 0.47$, respectively.  
The scatter is even larger than that for $\xi_{24}$, which may arise from the fact that a calibration such as this depends on the range of ages among individual star formation sites in each star-forming complex covered by the GBT 33\,GHz beam.  
Thus, we expect increasing uncertainties in such a calibration when applied to a single star-forming region depending on how well the IR and UV are measuring emission from recent star formation, not to mention the fact that the IR and UV are sensitive to star formation on different timescales (i.e., $\sim$10\,Myr versus $\sim$$10-100$\,Myr, respectively).  


\begin{figure}[tc]
\begin{center}
\epsscale{1.1}
\plotone{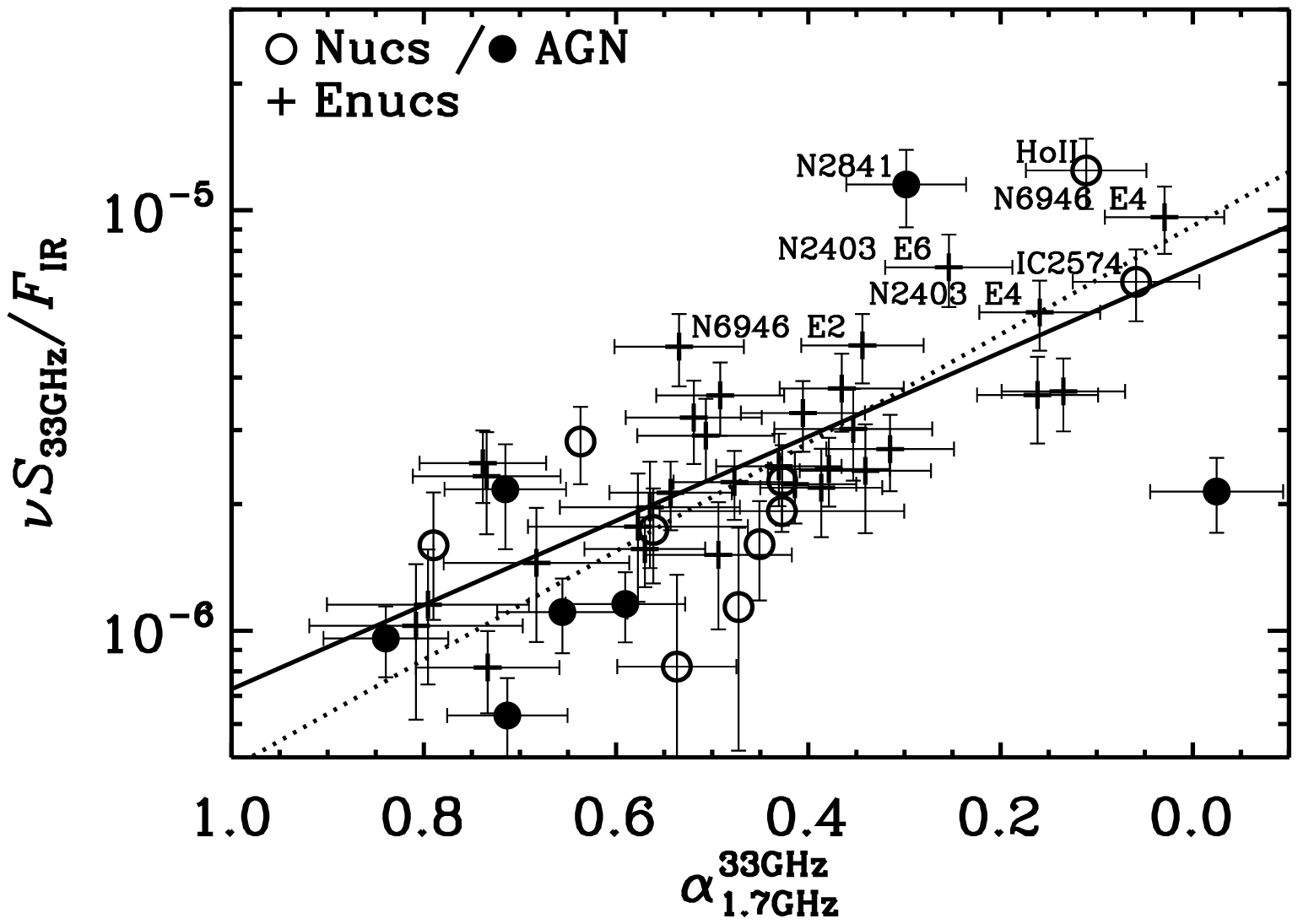}
\end{center}
\caption{
The ratio of the 33\,GHz to total infrared (IR; $8-1000\,\mu$m) fluxes plotted as a function of radio spectral index measured between 1.7 and 33\,GHz.  
Nuclei and extranuclear regions are shown by circles and crosses, respectively, and AGN are identified by filled circles.  
The solid line is the ordinary least squares fit to the data, while the dotted line illustrates the expected trend for a fixed ratio of $\nu S_{\rm 1.7\,GHz}/F_{\rm IR}$.  
We show the names for sources having both flux ratios $\nu S_{\rm 33\,GHz}/F_{\rm IR} >4\times10^{-6}$ and spectral indices $\alpha_{\rm 1.7GHz}^{\rm 33GHz} < 0.5$ as these sources lie near the position of NGC\,6946~Enuc.\,4 \citep{ejm10,as10} in this plot, and {\it may} be good candidates for containing anomalous dust emission.  
}
\label{fig-spx-s33fir}
\end{figure}

\begin{figure}[tc]
\begin{center}
\epsscale{1.1}
\plotone{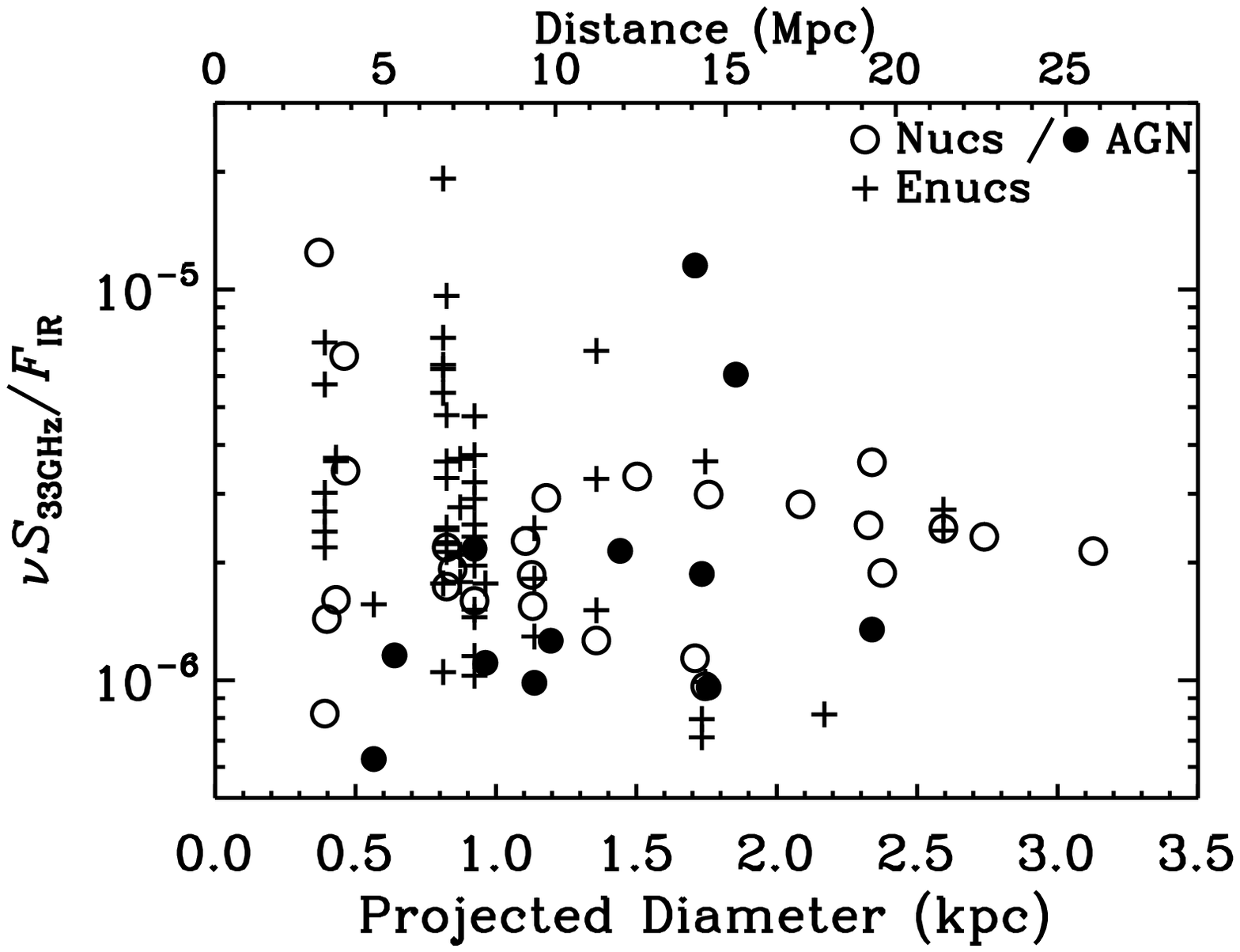}
\end{center}
\caption{
The ratio of the 33\,GHz to total IR fluxes plotted as a function of projected area of the GBT 33\,GHz beam.  
Nuclei and extranuclear regions are shown by circles and crosses, respectively, and AGN are identified by filled circles.  
There does not appear to be a strong trend between the flux ratio and increasing projected area.  
}
\label{fig-dist-s33fir}
\end{figure}

\section{Discussion}
We have used new GBT 33\,GHz radio continuum data to measure free-free emission toward 103 nearby galaxy nuclei and extranuclear star-forming regions.  
We find that the 33\,GHz emission is typically dominated by free-free emission, making it a sensitive measure of the current star formation rate.  
Here we discuss the results of the comparisons between the 33\,GHz star formation rates and other dust-sensitive estimators.  
We additionally discuss some instances where the 33\,GHz emission may contain a contribution from anomalous dust emission and how having high frequency radio data can help infer physical properties of the star-forming complexes.  

\subsection{Dust-Dependent Star Formation Rates}

Using the 33\,GHz star formation rates as a reference, we have investigated a number of dust-sensitive star formation rate indicators used in the literature.  
In $\S$\ref{sec-sfr24} we find a relation between the 33\,GHz star formation rates and 24\,$\mu$m luminosity that is consistent with others in the literature.  
However, while similar, we do find significantly larger scatter.  
We believe that the increased scatter most likely arises due to a combination of several effects.  
First, properly matching the single-beam GBT photometry to that from the ancillary images of resolved sources is difficult, and the associated uncertainties will increase the dispersion of any trends.  
Additionally, the area covered by GBT beam covers a range of physical sizes in the sample, which will also inject scatter into any observed correlations.  
And, of course, the corrections applied in cases of potential over subtraction of sky due to reference nods landing on bright regions of the targeted galaxy add another source of scatter.  


In looking at the relation between the 33\,GHz star formation rates and 24\,$\mu$m spectral luminosities for galaxy nuclei and extranuclear regions independently, we find them to be consistent.  
This is also true for the relation between  33\,GHz star formation rates and total IR luminosities.  
Thus, on average, there does not appear to be excess infrared emission per unit star formation for galaxy nuclei relative to the extranuclear star-forming complexes on the physical scales being investigated.  

In $\S$\ref{sec-sfrha24} and \ref{sec-sfruvir}, we investigate two hybrid star formation diagnostics that attempt to account for the unobscured and obscured star formation components (i.e., H$\alpha + 24\,\mu$m and UV$ + $IR).  
While we find scaling coefficients for the 24\,$\mu$m and total IR luminosities that are similar to those reported in the literature, albeit with a large scatter most likely due to the reasons mentioned above, it is worth noting that the values reported are similar to those found from measurements of {\it entire} galaxies \citep[e.g.,][]{rck09,cnh11}.  
This agreement most likely arises because of the large area covered by the GBT beam at the distances of the sample galaxies, being roughly $\sim$1\,kpc, on average.  
Thus, on these physical scales, and by not including a local background subtraction, it appears that the star formation diagnostics presented here are applicable for globally integrated measurements of galaxies.  

We summarize the final relations for each of these four dust-dependent star formation rate estimates, along with all theoretically motivated calibrations discussed in \S\ref{sec-sfr}, in Appendix Table \ref{tbl-A3}.  


\subsection{Variations on the Thermal Fraction and Non-Thermal Spectral Index}
In Figure \ref{fig-spx-s33fir} we plot the ratio of the 33\,GHz to total IR fluxes as a function of the radio spectral index measured between 1.7 and 33\,GHz.  
There is a clear trend of increasing flux ratio with flatter spectral index.  
This trend does not seem to arise due to differences in the projected physical area of the 33\,GHz beam among the sources (see Figure \ref{fig-dist-s33fir}).   
In fact, this trend is opposite the expectation for distance effects:  
as the physical area subtended by the GBT beam increases, one would expect a larger contribution of diffuse non-thermal emission from a galaxy disk relative to additional free-free emission, thus resulting in a steeper spectral index and an increase in 33\,GHz to IR flux ratios.  

The ordinary least squares fit to the non-AGN sources (solid line) appears generally consistent with, albeit slightly flatter than, the expected trend assuming a fixed ratio of $\nu S_{\rm 1.7\,GHz}/F_{\rm IR}$; 
the dotted line is given by scaling the median ratio $\nu S_{\rm 1.7\,GHz}/F_{\rm IR}$ by  $(33/1.7)^{1-\alpha}$.  
By only having radio data at two well-spaced frequencies, it is not possible to distinguish whether the spectral flattening is being driven by an increase in the thermal fraction or flattening of the non-thermal radio spectral indices.  
If we are to assume that this trend is in fact dominated by an increase in the thermal fraction, we can use the scatter about the ordinary least squares fit to the data to set the {\it maximum} dispersion for the non-thermal spectral indices among the non-AGN sources, which is found to be $\sigma_{\alpha^{\rm NT}} \la 0.13$.  
Interestingly, this is similar to the dispersion in global non-thermal spectral indices measured by \citet[][$ \sigma_{\alpha^{\rm NT}} = 0.13$]{nkw97}.  

Again, assuming that the trend is in fact driven by a change in thermal fraction among the sample, suggests that the non-thermal emission component may be quite similar among each star-forming region, with a nearly uniform CR electron injection spectrum and similar spectral steepening from associated energy losses (e.g., synchrotron, inverse Compton, bremsstrahlung, and ionization processes).  
More specifically, there must only be moderate variability in the ratio of synchrotron to total CR electron cooling processes among each of the star-forming complexes being investigated.  

\subsection{Anomalous Dust Candidates}
Cosmic Microwave Background experiments were the first to discover the presence of an ``anomalous" dust-correlated emission component at frequencies between $\sim10 - 90$\,GHz \citep[e.g.,][]{eml97}. 
In the original GBT pilot study of NGC\,6946, \citet{ejm10} identified an extranuclear star-forming complex that had a significant amount of excess emission at 33\,GHz relative to what was expected by extrapolating from multifrequency measurements at lower frequencies (i.e., $<$10\,GHz).   
While anomalous dust emission has been observed in excess of synchrotron and free-free components in Galactic H{\sc ii} regions \citep[e.g.,][]{cd09}, this observation of excess 33\,GHz emission towards a star-forming complex in NGC\,6946 is identified as the first detection of anomalous dust emission outside of the Milky Way.    
Now, using the full sample data, we can look for other potential extragalactic anomalous dust emitting candidates.  

As stated above, Figure  \ref{fig-spx-s33fir}  shows a clear trend in which the ratio of  33\,GHz to total IR fluxes increases as the radio spectral index measured between 1.7 and 33\,GHz flattens.  
In Figure \ref{fig-spx-s33fir} we also show the location of NGC\,6946~Enuc.\,4, identified as containing anomalous dust emission \citep{ejm10, as10} that accounts for $\approx$50\% of the observed 33\,GHz flux density.    
However, this source does not appear to be significantly discrepant from the expected trend in this plot.  
It therefore appears that identifying anomalous dust candidates with such coarse radio spectral resolution may not be possible, and a much finer sampling (i.e., better than a factor of 2 in frequency steps) is necessary.  

Besides NGC\,6946~Enuc.\,4, there are a number of other sources that both show flat radio spectral indices with large 33\,GHz to infrared flux ratios; 
sources having flux ratios $\nu S_{\rm 33\,GHz}/F_{\rm IR} > 4\times10^{-6}$ and observed radio spectral indices $\alpha_{\rm 1.7GHz}^{\rm 33GHz} < 0.5$ have been highlighted as potential anomalous dust candidates in Figure \ref{fig-spx-s33fir}.  
However, a detailed investigation of these regions to confirm or deny the presence of anomalous dust emission is outside the scope of this paper, and really requires additional radio data at finely spaced frequencies much closer to 33\,GHz (e.g., between $\sim$15 and 90\,GHz) to confirm a peaked spectrum.

\subsection{The Nascent Starburst NGC\,1377}
The archetypal nascent starburst NGC\,1377, which is thought to be within a few Myr of the onset of an intense star formation episode after being quiescent for at least $\sim$100\,Myr, remains undetected in the radio even at the depth of our 33\,GHz observations.  
Given its measured flux density at 24\,$\mu$m and the 3\,$\sigma$ upper limit of 0.44\,mJy at 33\,GHz, we would have expected a $\sim$70\,$\sigma$ detection.  
As has already been shown by \citet{hr03}, this non-detection in the radio most likely does not arise from the source being optically thick.  
Using the 33\,GHz upper limit, we can set even more stringent constraints on this.  
Assuming an electron temperature of $\sim$10$^{4}$\,K and $\tau_{\rm 33GHz} \sim 1$ sets the expected brightness temperature and emission measure of the source to be $\sim$6000\,K and $\sim5\times10^{9}\,{\rm pc\,cm^{-6}}$, respectively.  
The corresponding size of the source would therefore have to be $<$1\,pc with an electron density of $>7\times10^5\,{\rm cm^{-3}}$.  
These values are significantly more extreme than the size and electron density limits of $<$28\,pc and $>$2700\,cm$^{-3}$ inferred by \citet{hr03}.  

Given the apparent absence of AGN activity in NGC\,1377, any free-free emission associated with ongoing star formation powering the bright dust continuum appears significantly suppressed.  
If the nascent starburst scenario suggested by \citet{hr03} is true, the lack of free-free emission detected at 33\,GHz could very well arise from having extremely young, deeply embedding star formation in which the dust is absorbing $\ga$95\% of the ionizing photons.  
The absorption of such a high fraction of ionizing photons by dust would then lead to star formation rates that are overestimated by a factor of $\sim$2 by using calibrations relying on dust emission.

\section{Conclusions}
In the paper we have presented GBT 33\,GHz photometry for 103 galaxy nuclei and extranuclear star-forming regions included as part of the Star Formation in Radio Survey.   
The sample galaxies are included in the {\it Spitzer}-SINGS and {\it Herschel}-KINGFISH legacy programs, and therefore have large amounts of ancillary data for future followup, including mid- and far-infrared spectral mappings.  
Here we summarize our main conclusions from this initial investigation.  

\begin{enumerate}

\item Among the non-AGN sources having lower frequency radio data, we find a median thermal fraction at 33\,GHz of $\approx$76\% with a dispersion of $\approx$24\%.  
For all sources resolved on scales $\la$0.5\,kpc, the thermal fraction is even larger, being $\ga$90\%, on average, however, there is very little lower frequency radio data for galaxies at the large distance end of the sample.  
This suggests that the rest-frame 33\,GHz emission provides a sensitive measure of the ionizing photon rate from young star-forming regions, thus making it a robust star formation rate indicator, especially at high spatial resolution.  

\item We find an increase in the 33\,GHz to total IR flux ratios as the radio spectral index flattens, which does not appear to correlate with the projected area of the GBT beam, but is instead consistent with the expectation from an increase in the thermal fraction among the sources.  
Consequently, the ratio of non-thermal to total IR emission appears relatively constant among star-forming regions, suggesting only moderate variability in the CR electron injection spectrum and ratio of synchrotron to total cooling processes.  
Given the scatter in this trend, the maximum dispersion in the non-thermal radio spectral indices among these non-AGN sources is $\sigma_{\alpha^{\rm NT}} \la 0.13$.  


\item Using the 33\,GHz star formation rates as a reference, we derive scaling coefficients for the following recipes that attempt to sum the obscured and unobscured star formation components: H$\alpha + 24\,\mu$m and UV$ + $IR.  
Although we have targeted galaxy nuclei and extranuclear star-forming complexes, with a median resolution of $\sim$1\,kpc, the hybrid scaling coefficients are consistent with those in the literature derived for globally integrated measurements of galaxies.  

\item Identifying anomalous dust candidates based on a coarse sampling of the radio spectrum, even with a large lever-arm spanning 1.7,  and 33\,GHz, is inconclusive.  
A much finer (i.e., better than a factor of 2) sampling in frequency space, spanning $\sim$15 to 90\,GHz to actually measure the peak, appears necessary for conclusive detections.  

\item Given the depth of our 33\,GHz photometry, we are able to put greater doubt on the possibility that NGC\,1377 is radio faint due to being optically-thick at radio frequencies; 
the source size would need to be $<$1\,pc with an electron density of $>7\times10^5\,{\rm cm^{-3}}$.  
Assuming that the source is at the onset of a starburst, the lack of detectable free-free emission at 33\,GHz would require dust to absorb $\ga$95\% of the ionizing photons, leading infrared-derived star formation rates to be overestimated by as much as a factor of 2.  

\end{enumerate}

We are currently in the process of collecting 1\farcm4-sized 33\,GHz maps using the VLA in the D-configuration for each of the sources targeted with the GBT, along with a few others at lower declination.  
These maps will have an angular resolution of $\sim$2\arcsec, allowing us to sample physical scales ranging between $30-300$\,pc given the distances of the galaxy sample.  
At this resolution, we will be able to probe the scales of individual giant H{\sc ii} regions, where we expect even larger thermal fractions for more accurate estimates of star formation rates.  
These maps will additionally enable us to isolate the location of potential extragalactic anomalous dust emitting sources with significantly greater accuracy.  
An initial round of data has been taken between October 2011 and January 2012.  

\acknowledgements
We thank the anonymous referee for useful suggestions that helped improve the presentation of the paper.  
E.J.M. thanks C. Leitherer and M. Seibert for useful discussions.  
The National Radio Astronomy Observatory is a facility of the National Science Foundation operated under cooperative agreement by Associated Universities, Inc.
This work is based in part on observations made with the {\it Spitzer} Space Telescope, which is operated by the Jet Propulsion Laboratory, California Institute of Technology under a contract with NASA.  
{\it Herschel} is an ESA space observatory with science instruments provided by European-led Principal Investigator consortia and with important participation from NASA.

\bibliography{aph}

\appendix
Here we present the more detailed results from the GBT observations, as well as summarize all star formation rate equations either utilized or empirically derived in the present analysis.  
For the GBT observations, we give the flux densities and errors reported from the four individual CCB ports, and provide details on how much time was spent on each source in Tables \ref{tbl-A1} and \ref{tbl-A2}.  
We again note that the flux densities of the individual channels reported here do not include any corrections.  
Upper limits for the individual channels are not given, but rather actual measured values and estimated errors are given.  

\input{tbl-A1.tex}

\input{tbl-A2.tex}

In Table \ref{tbl-A3} we provide a summary of the theoretical and empirically derived star formation rate calibrations from this paper.  
A thorough discussion of the theoretical derivations can be found in \S\ref{sec-sfr} and in \citet{ejm11b}, while details on the empirically derived, dust-dependent star formation rate estimates are given in \S\ref{sec-sfr24} and \ref{sec-sfrha24}.  
We also provide the timescale for the emission of each star formation rate diagnostic to fade to 50 and 5\% of its peak value after having had 100\,Myr of continuous star formation come to a halt.  
These timescales, which are sensitive to the exact star formation history, were estimated using Starburst99 \citep[][]{cl99} following the same assumptions that went into deriving the corresponding theoretical relations as described in \S\ref{sec-sfr} and in the table caption.  
Additional considerations for the radiating times of CR electrons were included when calculating the fading time for the non-thermal radio continuum emission, the details of which can be found in \citet{ejm09c}.  

\input{tbl-A3.tex}

\clearpage
\setcounter{figure}{0}
\begin{figure*}[h!H]
\begin{center}
\begin{tabular}{cc}
\epsfig{file=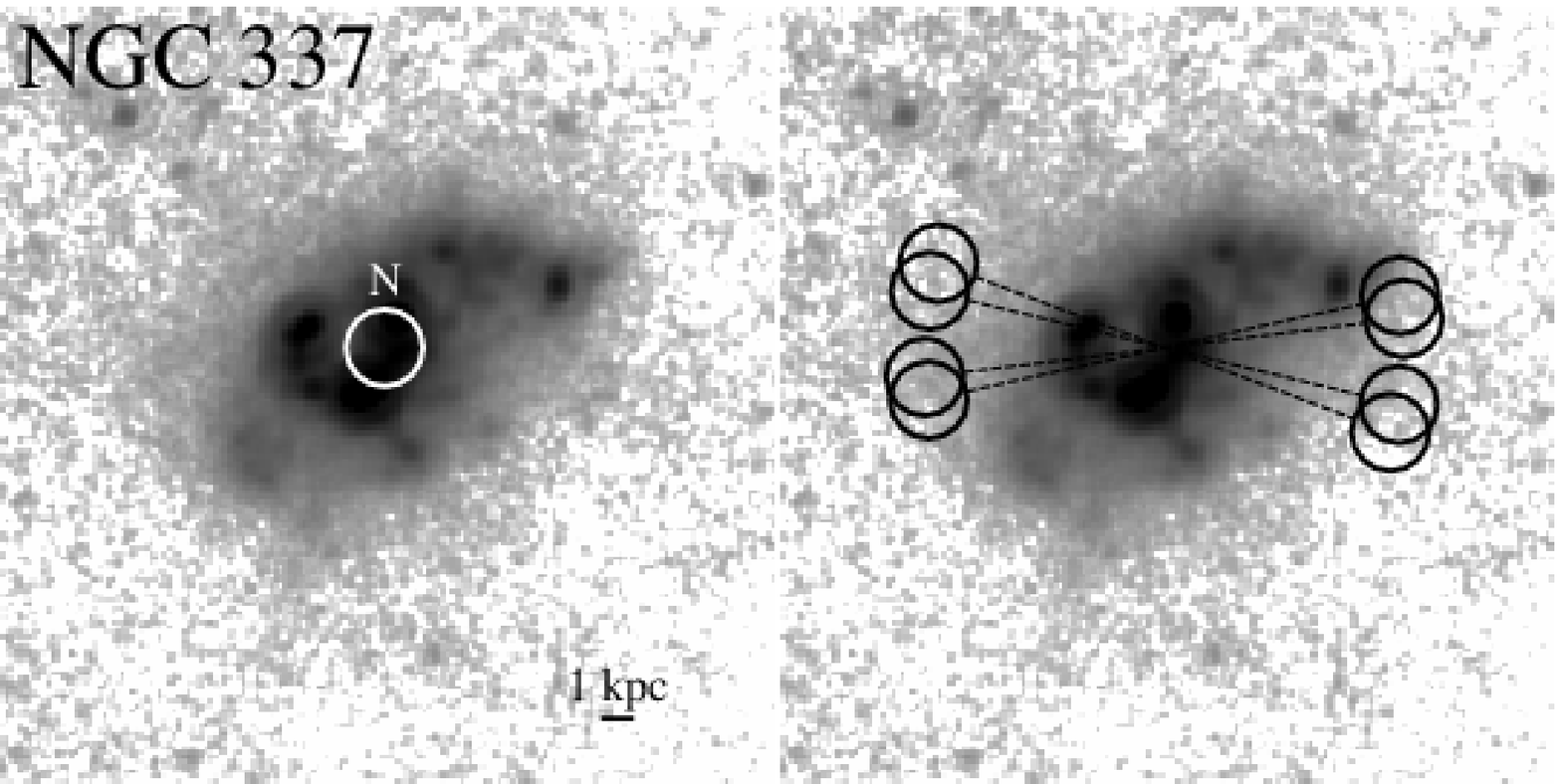,width=0.5\linewidth,clip=} & \epsfig{file=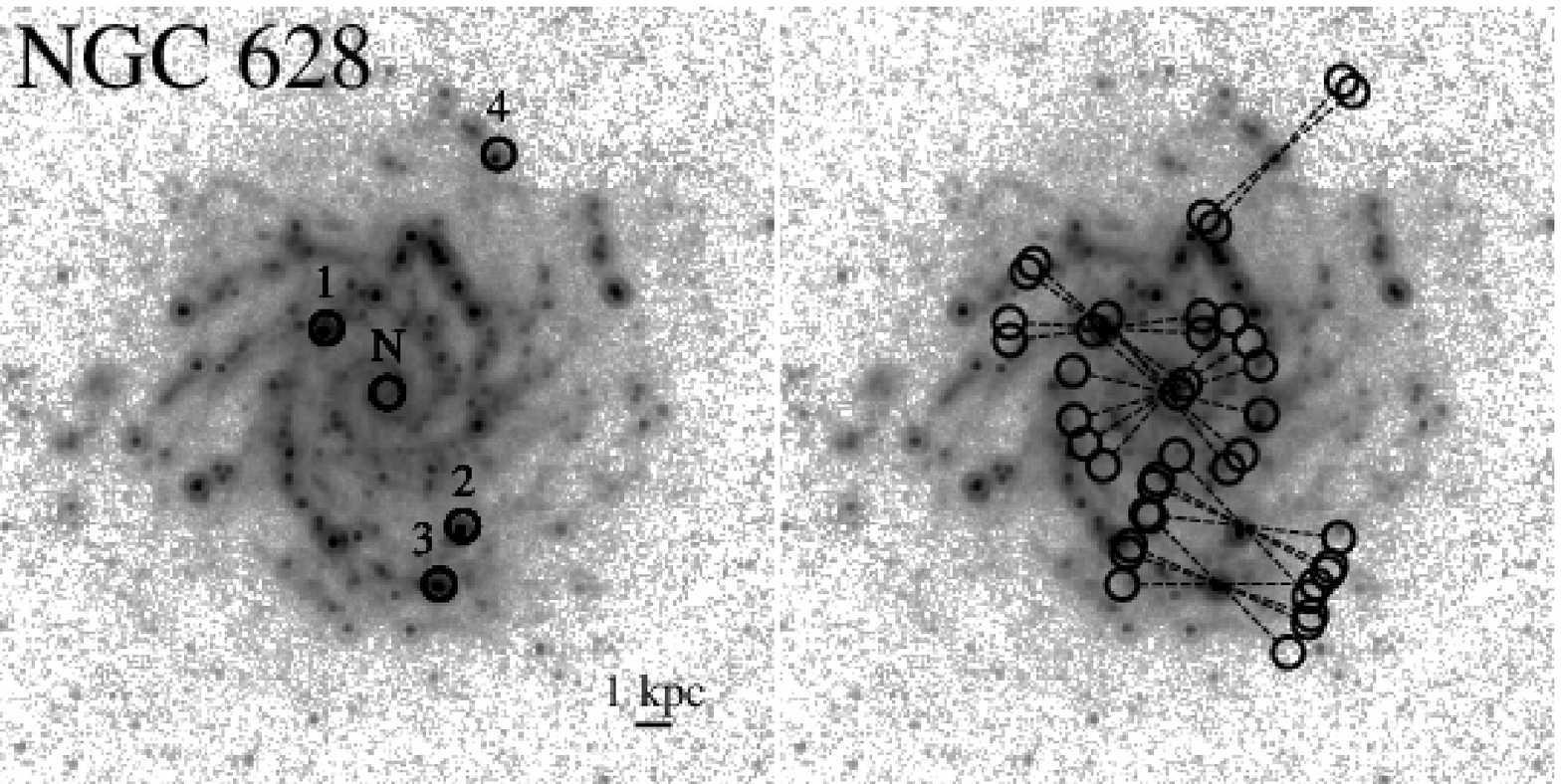,width=0.5\linewidth,clip=}\\
\epsfig{file=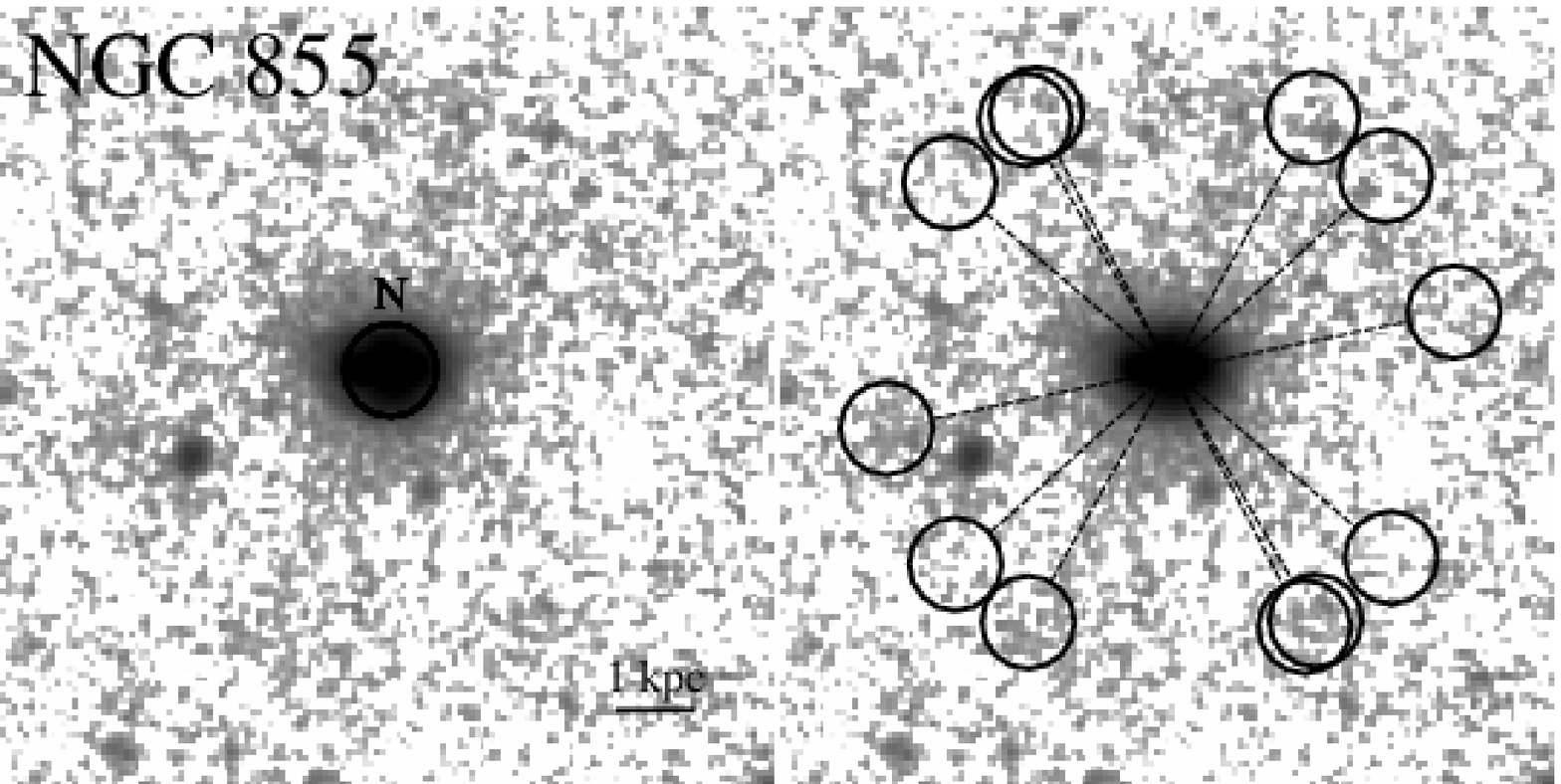,width=0.5\linewidth,clip=} & \epsfig{file=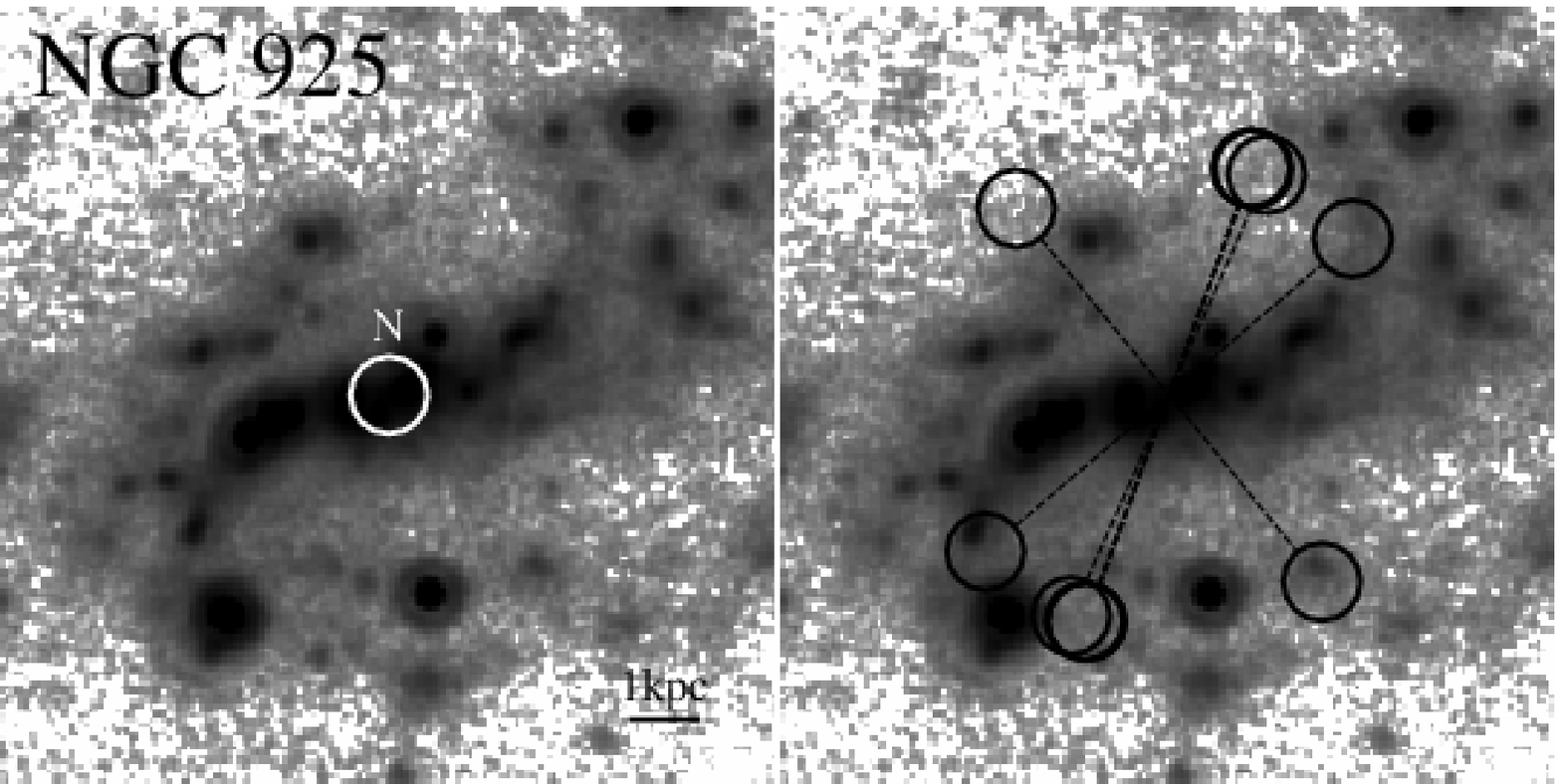,width=0.5\linewidth,clip=}\\
\epsfig{file=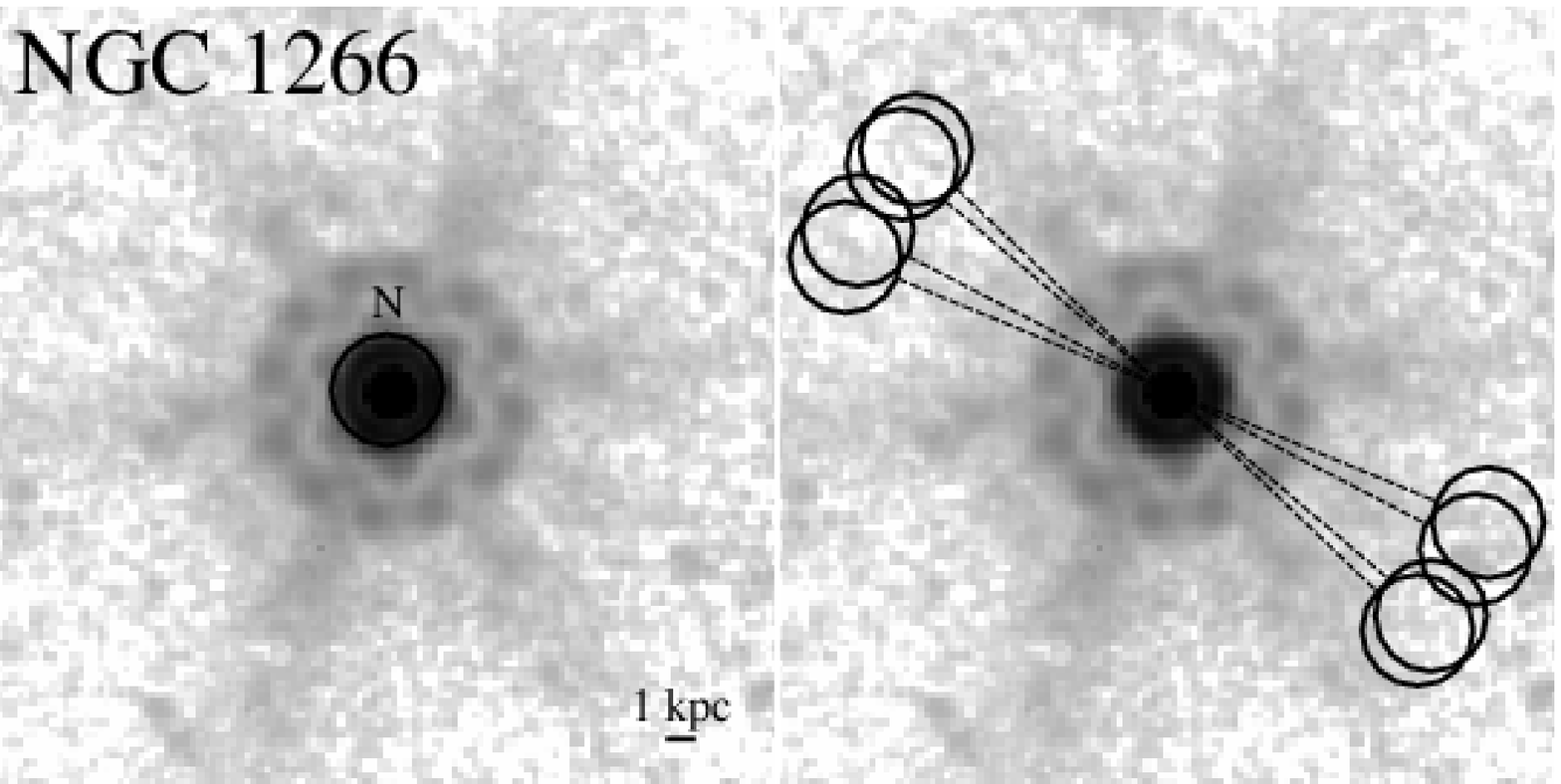,width=0.5\linewidth,clip=} & \epsfig{file=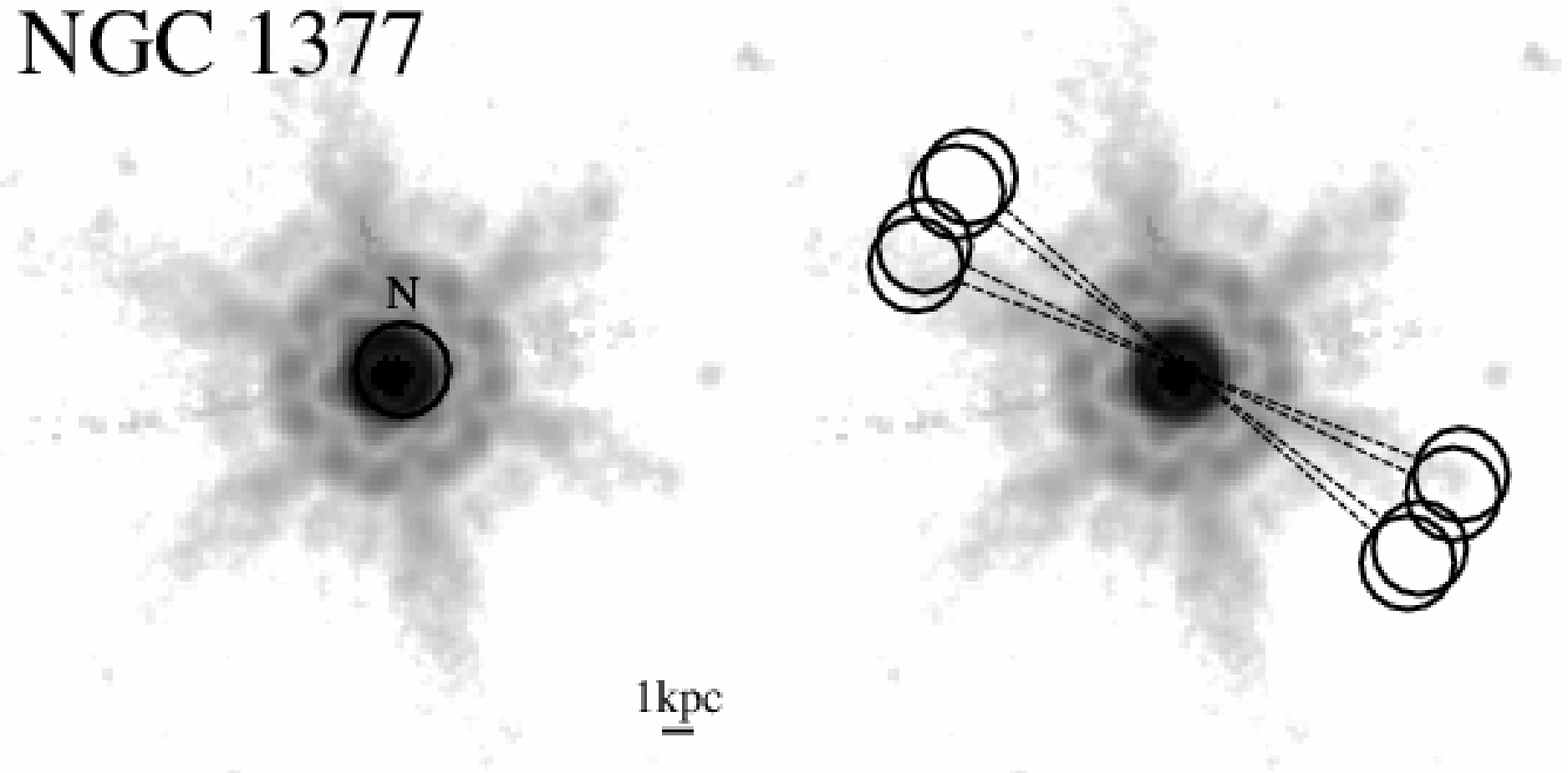,width=0.5\linewidth,clip=}\\
\epsfig{file=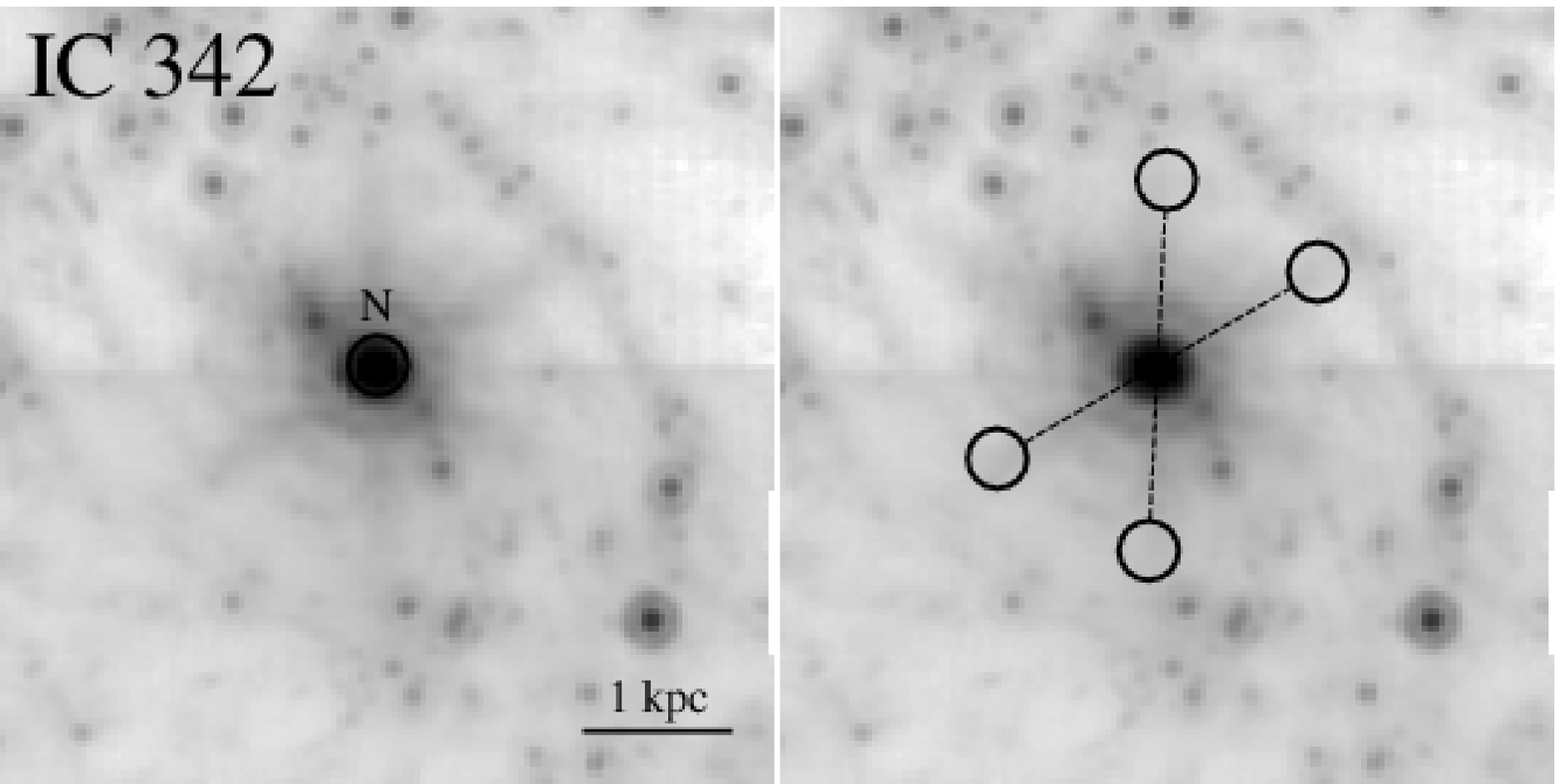,width=0.5\linewidth,clip=} & \epsfig{file=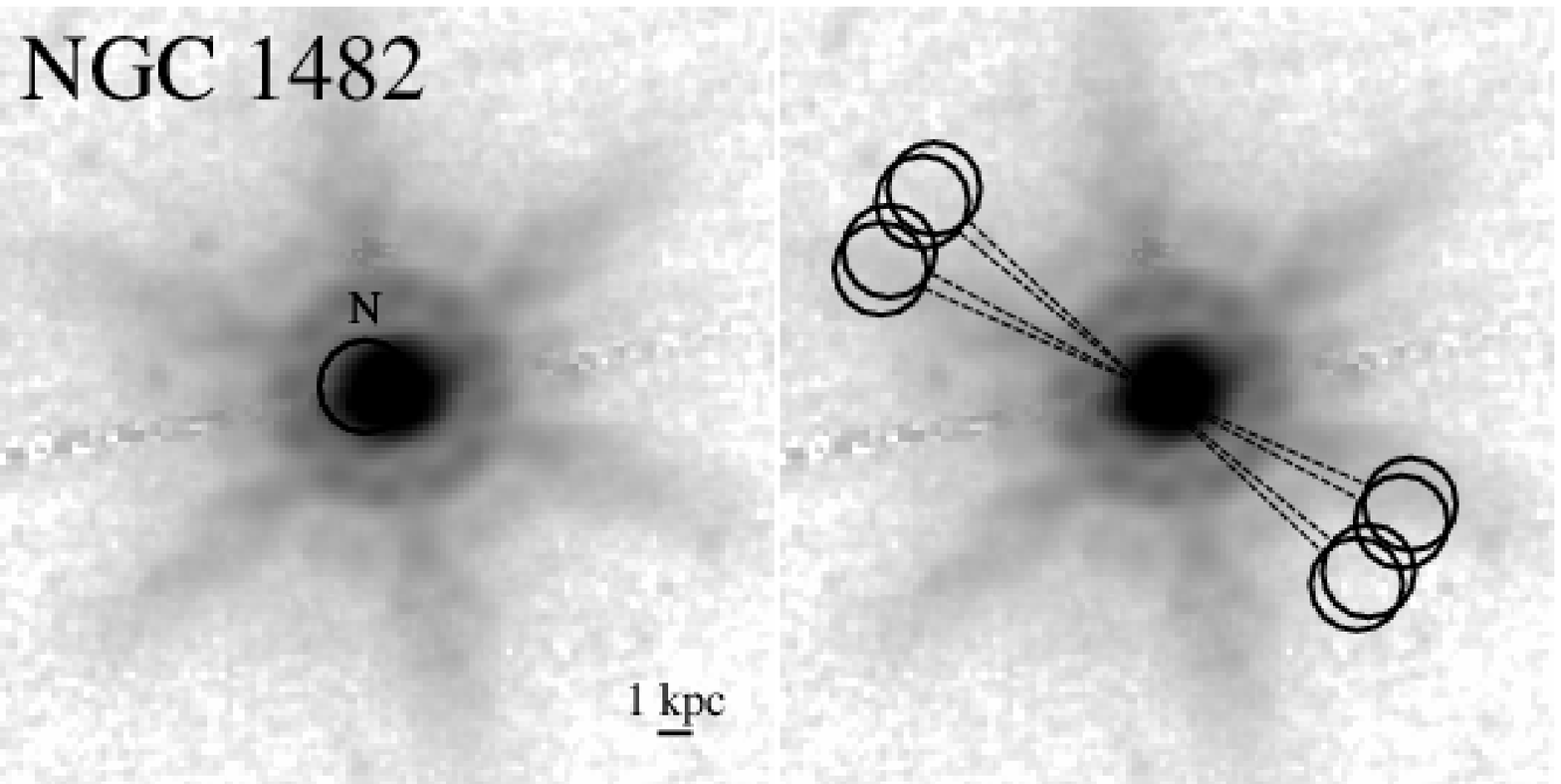,width=0.5\linewidth,clip=}\\
\epsfig{file=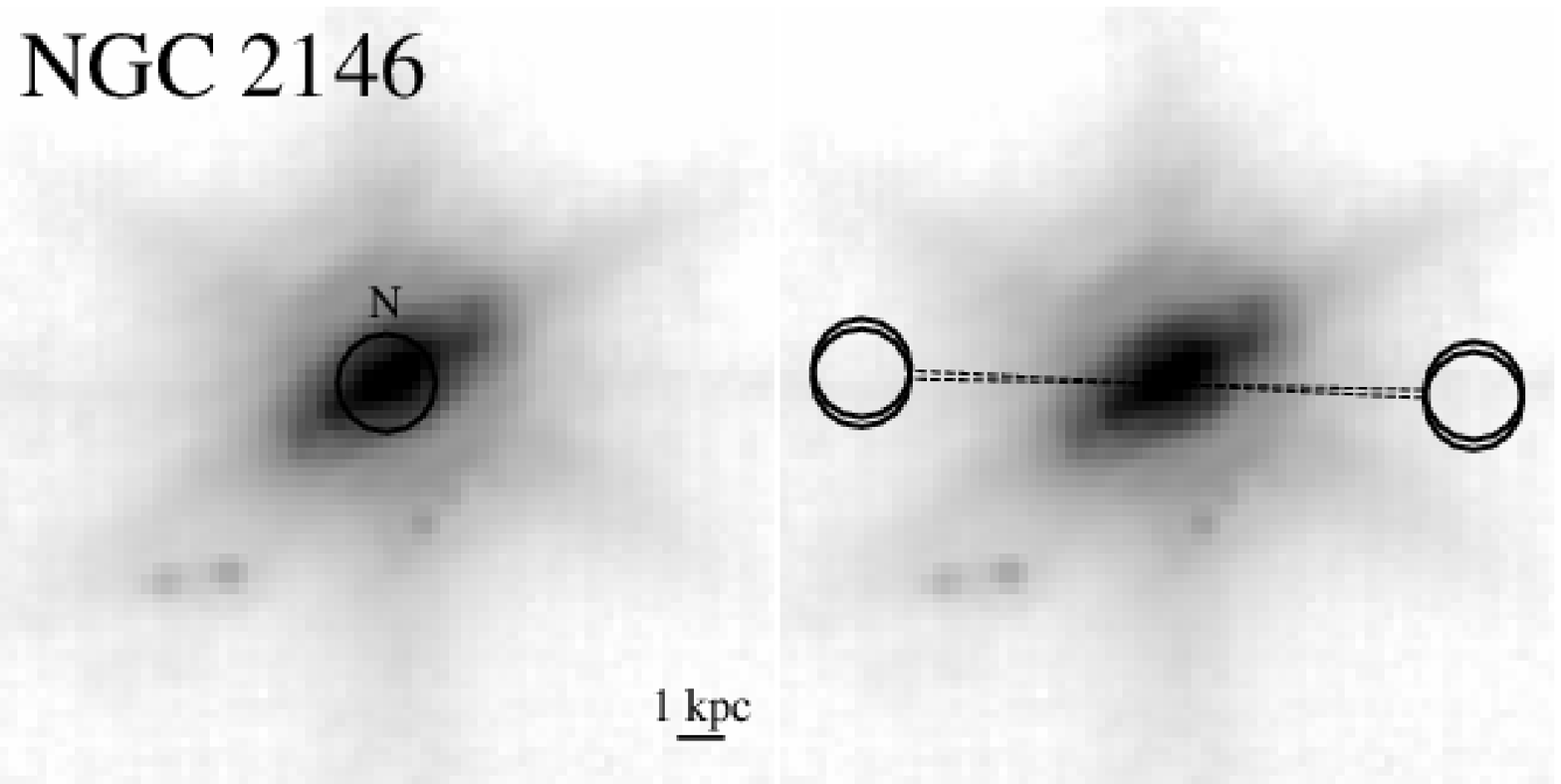,width=0.5\linewidth,clip=} & \epsfig{file=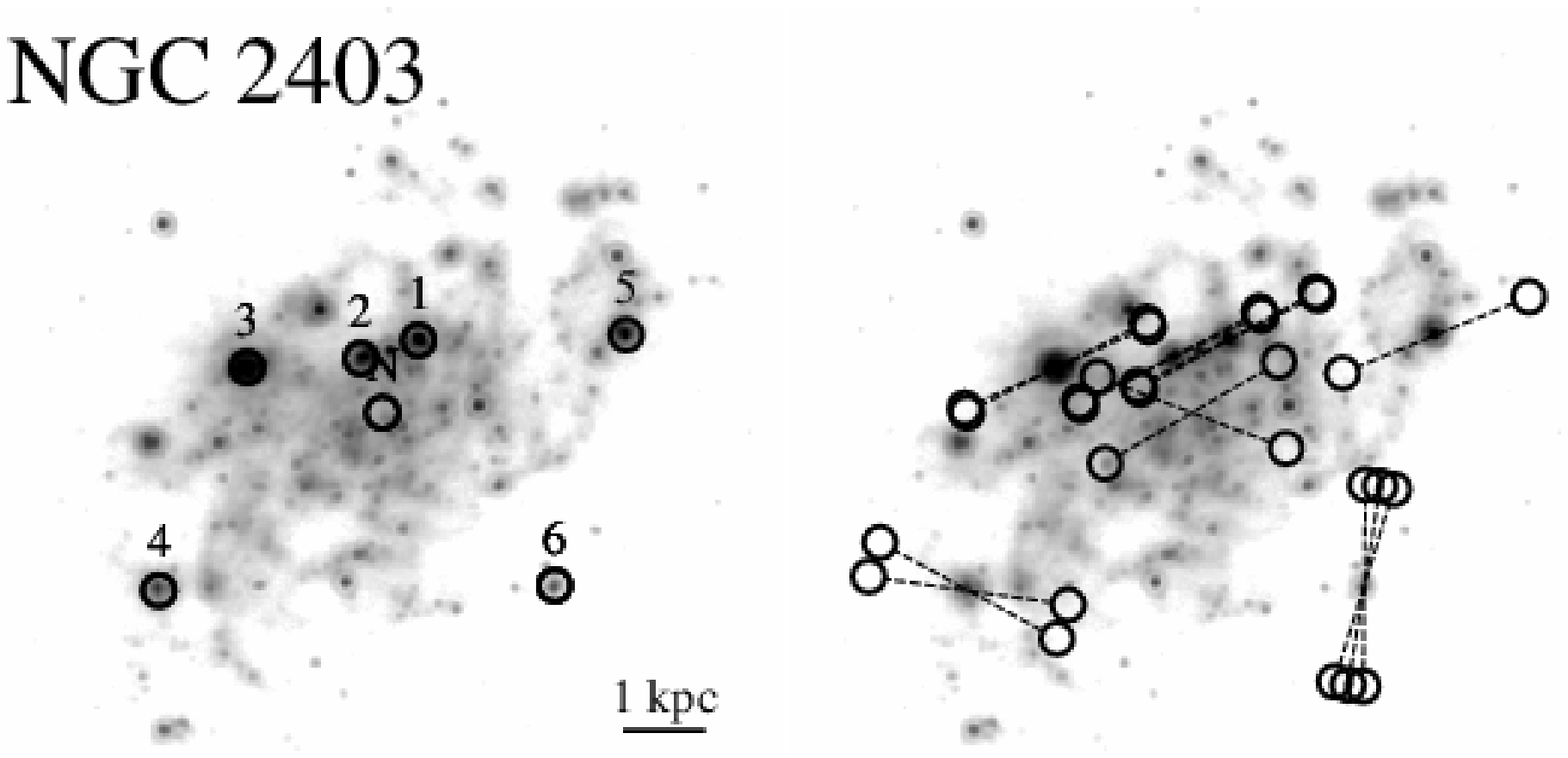,width=0.5\linewidth,clip=}
\end{tabular}
\end{center}
\caption{
For each galaxy we show 2 side-by-side panels illustrating the GBT observations on 24\,$\mu$m {\it Spitzer} images.     
The left panels indicate the location of each GBT 33\,GHz pointing, while the right panels show the location of the GBT reference nod positions (see \S\ref{sec-gbtobs}).  
The diameter of each circle is 25\arcsec, which corresponds to the typical FWHM of the GBT beam at 33\,GHz.  
The 1\,kpc linear scale bars assume the distances given in Table \ref{tbl-1}.  
}
\label{fig-gbtobs}
\end{figure*}

\clearpage
\setcounter{figure}{0}
\begin{figure*}[ht!]
\begin{center}
\begin{tabular}{cc}
\epsfig{file=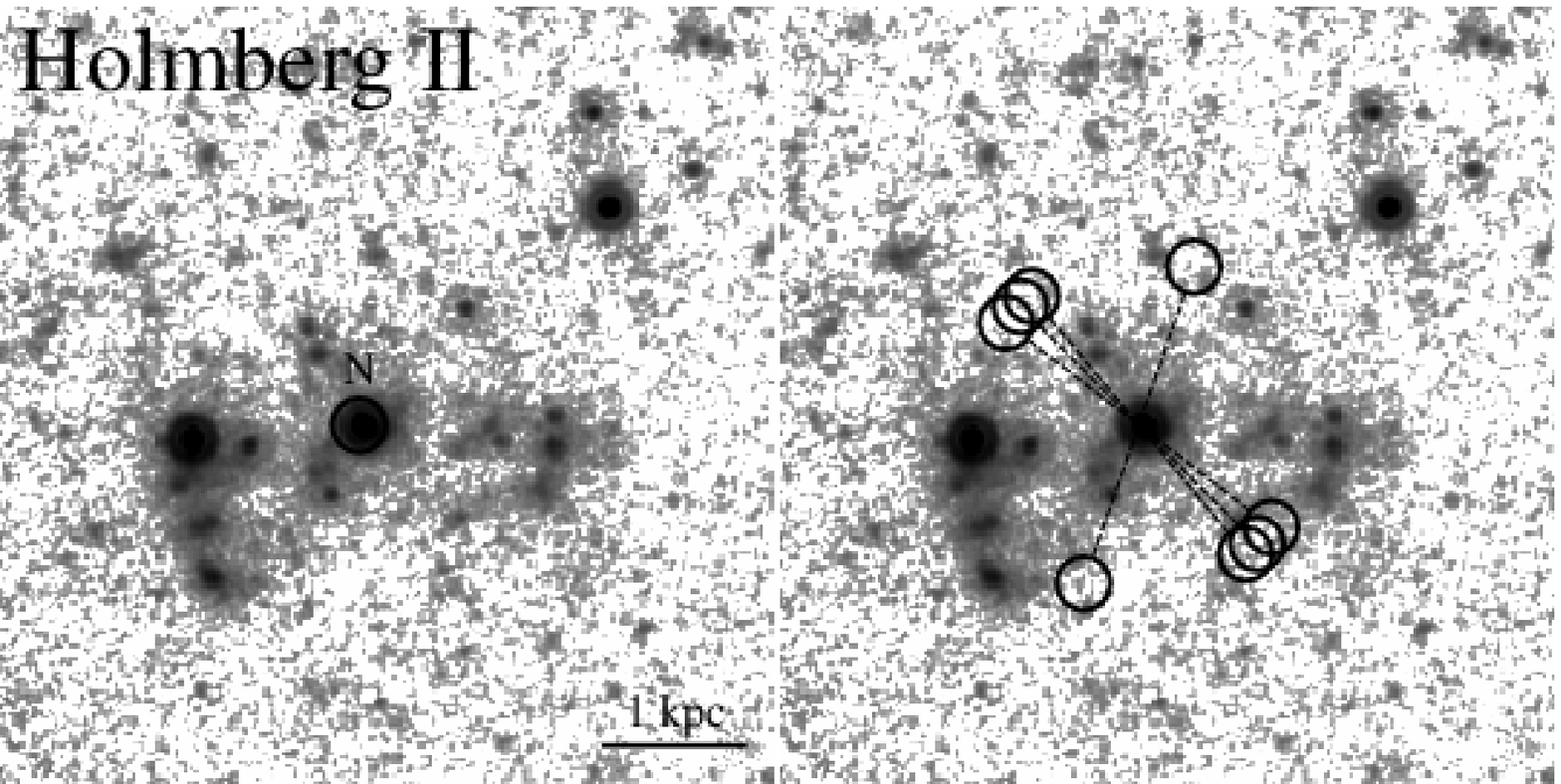,width=0.5\linewidth,clip=} & \epsfig{file=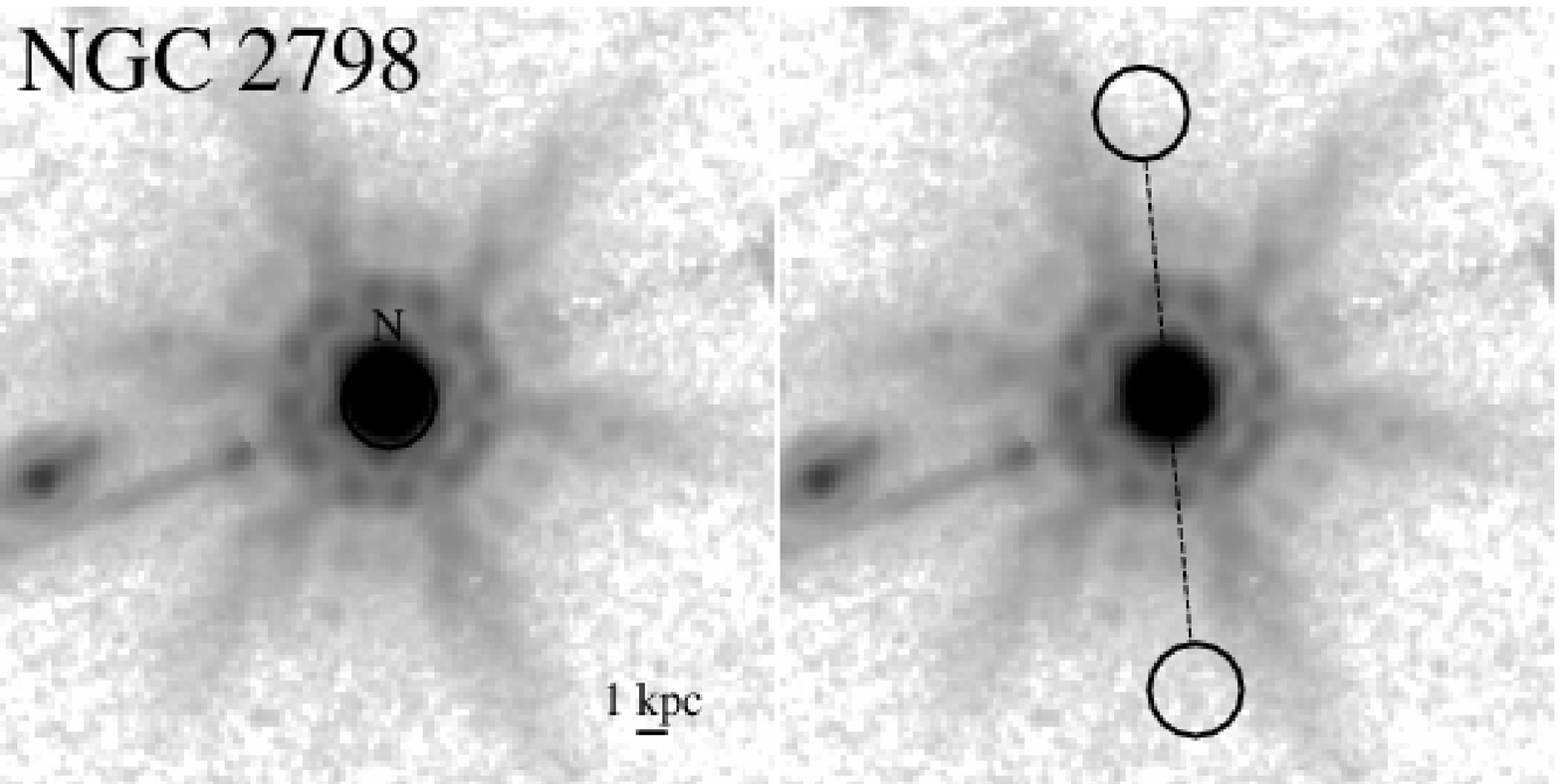,width=0.5\linewidth,clip=}\\
\epsfig{file=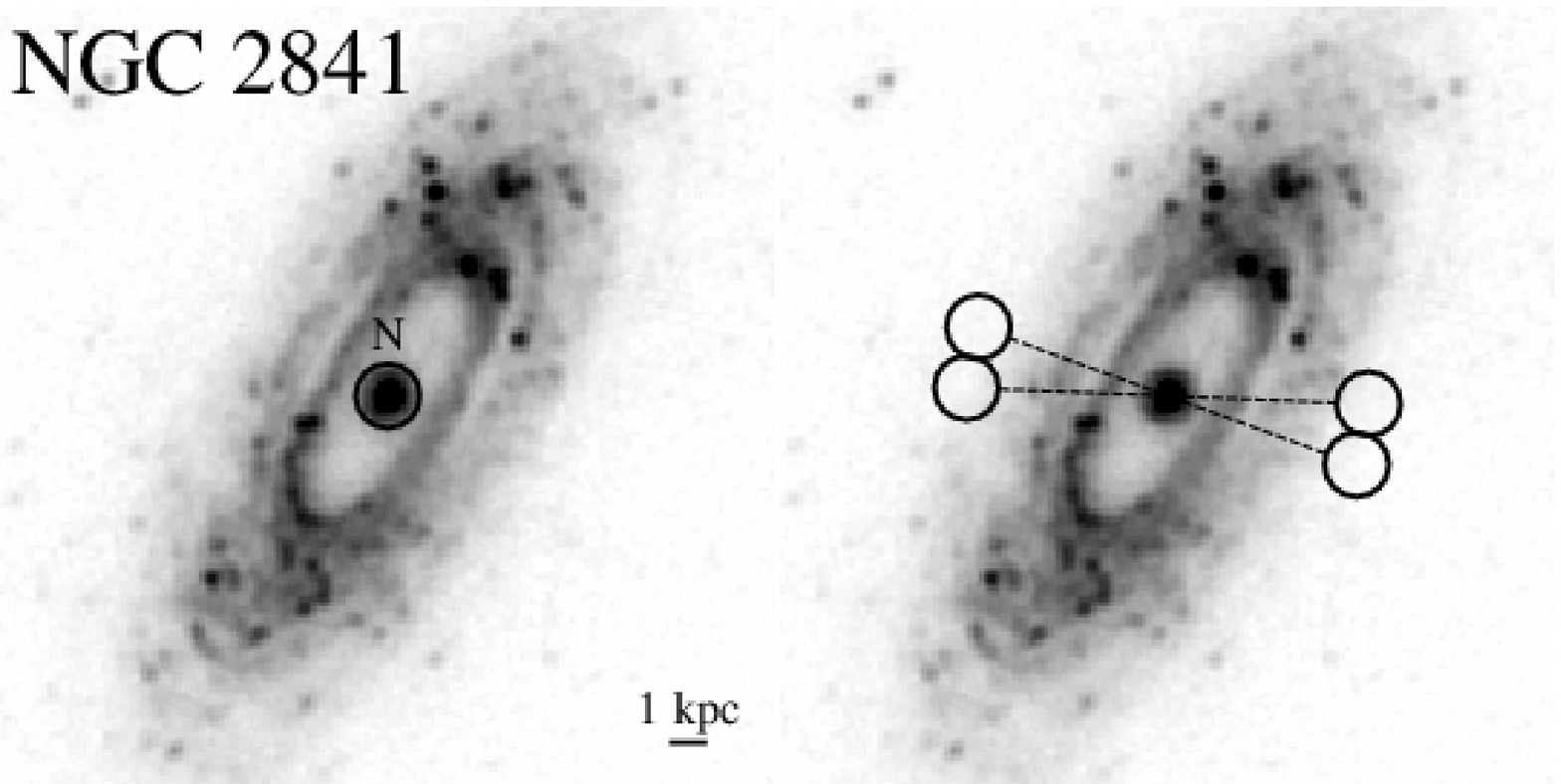,width=0.5\linewidth,clip=} & \epsfig{file=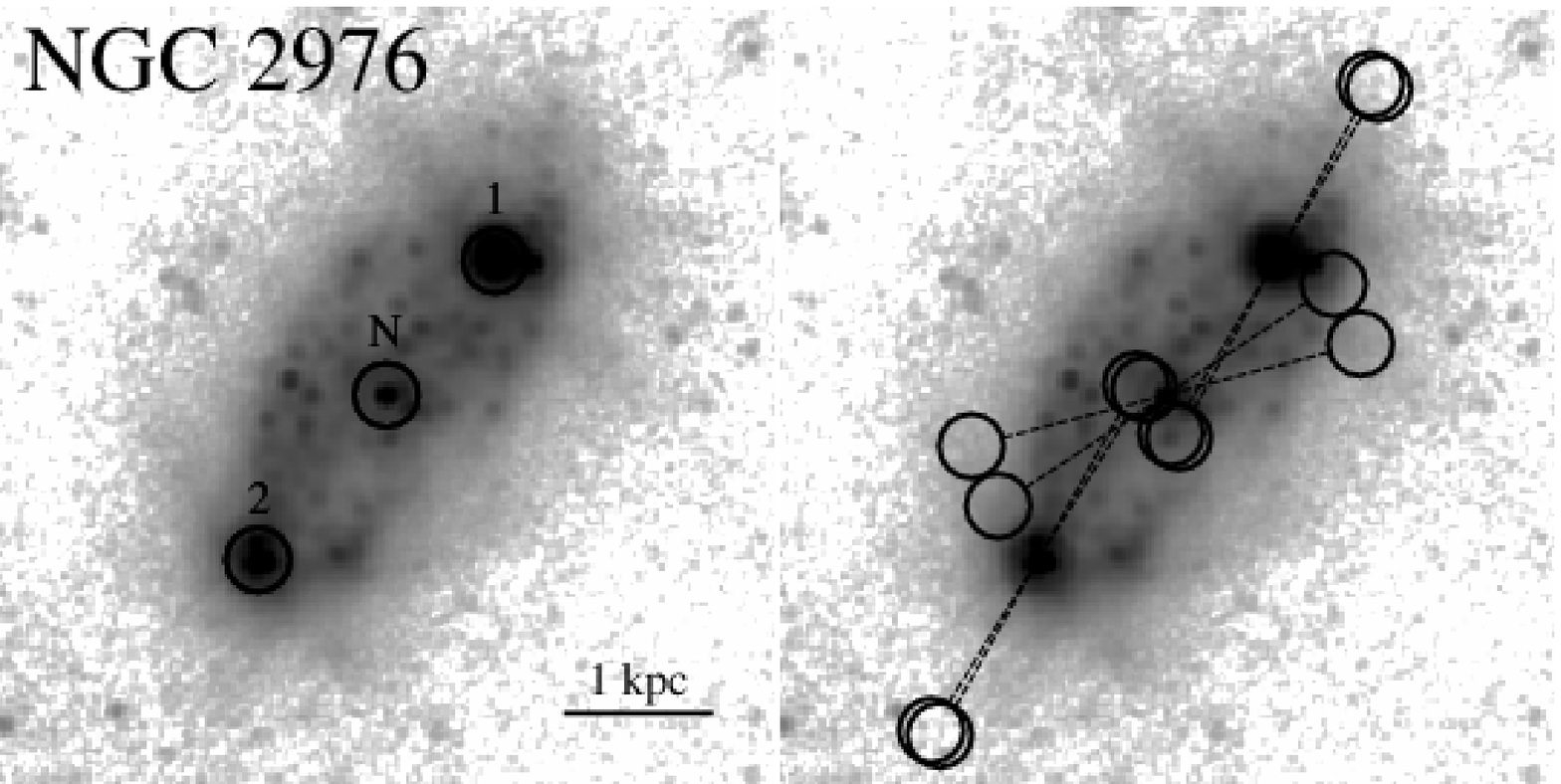,width=0.5\linewidth,clip=}\\
\epsfig{file=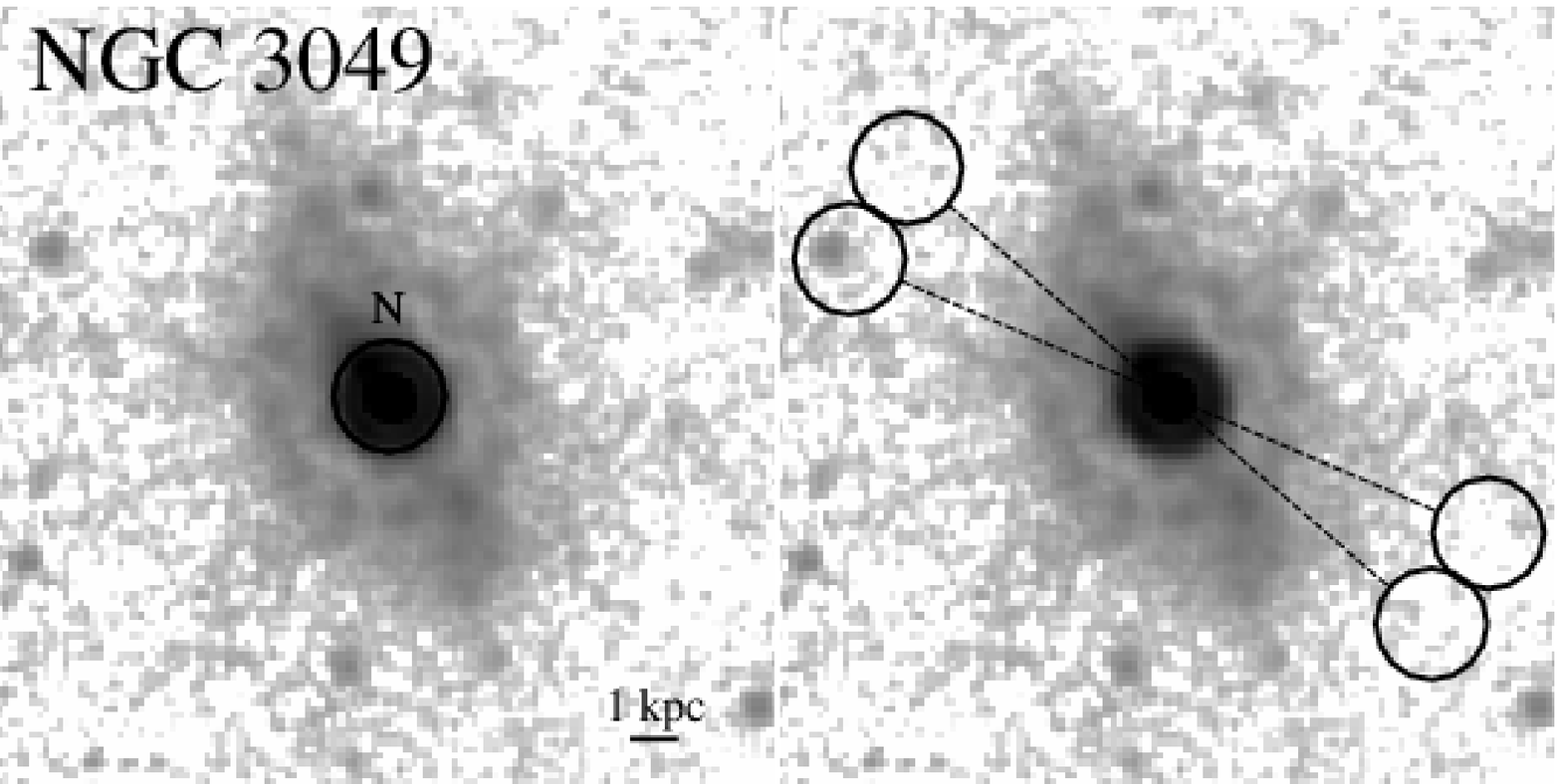,width=0.5\linewidth,clip=} & \epsfig{file=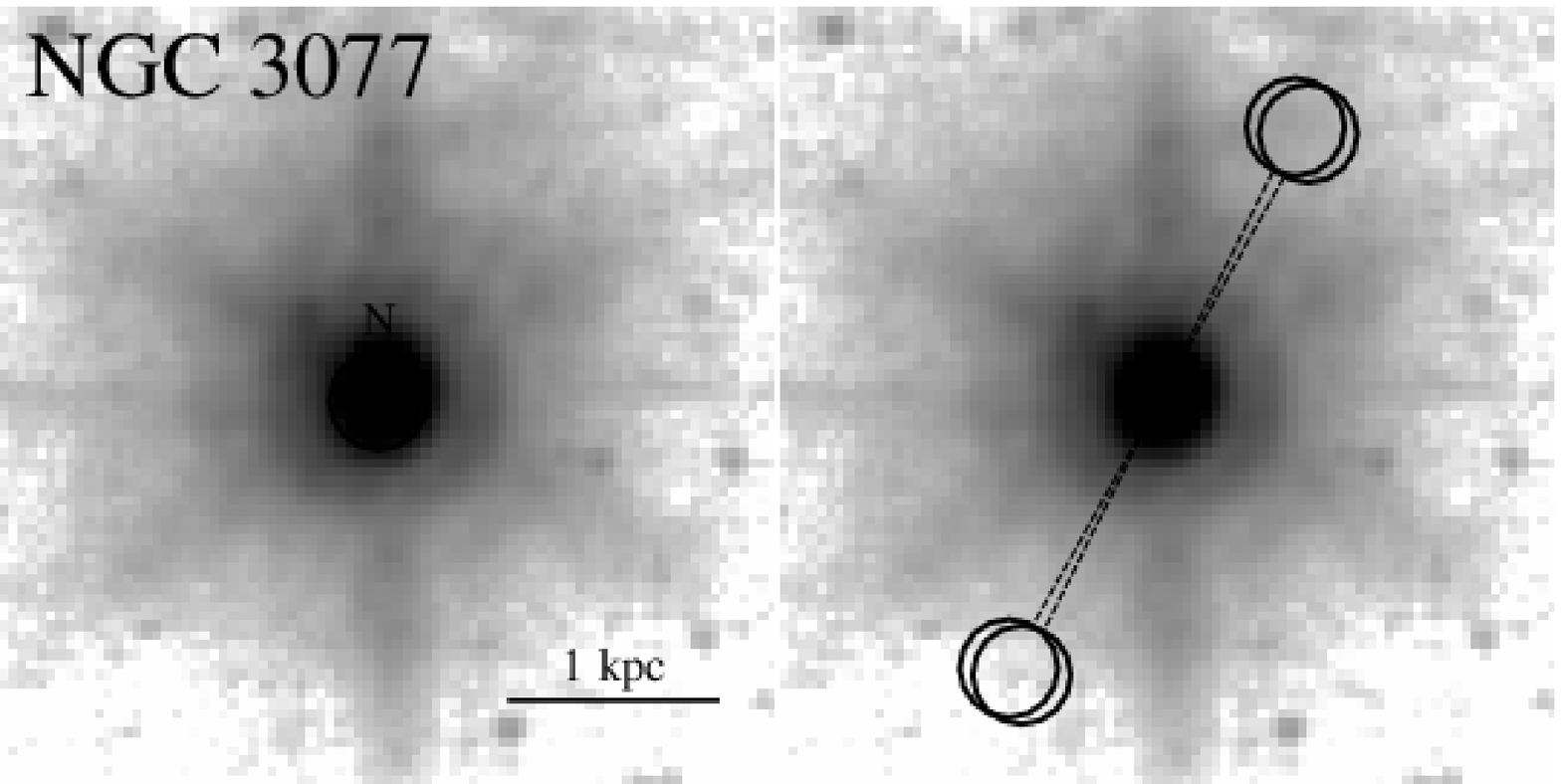,width=0.5\linewidth,clip=}\\
\epsfig{file=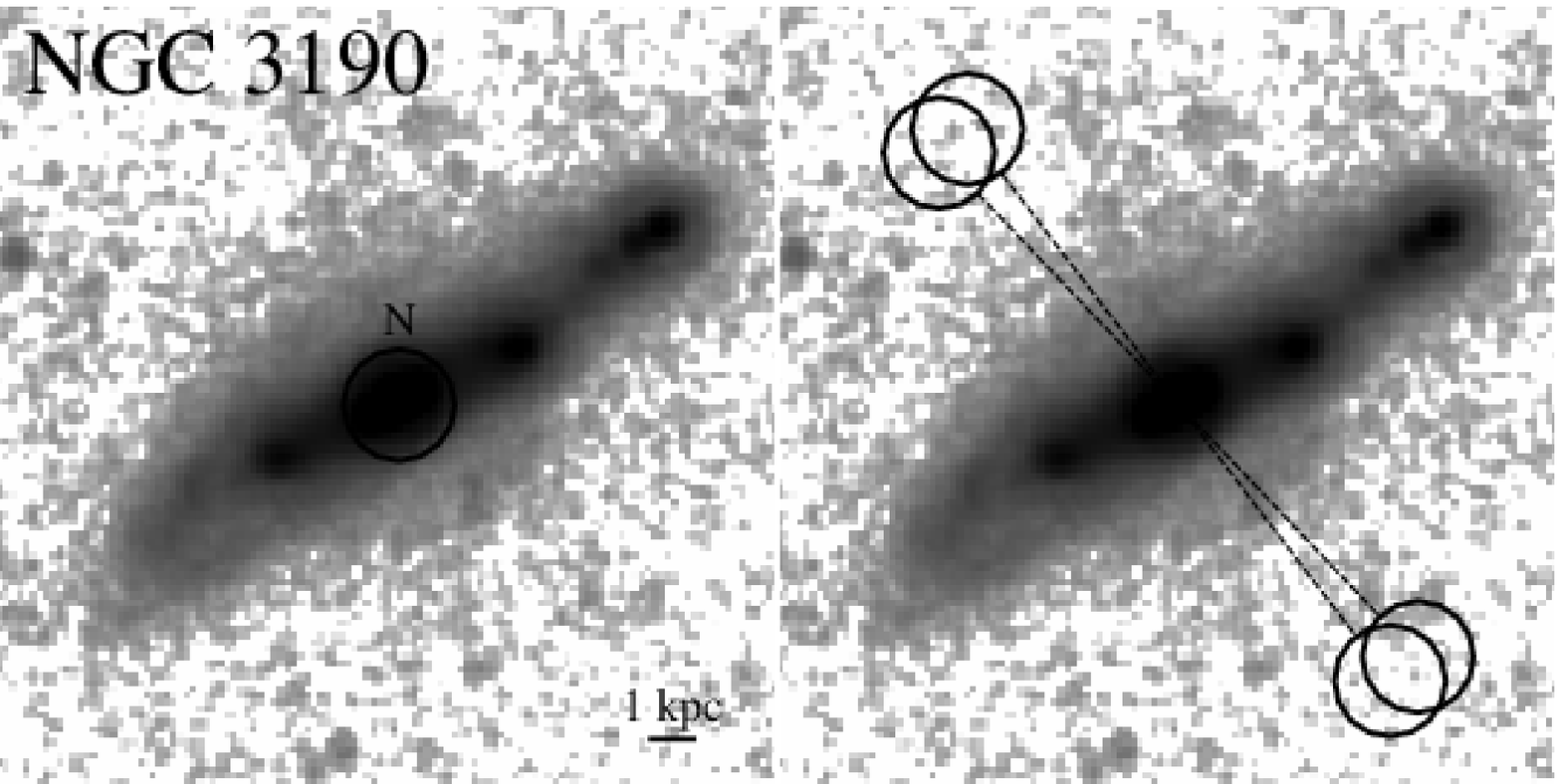,width=0.5\linewidth,clip=} & \epsfig{file=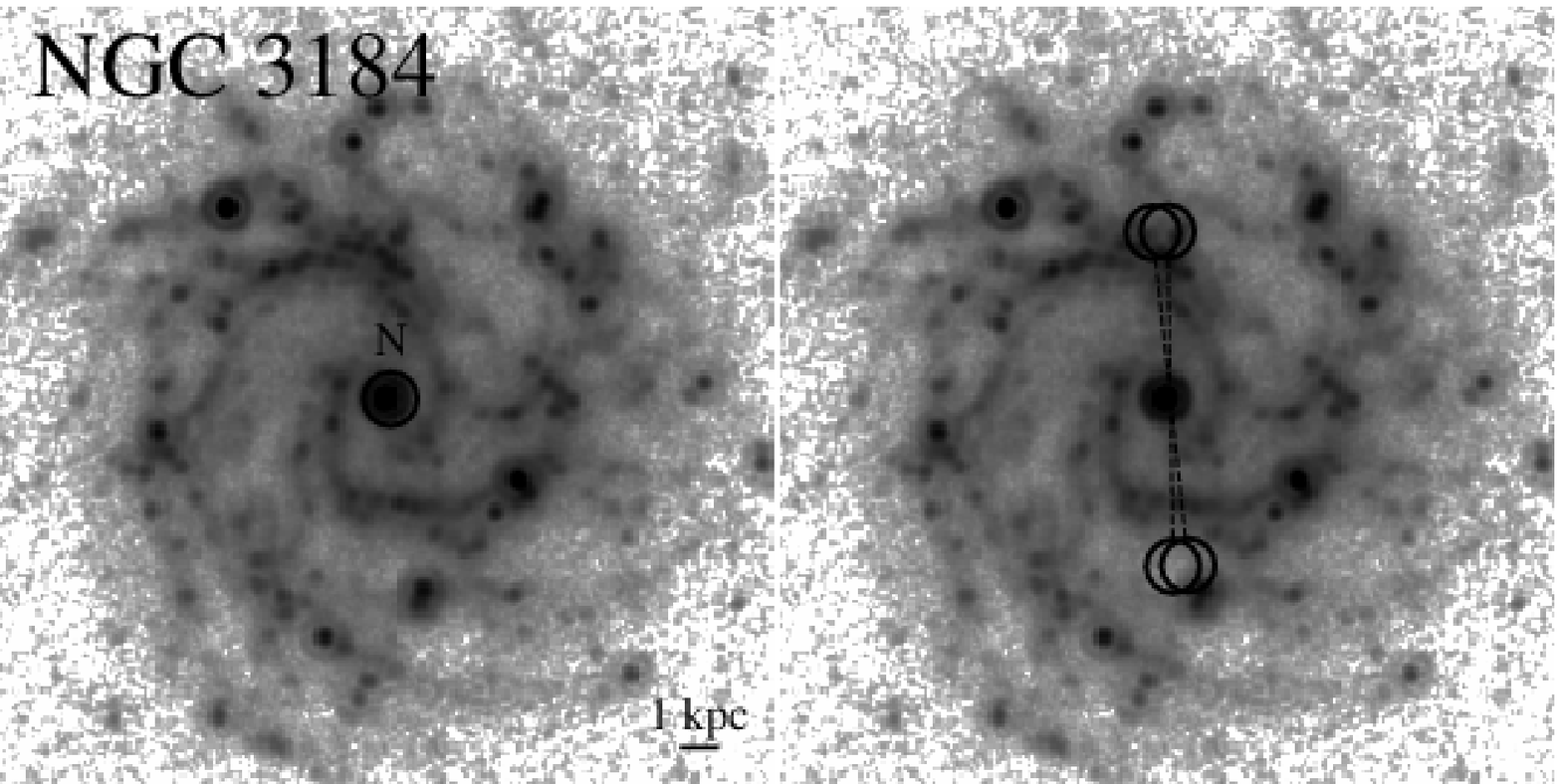,width=0.5\linewidth,clip=}\\
\epsfig{file=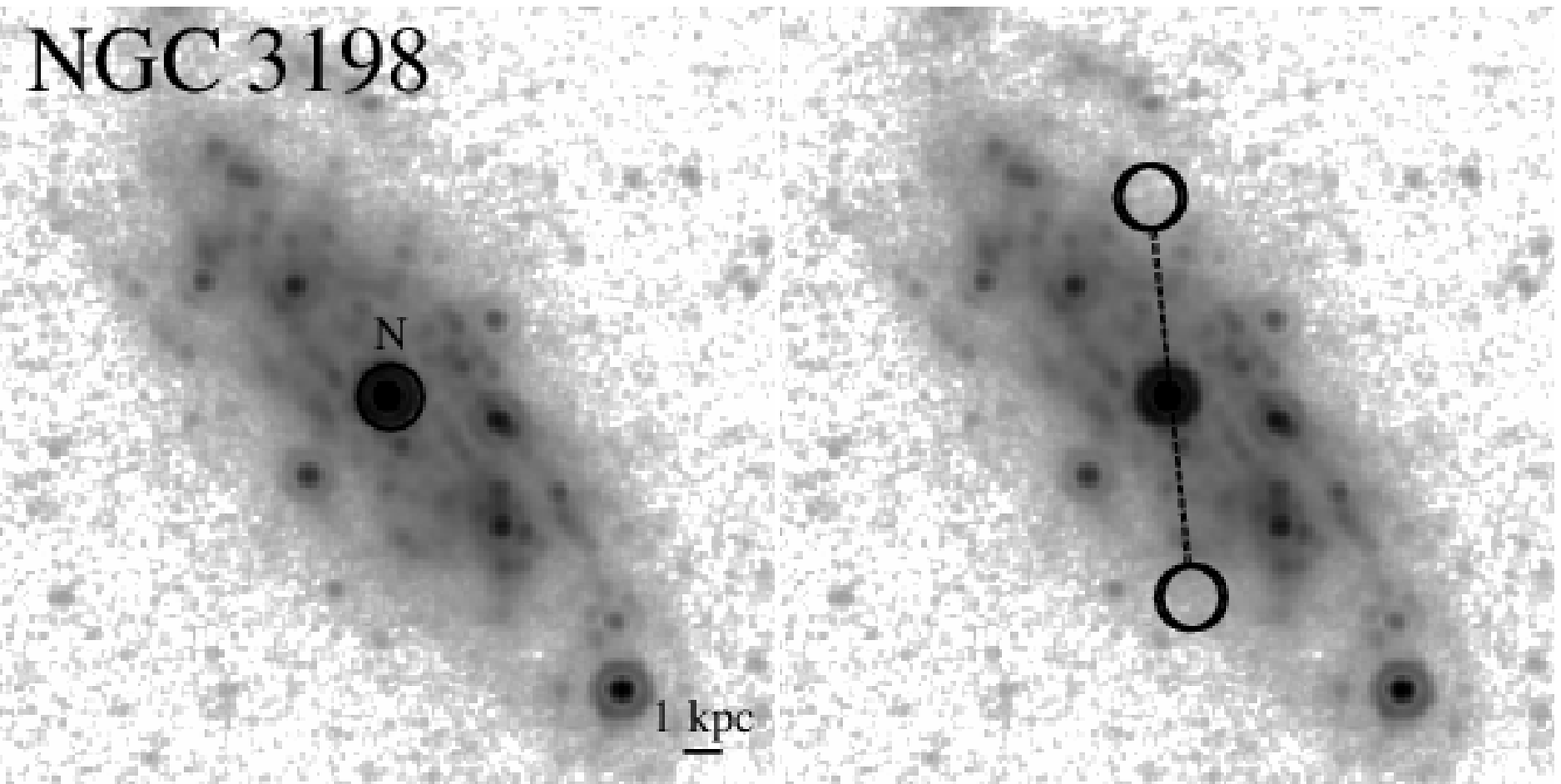,width=0.5\linewidth,clip=} & \epsfig{file=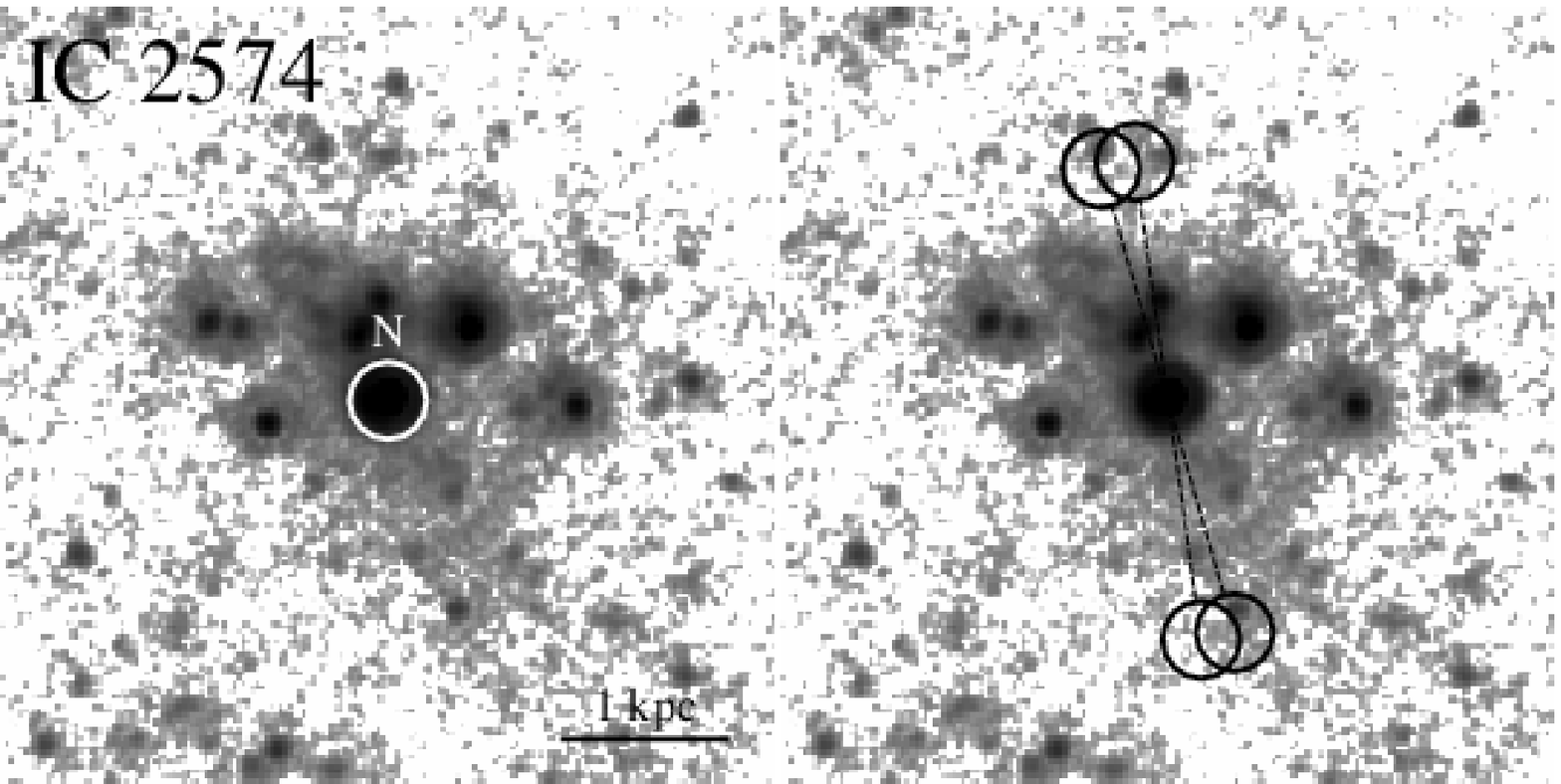,width=0.5\linewidth,clip=}
\end{tabular}
\end{center}
\caption{\it Continued}
\end{figure*}

\clearpage
\setcounter{figure}{0}
\begin{figure*}[ht!]
\begin{center}
\begin{tabular}{cc}
\epsfig{file=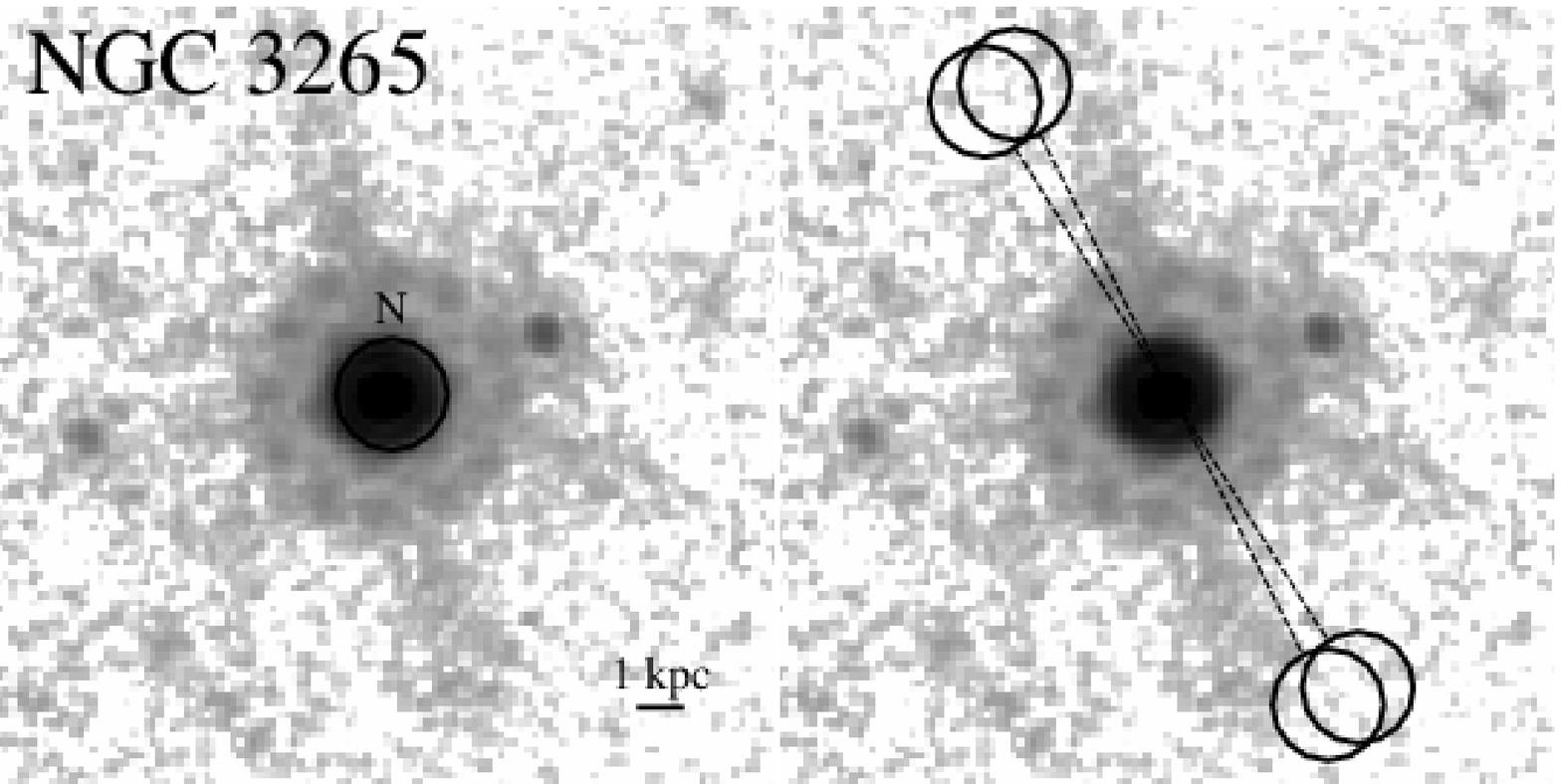,width=0.5\linewidth,clip=} & \epsfig{file=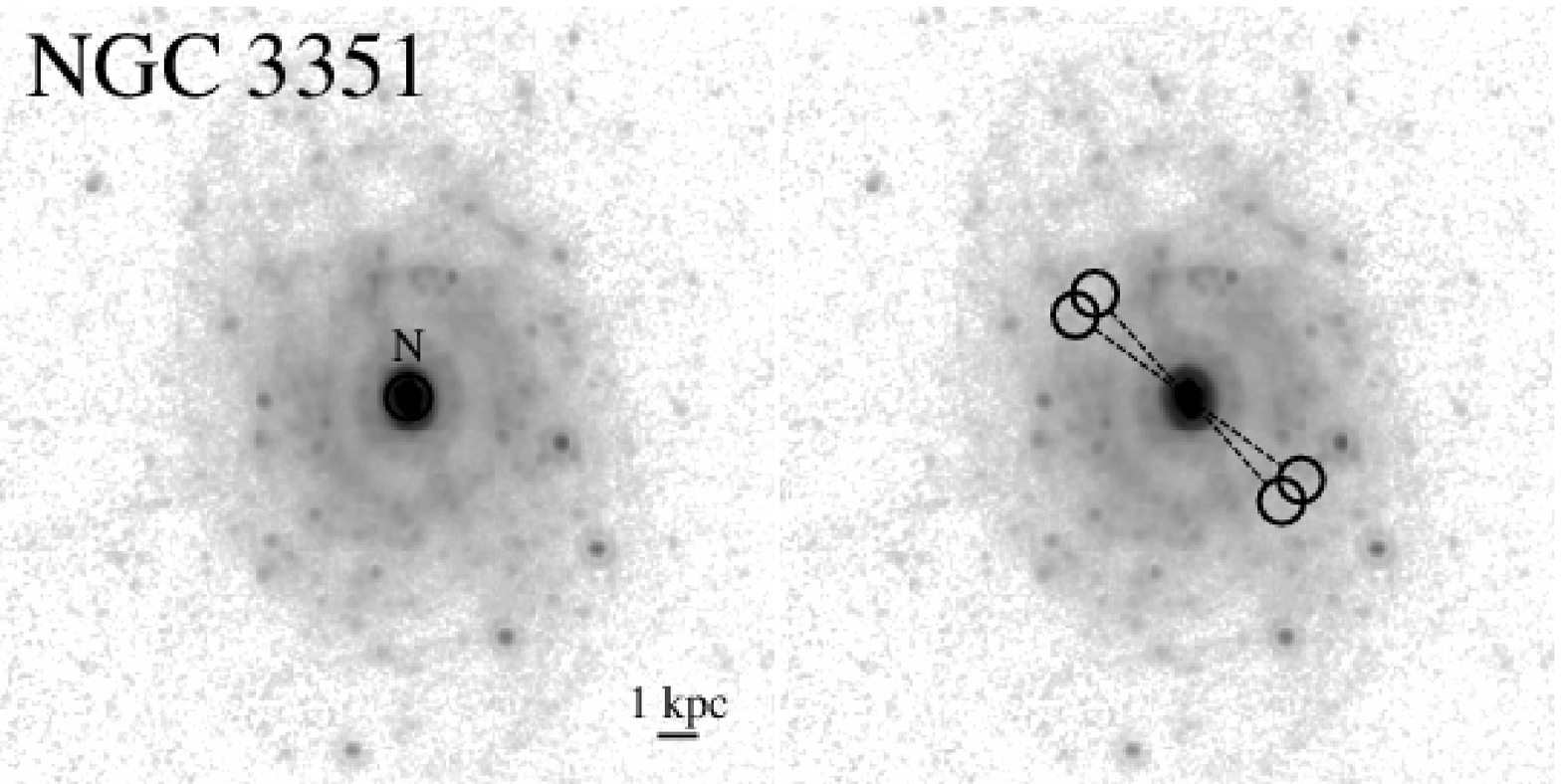,width=0.5\linewidth,clip=}\\
\epsfig{file=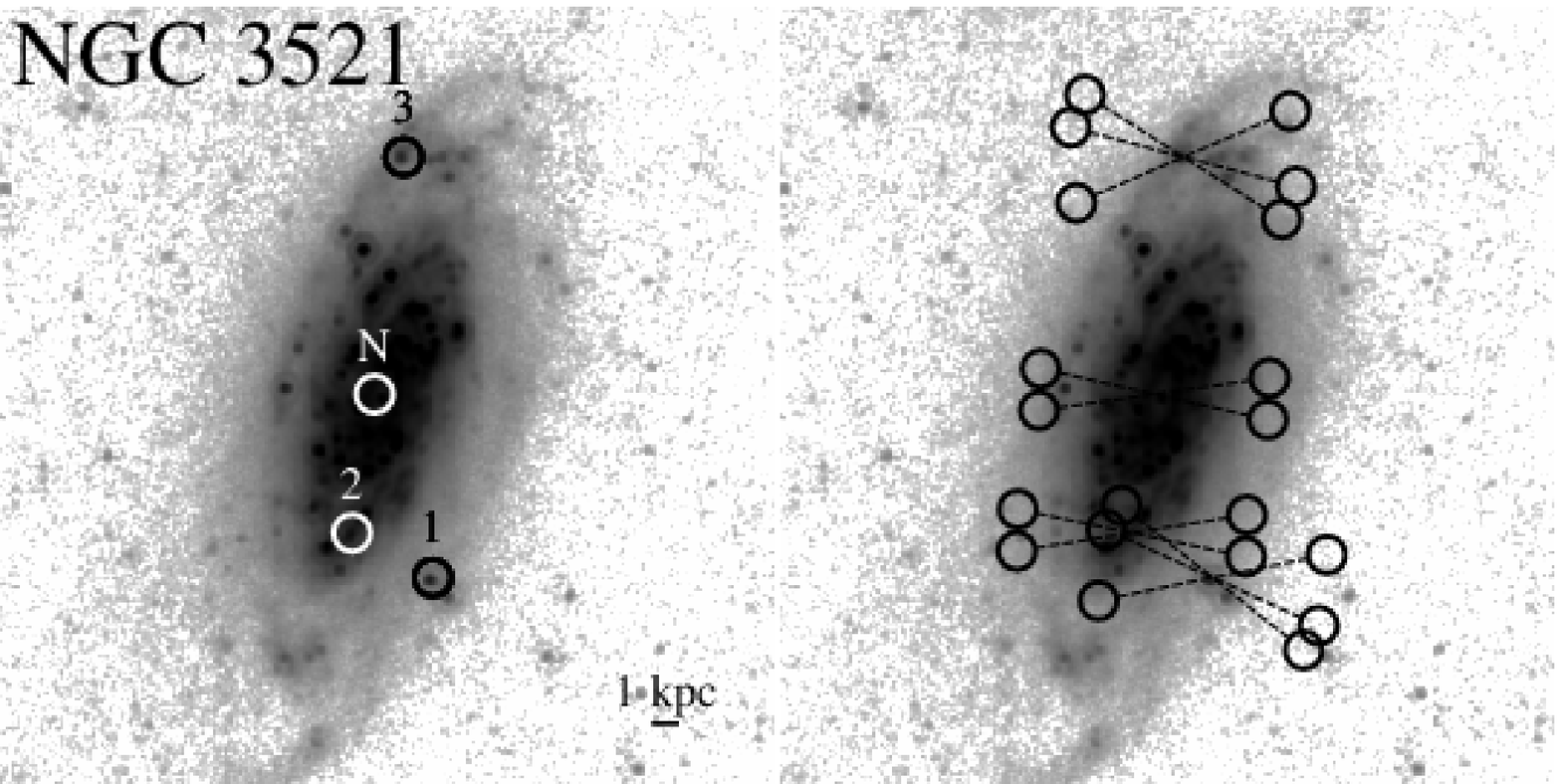,width=0.5\linewidth,clip=} & \epsfig{file=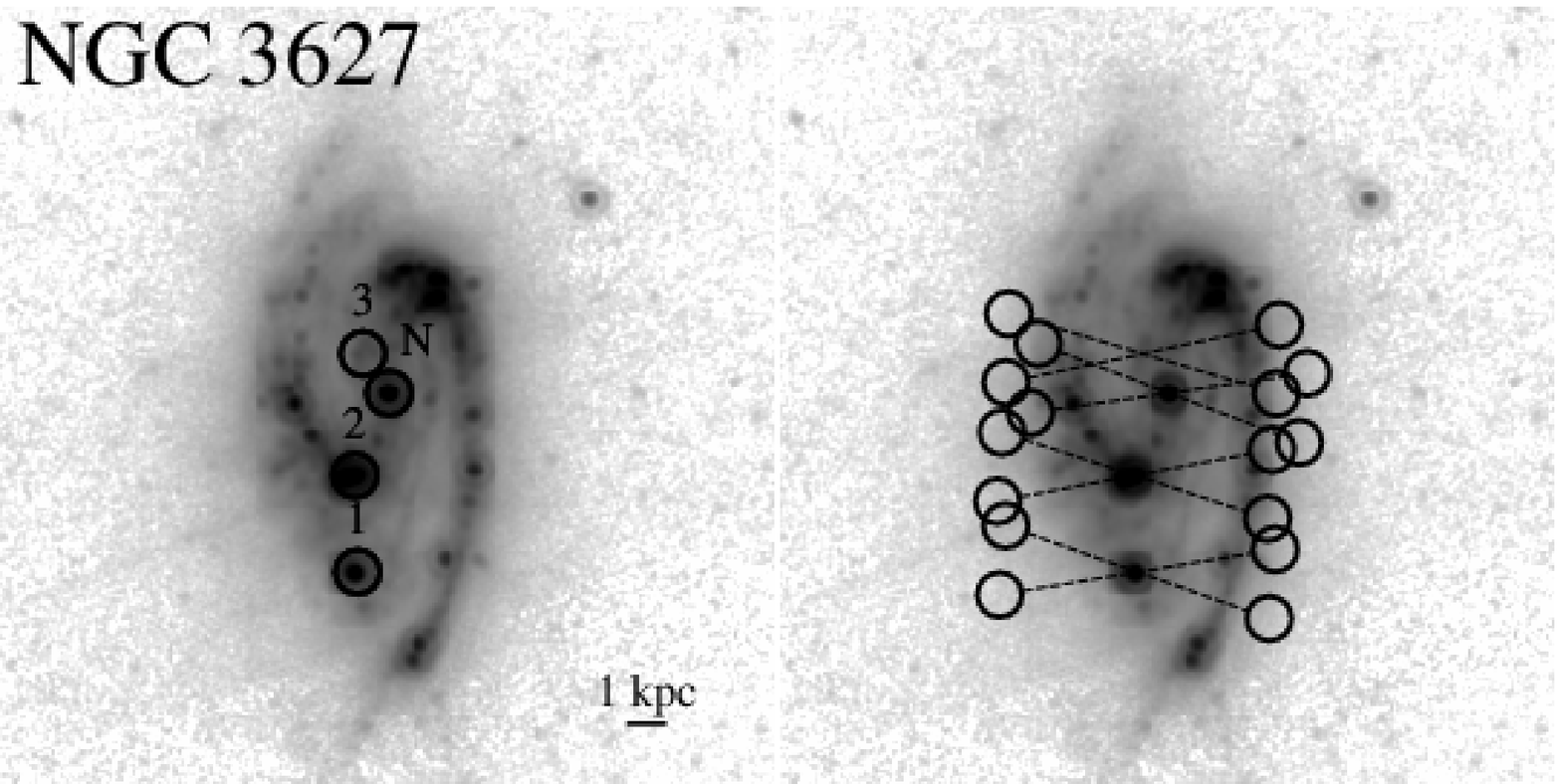,width=0.5\linewidth,clip=}\\
\epsfig{file=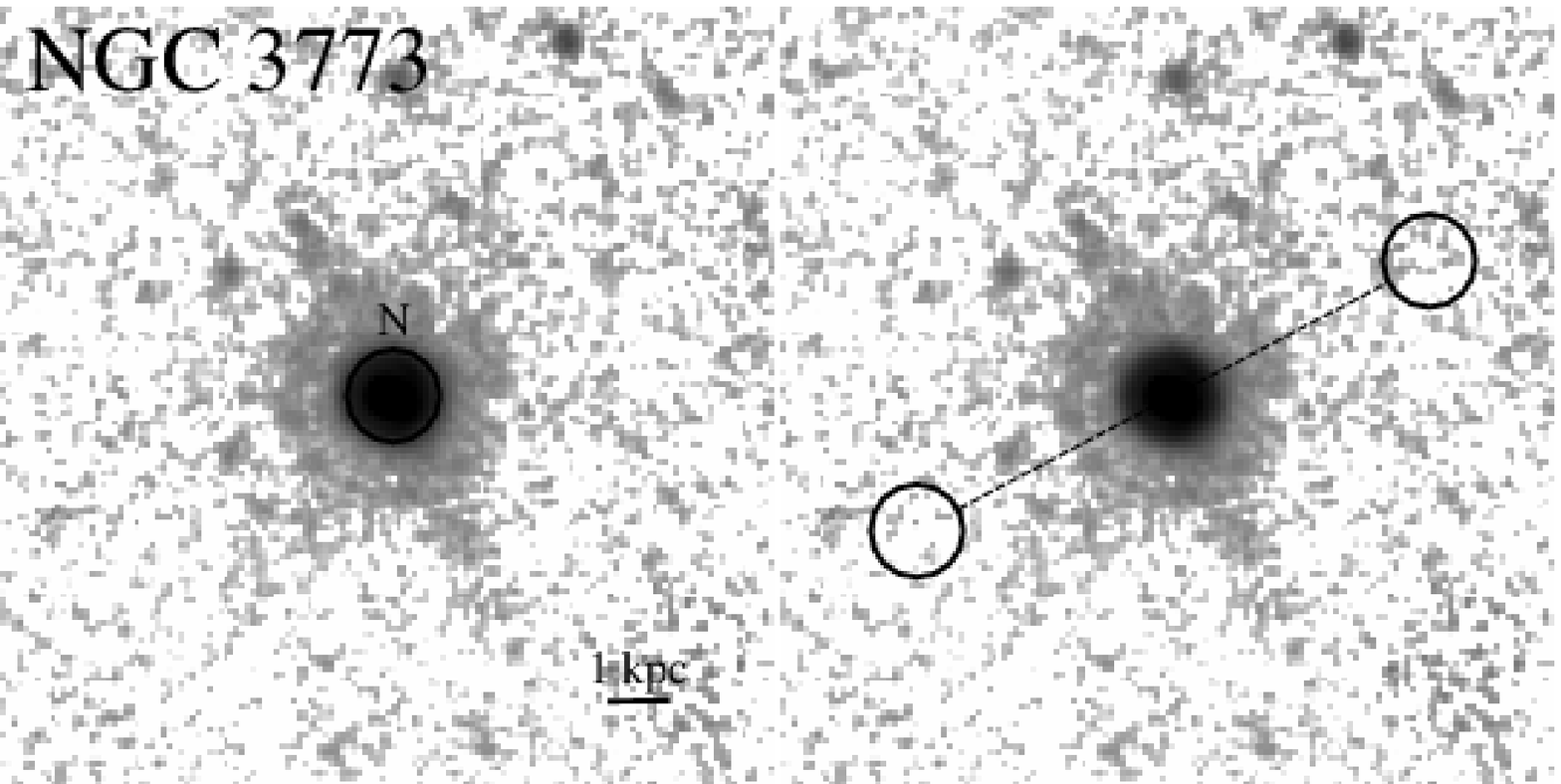,width=0.5\linewidth,clip=} & \epsfig{file=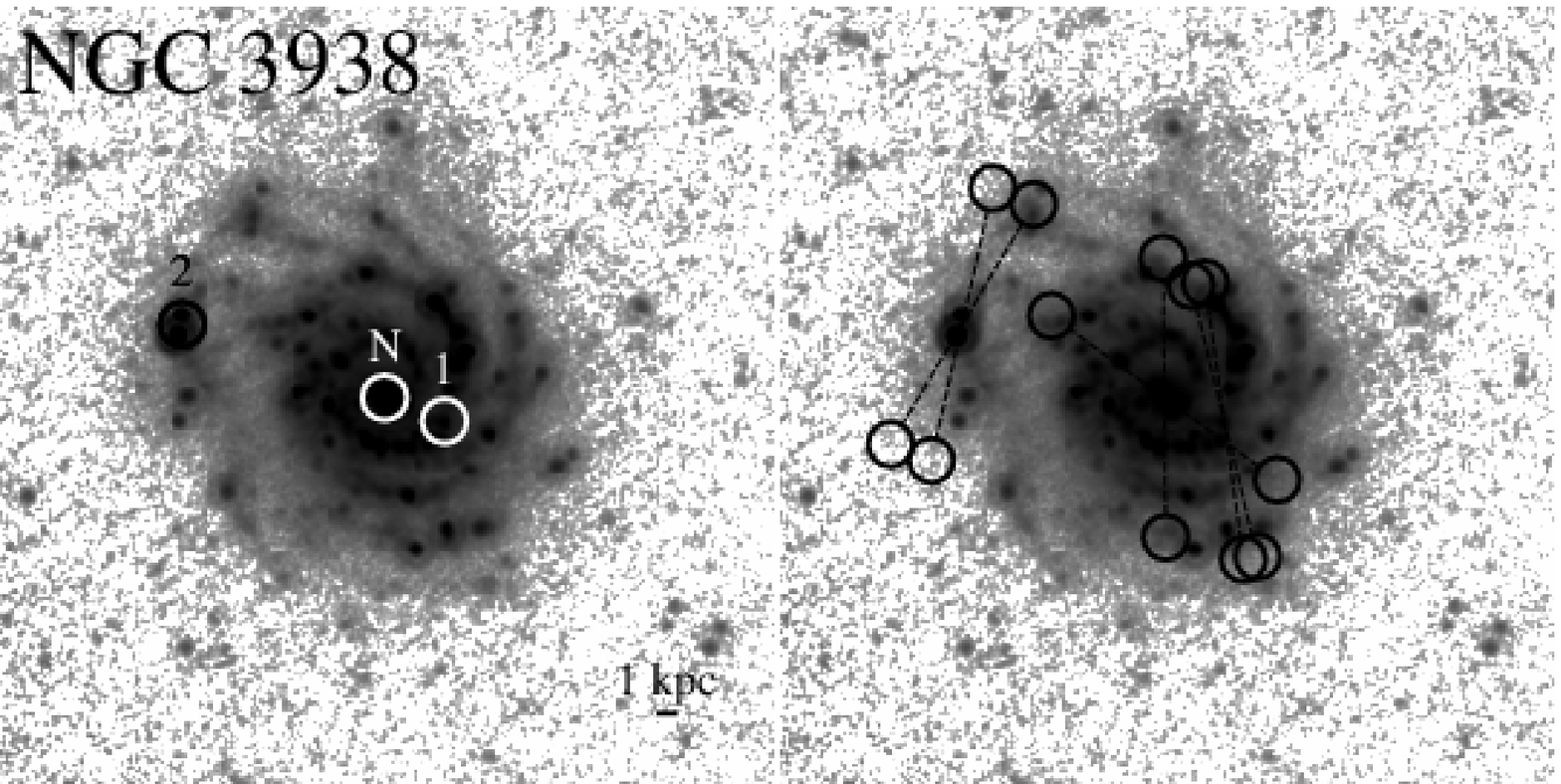,width=0.5\linewidth,clip=}\\
\epsfig{file=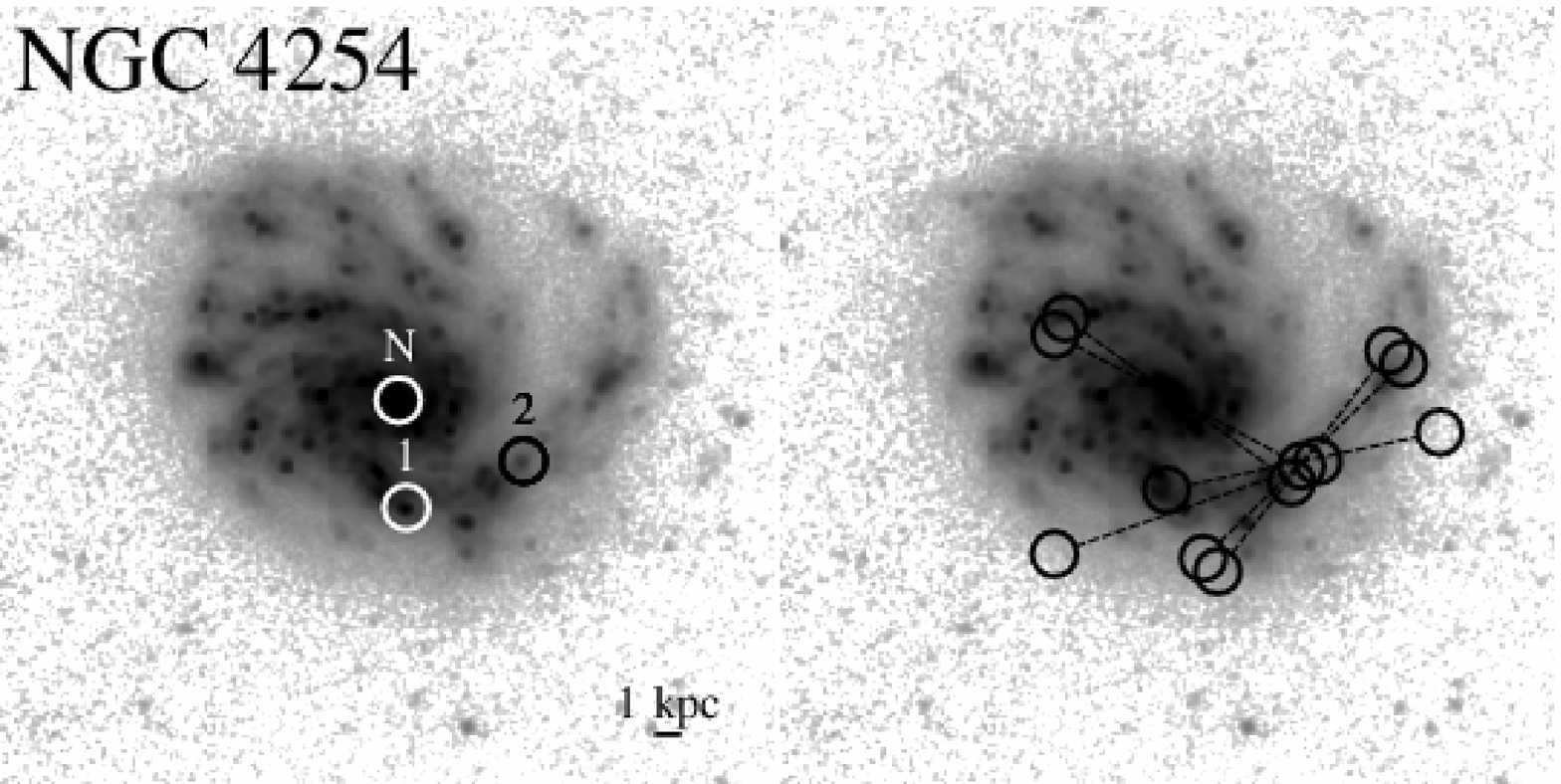,width=0.5\linewidth,clip=} & \epsfig{file=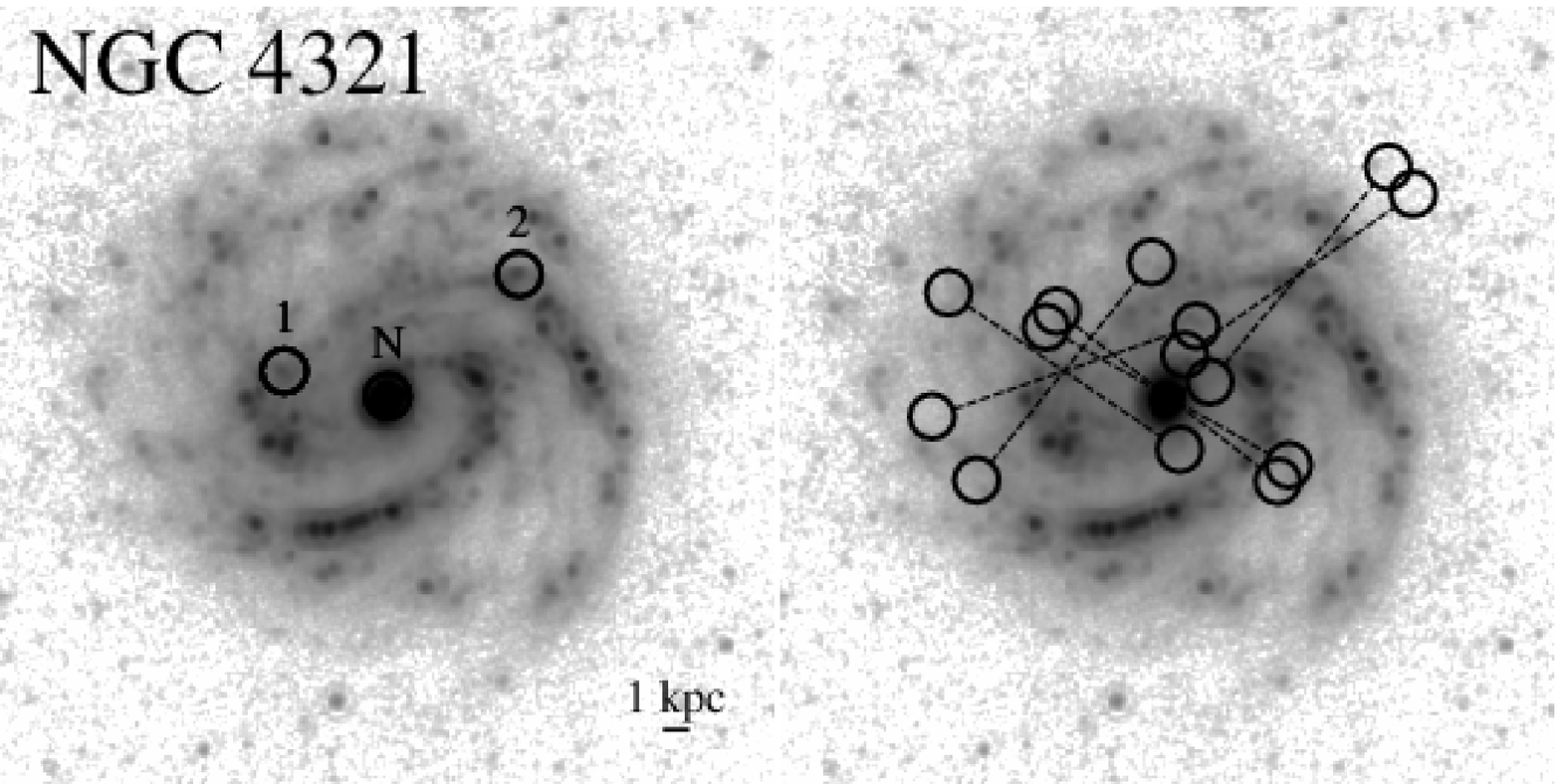,width=0.5\linewidth,clip=}\\
\epsfig{file=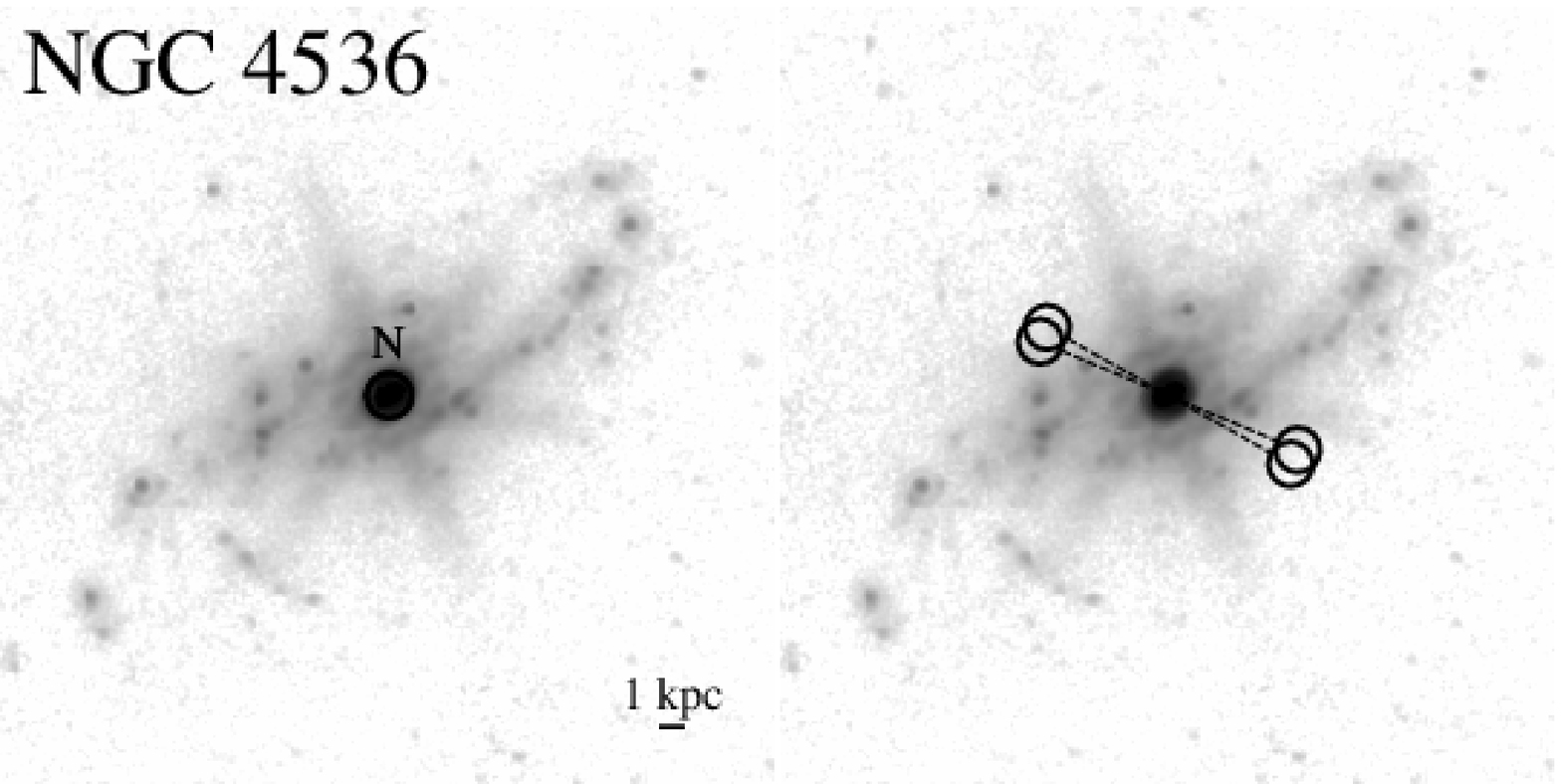,width=0.5\linewidth,clip=} & \epsfig{file=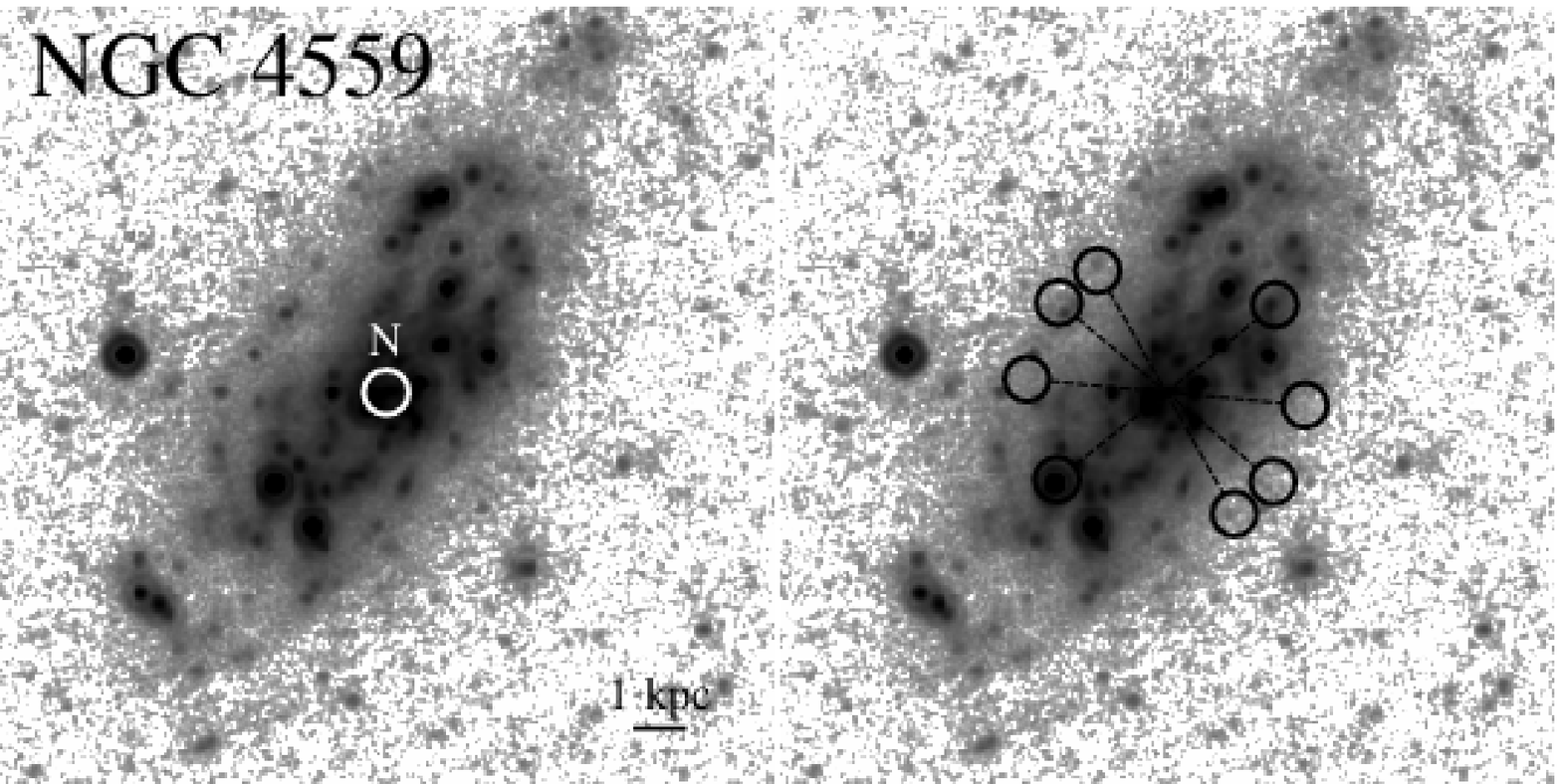,width=0.5\linewidth,clip=}
\end{tabular}
\end{center}
\caption{\it Continued}
\end{figure*}

\clearpage
\setcounter{figure}{0}
\begin{figure*}[ht!]
\begin{center}
\begin{tabular}{cc}
\epsfig{file=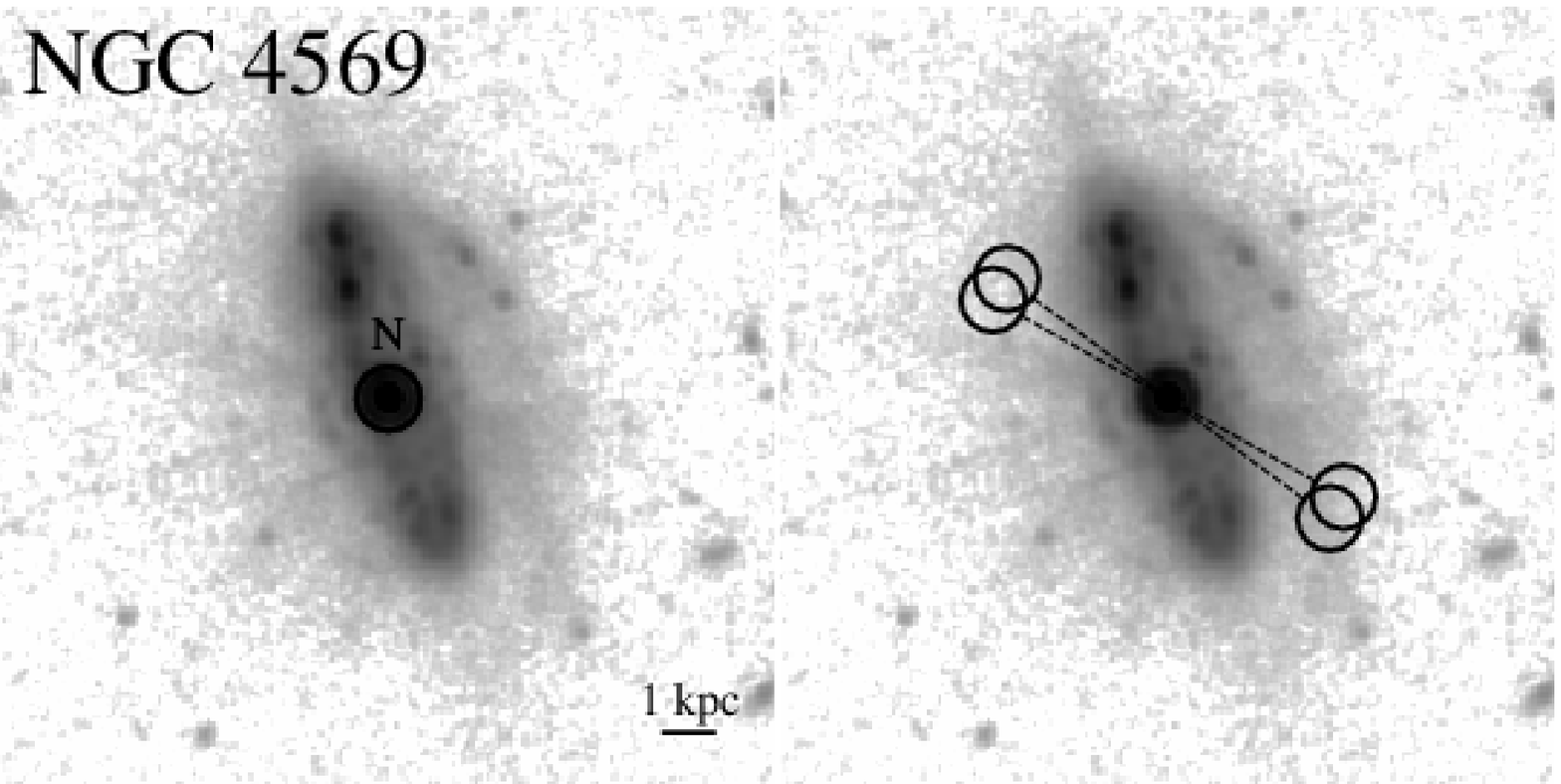,width=0.5\linewidth,clip=} & \epsfig{file=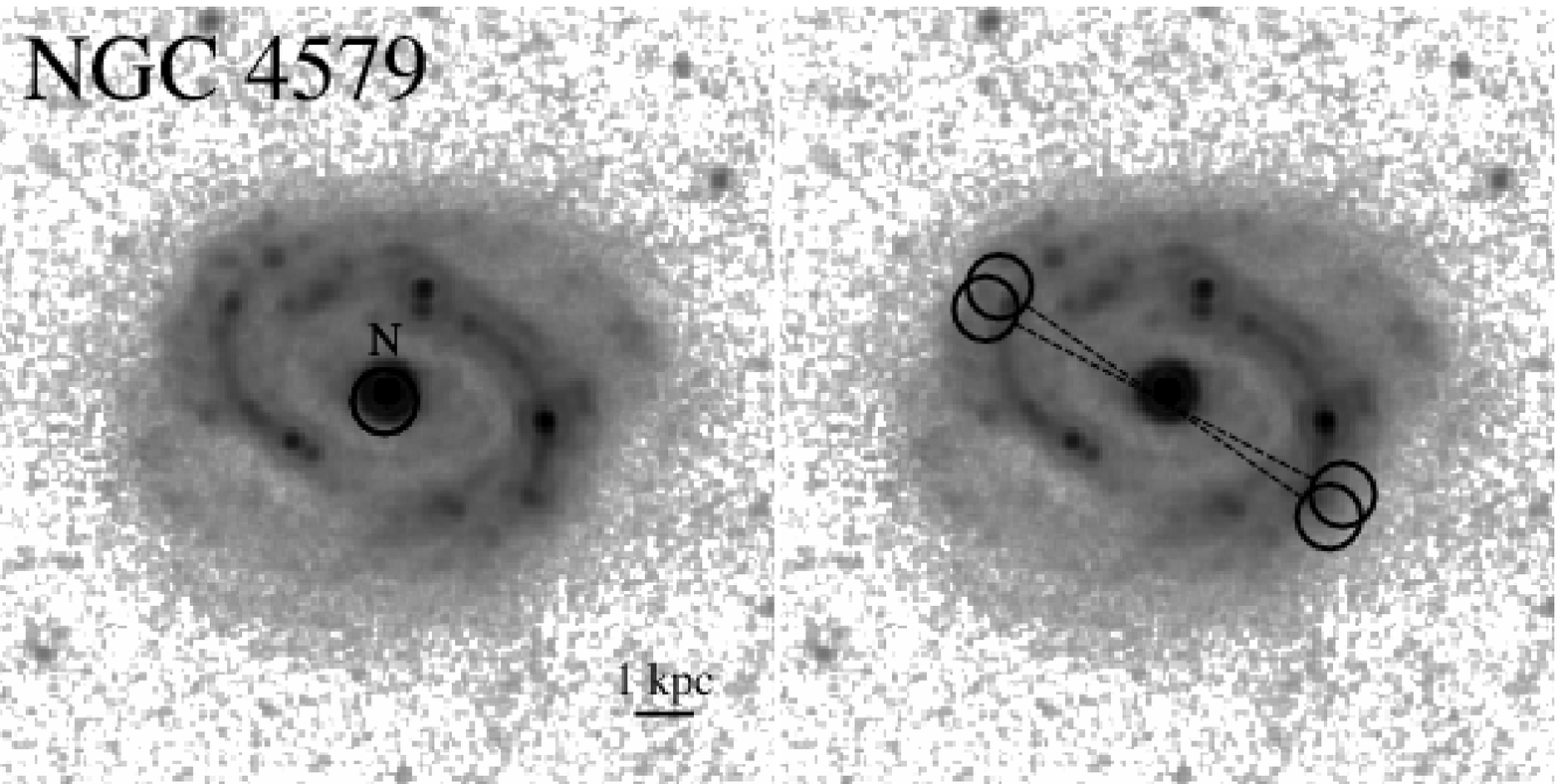,width=0.5\linewidth,clip=}\\
\epsfig{file=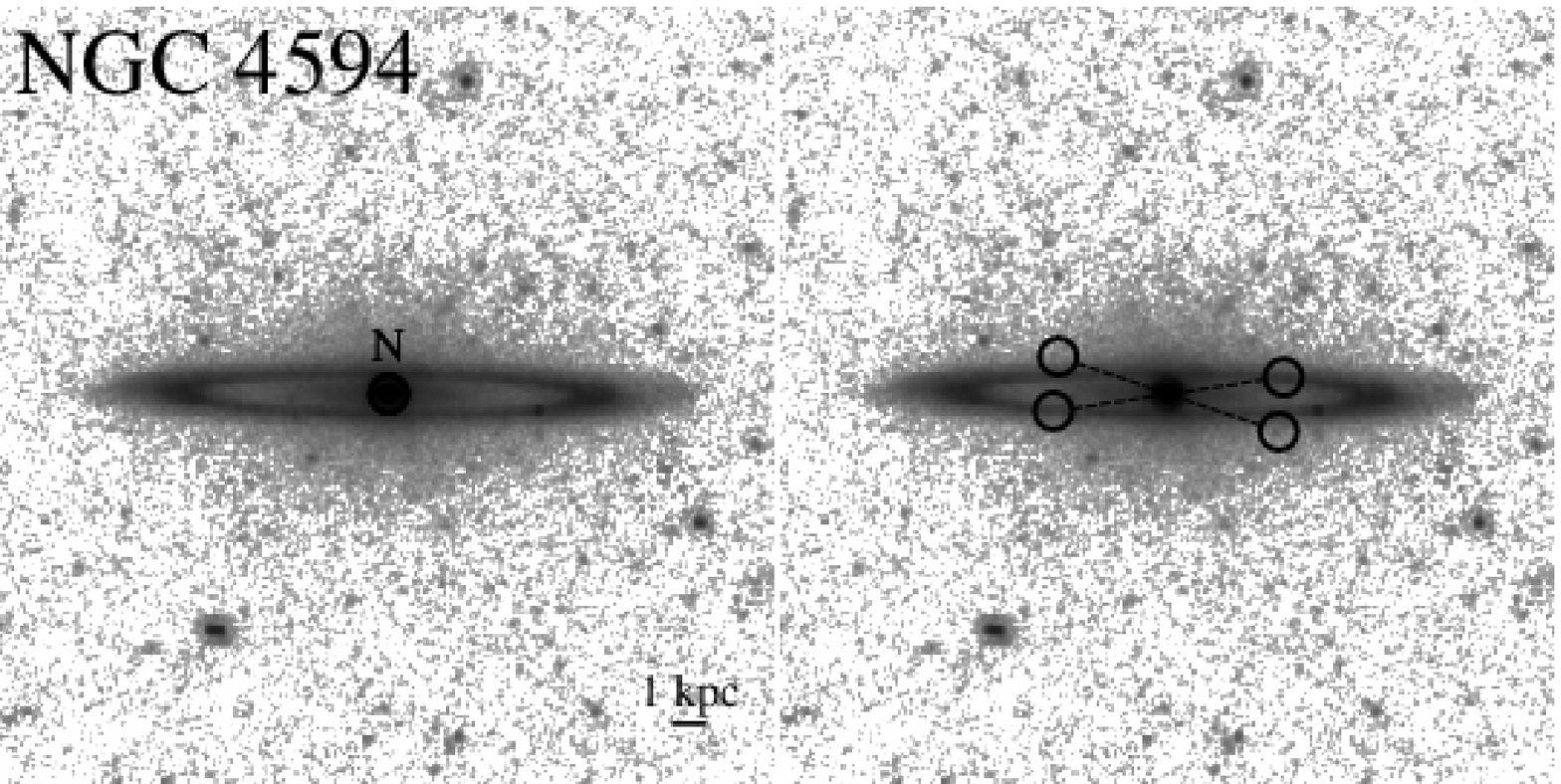,width=0.5\linewidth,clip=} & \epsfig{file=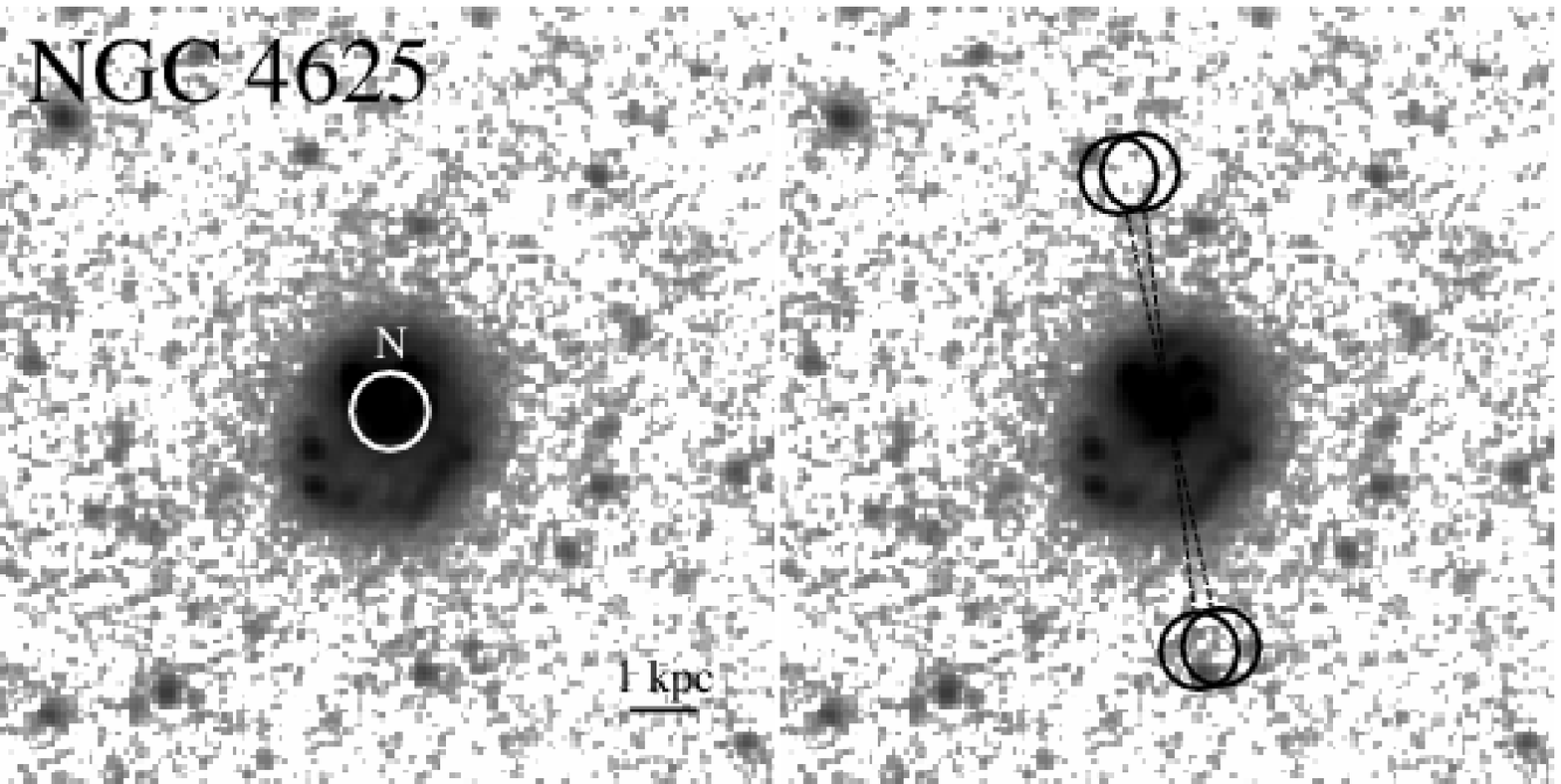,width=0.5\linewidth,clip=}\\
\epsfig{file=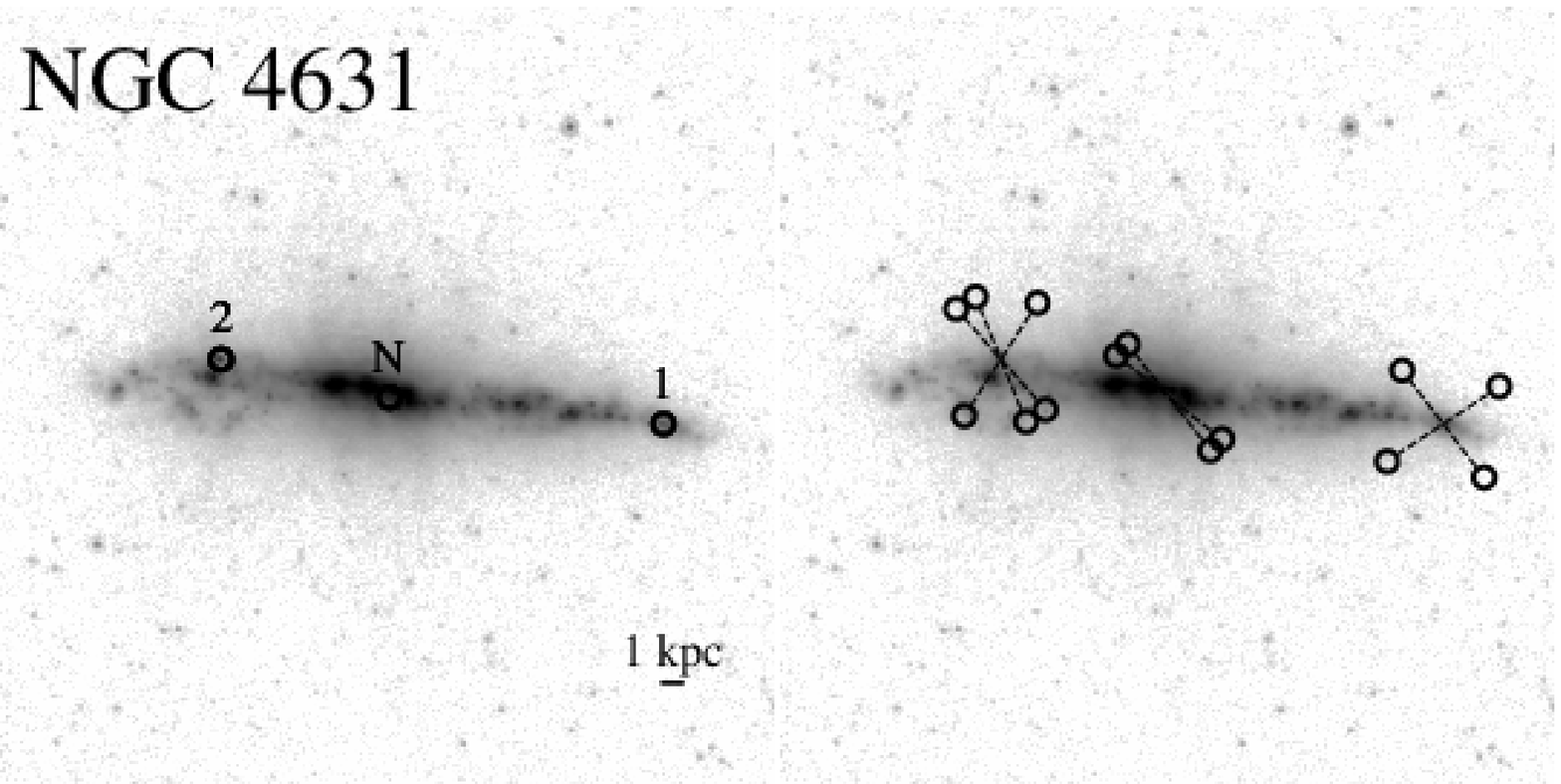,width=0.5\linewidth,clip=} & \epsfig{file=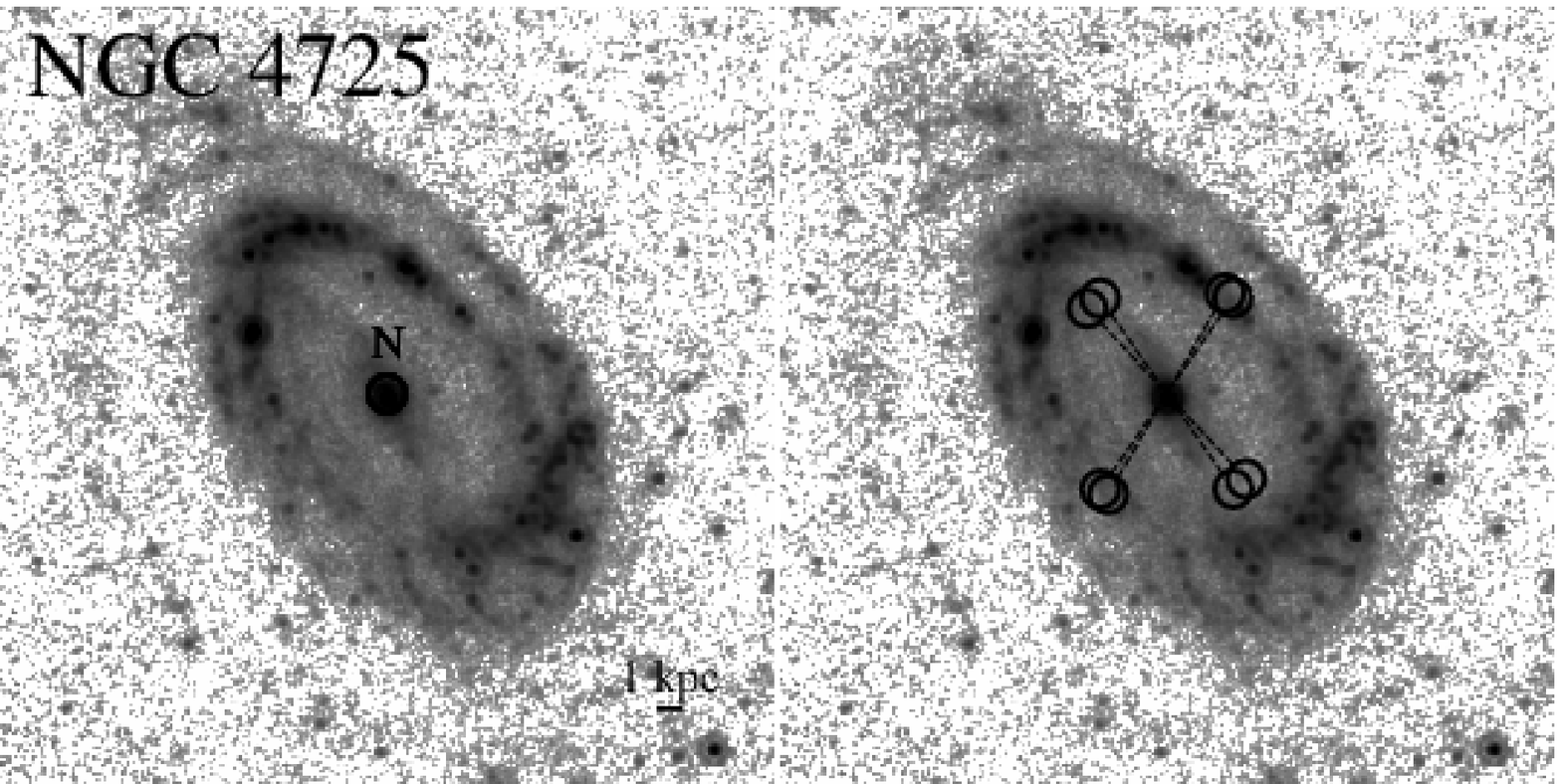,width=0.5\linewidth,clip=}\\
\epsfig{file=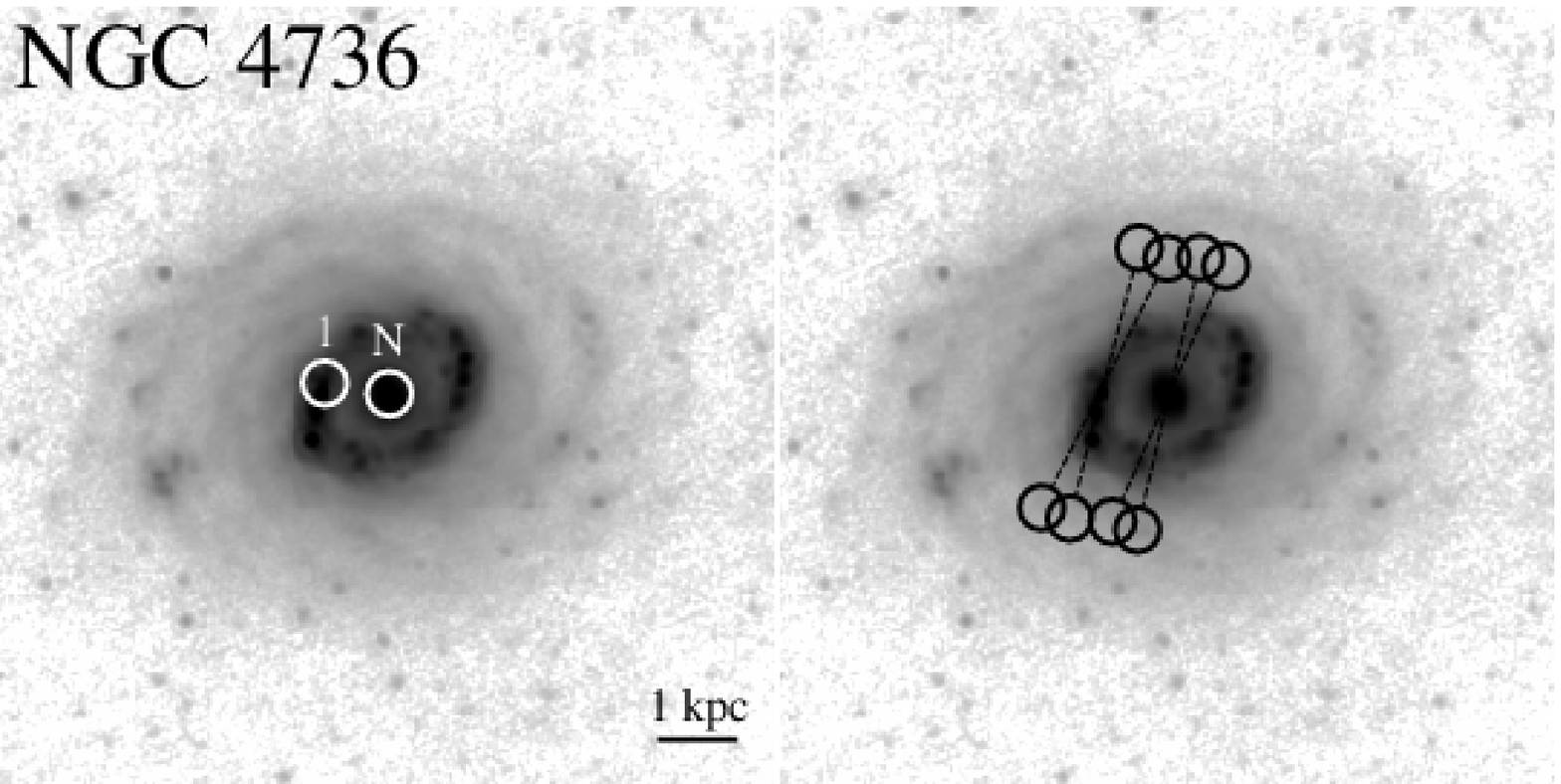,width=0.5\linewidth,clip=} & \epsfig{file=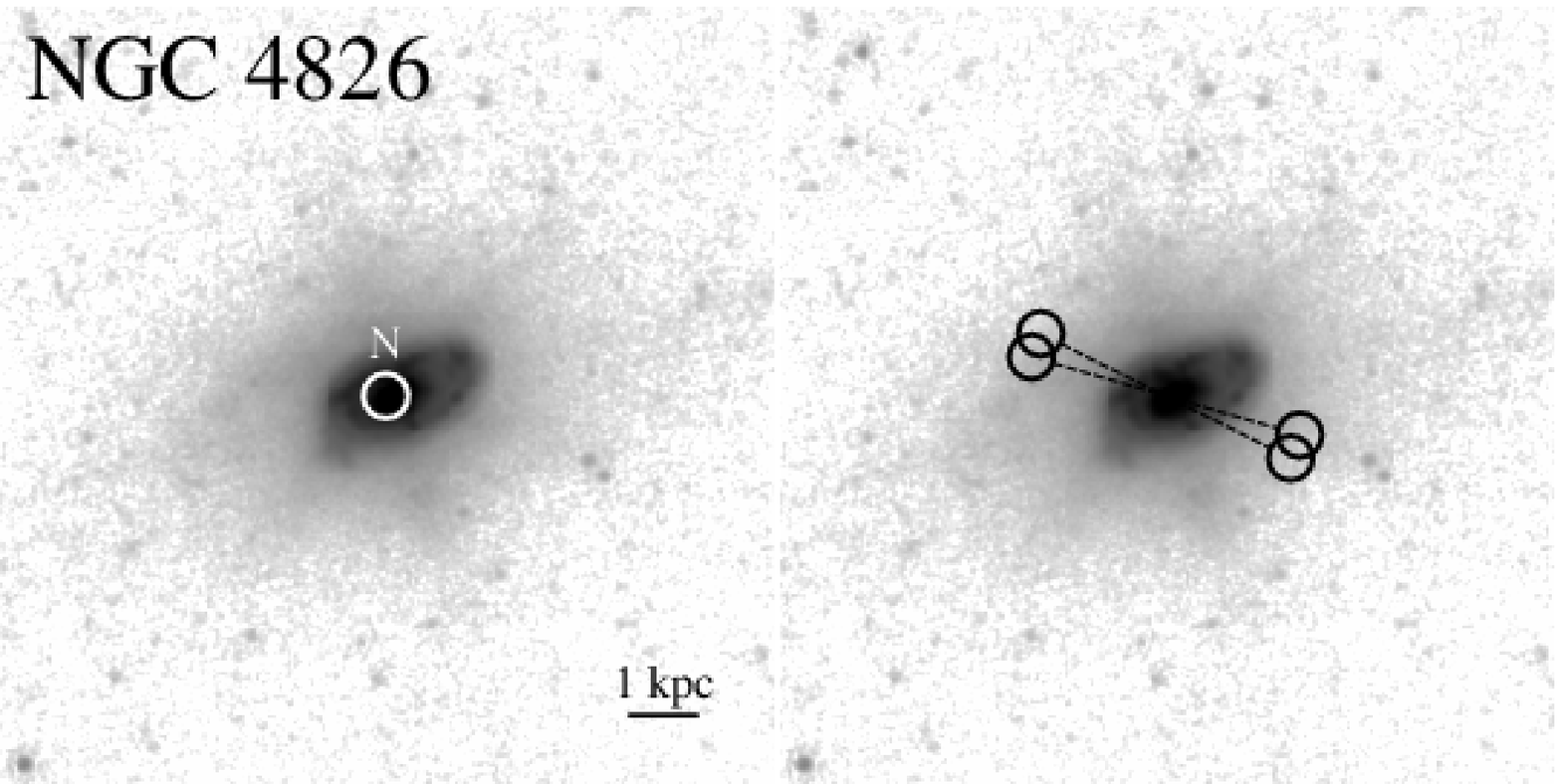,width=0.5\linewidth,clip=}\\
\epsfig{file=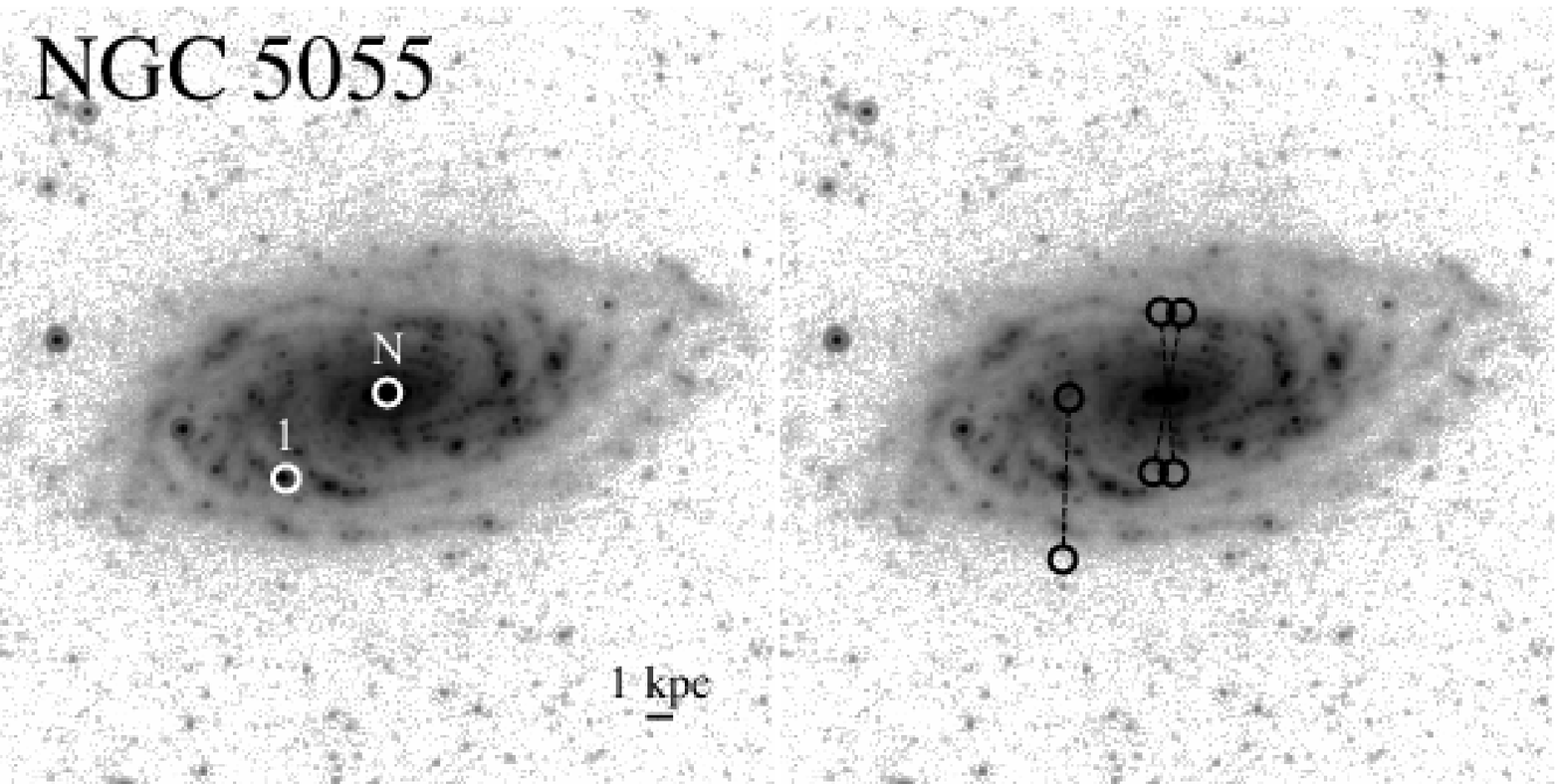,width=0.5\linewidth,clip=} & \epsfig{file=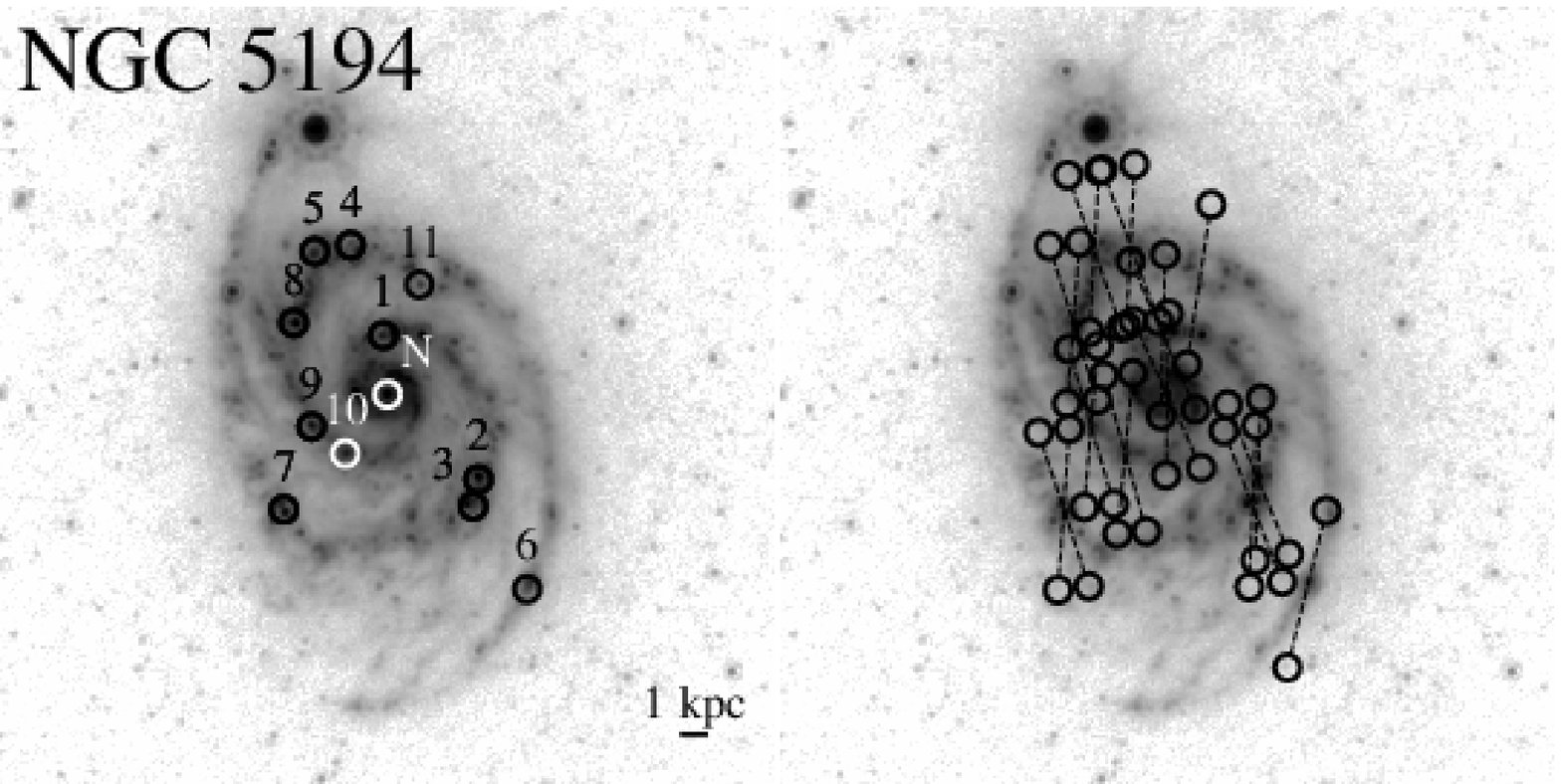,width=0.5\linewidth,clip=}
\end{tabular}
\end{center}
\caption{\it Continued}
\end{figure*}

\clearpage
\setcounter{figure}{0}
\begin{figure*}[ht!]
\begin{center}
\begin{tabular}{cc}
\epsfig{file=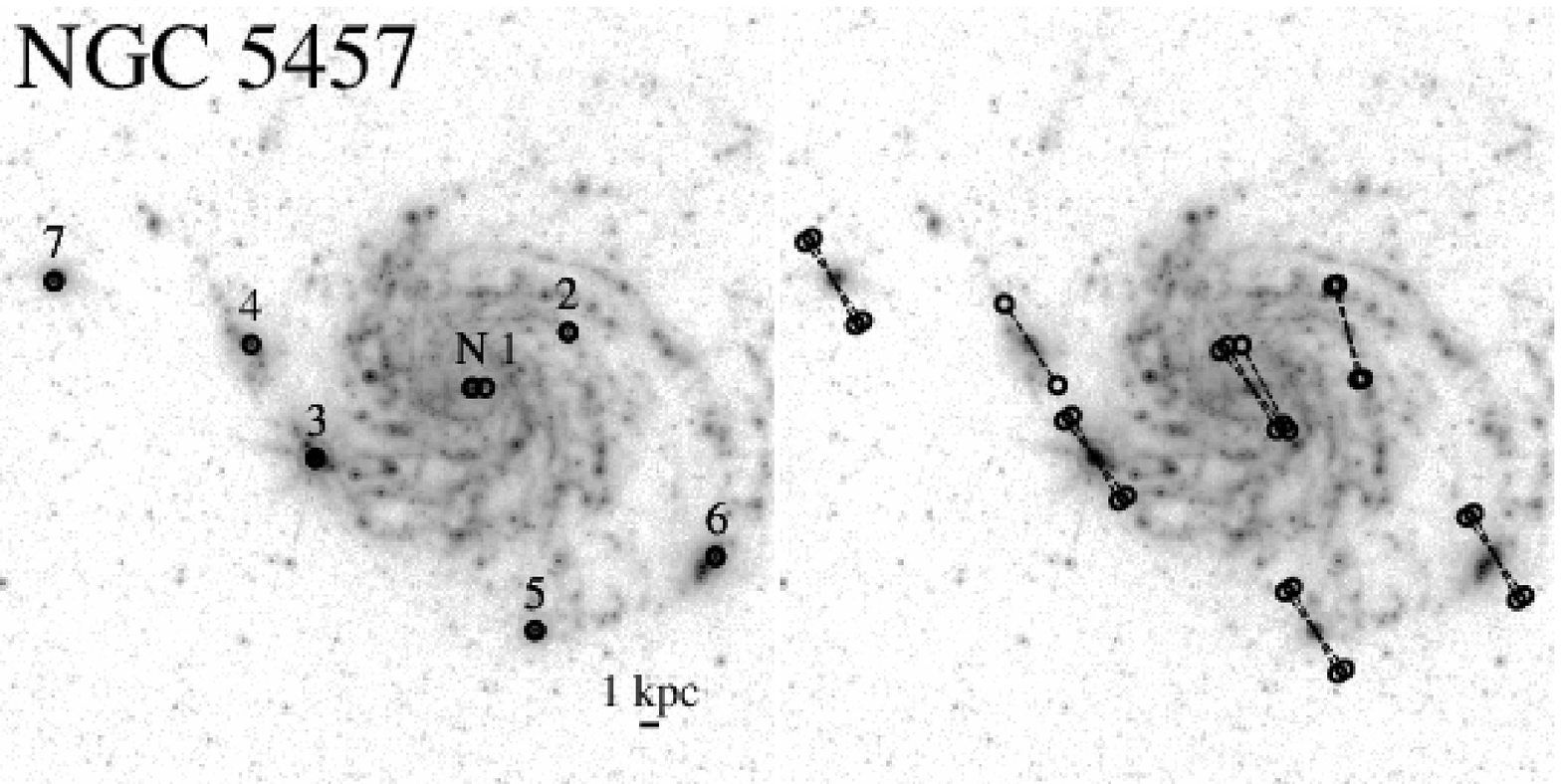,width=0.5\linewidth,clip=} & \epsfig{file=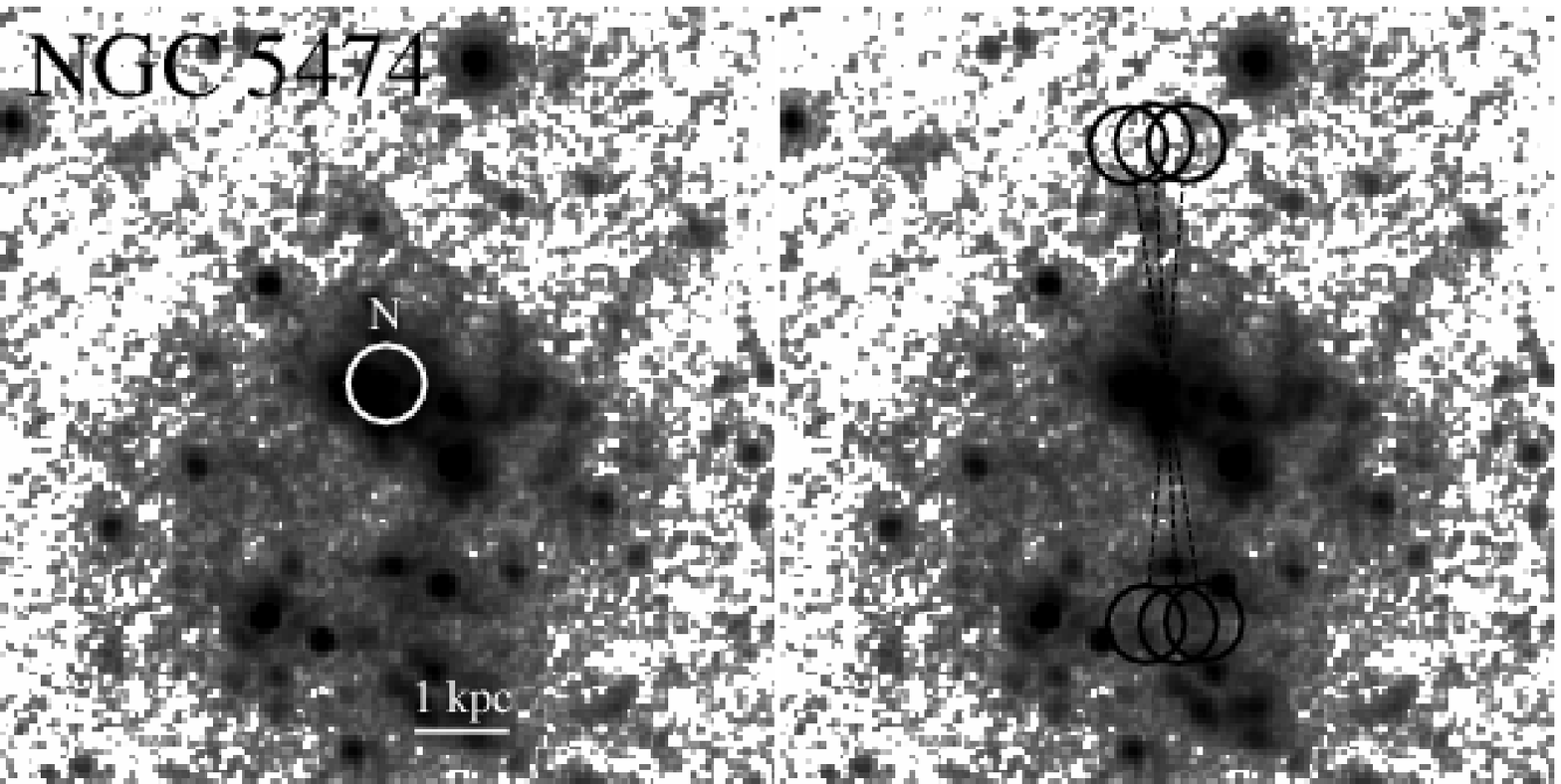,width=0.5\linewidth,clip=}\\
\epsfig{file=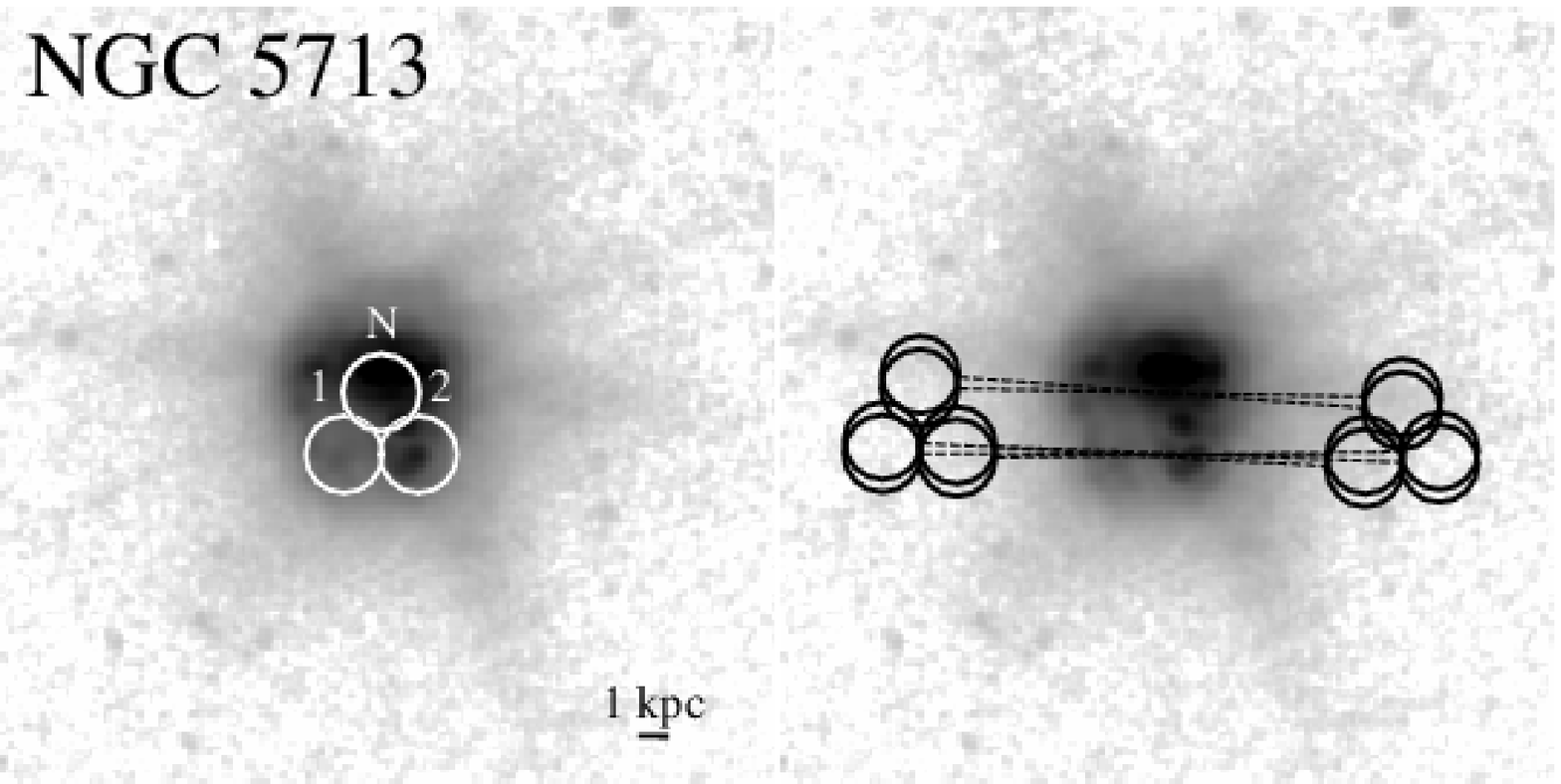,width=0.5\linewidth,clip=} & \epsfig{file=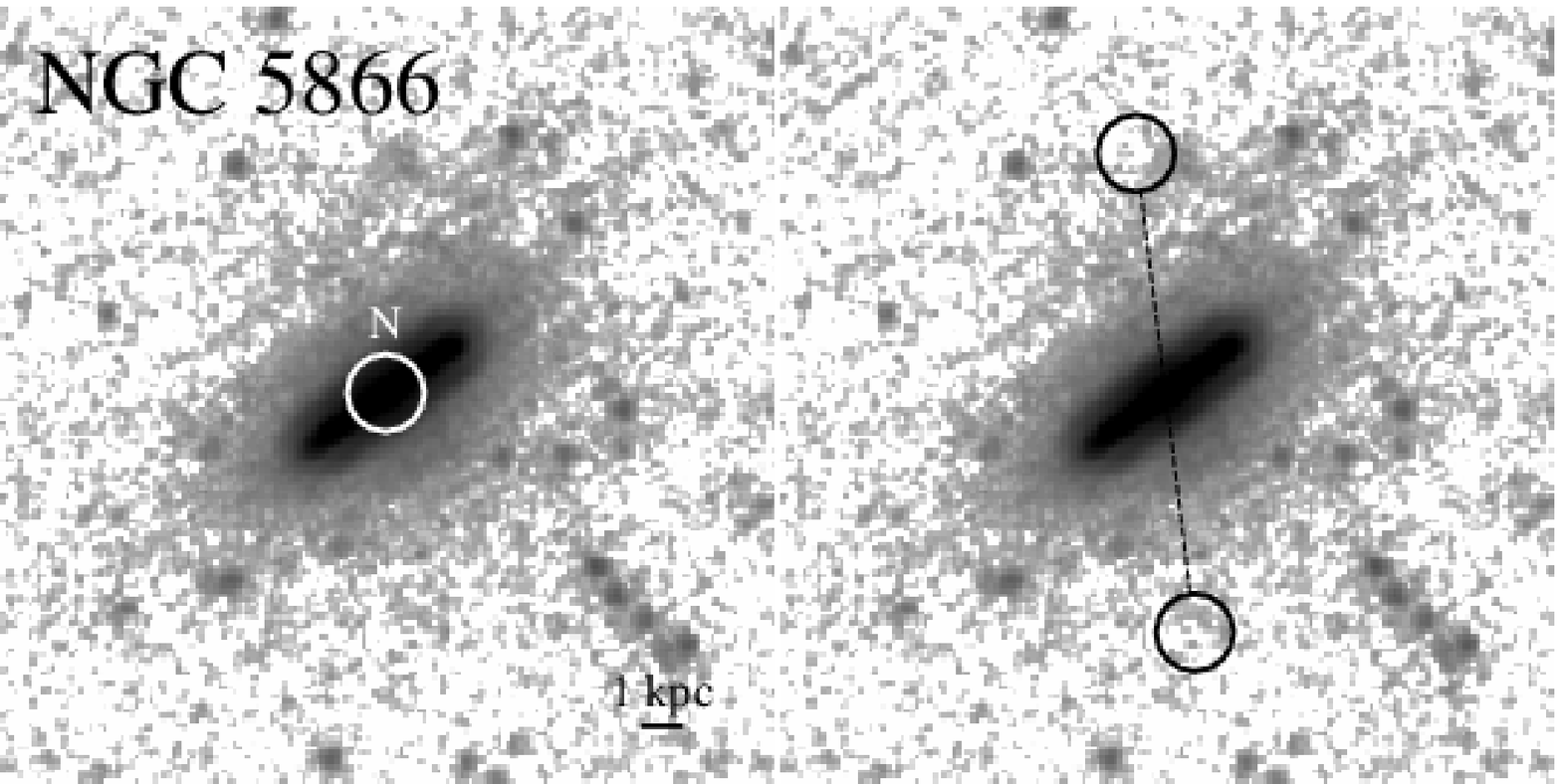,width=0.5\linewidth,clip=}\\
\epsfig{file=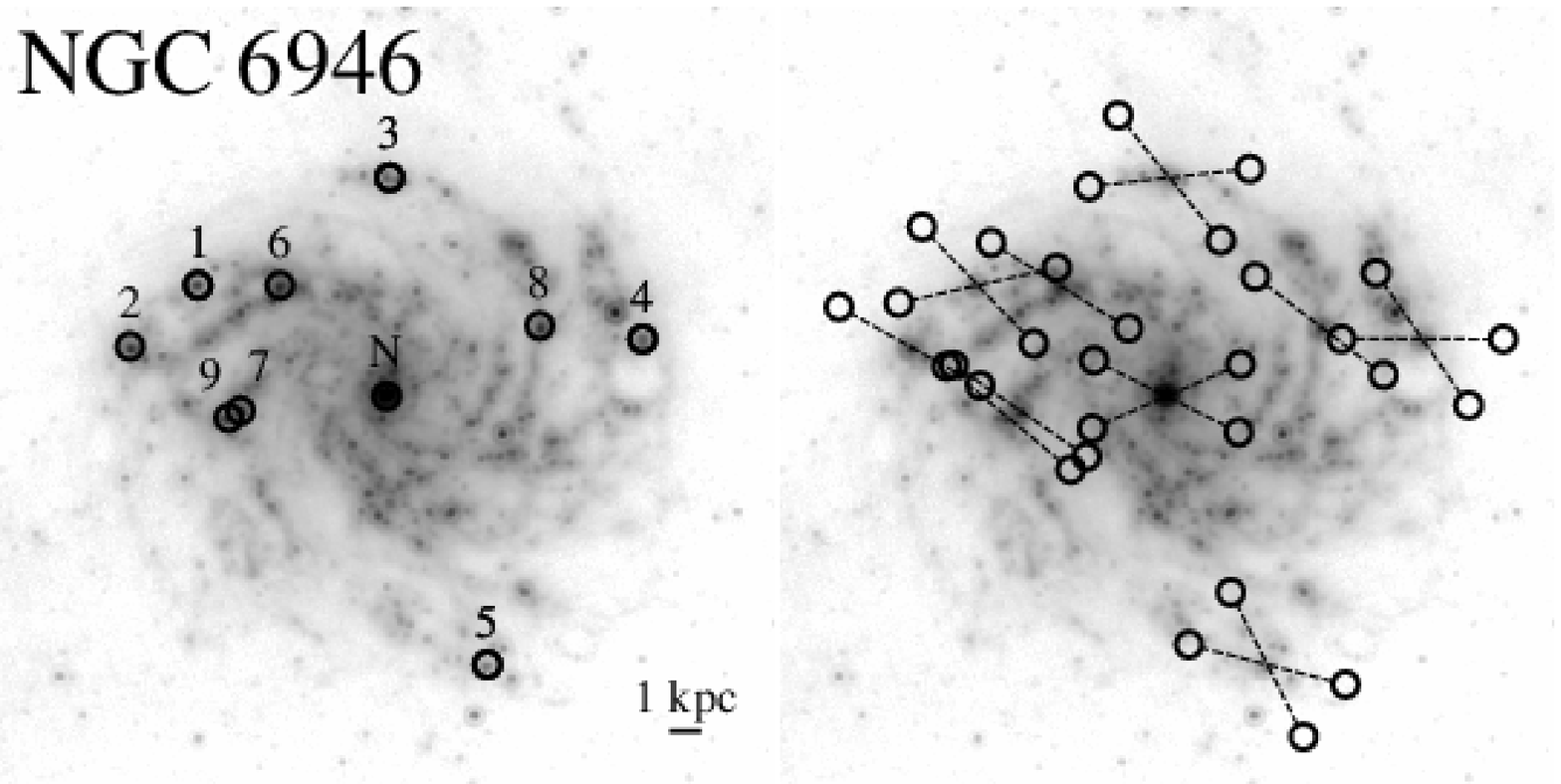,width=0.5\linewidth,clip=} & \epsfig{file=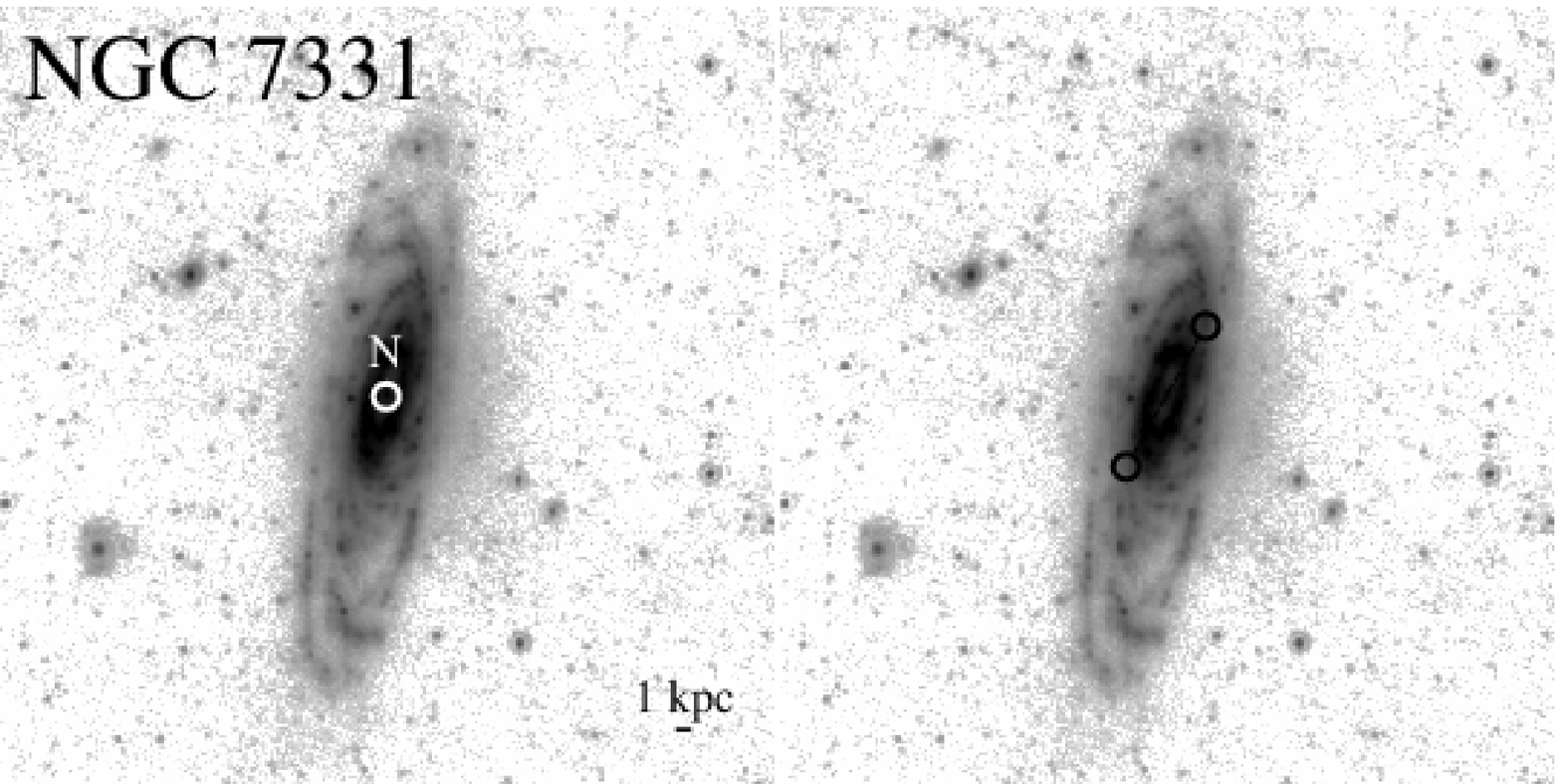,width=0.5\linewidth,clip=}
\end{tabular}
\end{center}
\caption{\it Continued}
\end{figure*}

\end{document}

%% file: tbl-1.tex
\begin{deluxetable*}{lccccccc}
\tablecaption{Nuclear Source Positions and 33\,GHz Photometry \label{tbl-1}}
\tabletypesize{\scriptsize}
\tablewidth{0pt}
\tablehead{
\colhead{Galaxy}  & \colhead{R.A.} & \colhead{Decl.} & \colhead{Type\tablenotemark{a}} & \colhead{Dist.\tablenotemark{b}} & \colhead{Nuc. Type\tablenotemark{c}} & \colhead{$S_{\rm 33\,GHz}$} & \colhead{$S_{\rm 33\,GHz}$\tablenotemark{d}}\\
\colhead{} & \colhead{(J2000)} & \colhead{(J2000)} & \colhead{} & \colhead{(Mpc)} & \colhead{} & \colhead{(mJy)} & \colhead{(mJy)}
}
    NGC\,337   &$~\, 0   ~59   ~50.3   $&$-~\, 7   ~34   ~44$  &      SBd  &19.3  &         SF  & 2.59\,$\pm$\,0.21  &           \nodata\\
    NGC\,628   &$~\, 1   ~36   ~41.7   $&$+   15   ~46   ~59$  &      SAc  & 7.2  &    \nodata  & 0.11\,$\pm$\,0.03  &         $<$\,0.09\\
    NGC\,855   &$~\, 2   ~14~~\, 3.7   $&$+   27   ~52   ~37$  &        E  &9.73  &         SF  & 0.68\,$\pm$\,0.05  &           \nodata\\
    NGC\,925   &$~\, 2   ~27   ~17.0   $&$+   33   ~34   ~42$  &     SABd  &9.12  &         SF  & 0.50\,$\pm$\,0.04  &           \nodata\\
   NGC\,1266   &$~\, 3   ~16~~\, 0.8   $&$-~\, 2   ~25   ~37$  &      SB0  &30.6  &        AGN  & 8.45\,$\pm$\,0.45  &           \nodata\\
   NGC\,1377   &$~\, 3   ~36   ~38.9   $&$-   20   ~54~~\, 6$  &       S0  &24.6  &    \nodata  &         $<$\,0.44  &           \nodata\\
     IC\,342   &$~\, 3   ~46   ~48.5   $&$+   68~~\, 5   ~45$  &    SABcd  &3.28  &      SF(*)  &25.39\,$\pm$\,1.27  &           \nodata\\
   NGC\,1482   &$~\, 3   ~54   ~39.5   $&$-   20   ~30~~\, 6$  &      SA0  &22.6  &         SF  &10.78\,$\pm$\,0.56  &           \nodata\\
   NGC\,2146   &$~\, 6   ~18   ~37.7   $&$+   78   ~21   ~24$  &     Sbab  &17.2  &      SF(*)  &76.16\,$\pm$\,3.83  &           \nodata\\
   NGC\,2403   &$~\, 7   ~36   ~50.0   $&$+   65   ~36~~\, 3$  &    SABcd  &3.22  &      SF(*)  & 0.28\,$\pm$\,0.03  & 0.15\,$\pm$\,0.03\\
Holmberg\,II   &$~\, 8   ~19   ~13.3   $&$+   70   ~43~~\, 8$  &       Im  &3.05  &    \nodata  & 0.69\,$\pm$\,0.04  &           \nodata\\
   NGC\,2798   &$~\, 9   ~17   ~22.8   $&$+   41   ~59   ~57$  &      SBa  &25.8  &     SF/AGN  & 7.82\,$\pm$\,0.52  &           \nodata\\
   NGC\,2841   &$~\, 9   ~22~~\, 2.7   $&$+   50   ~58   ~36$  &      SAb  &14.1  &        AGN  & 1.53\,$\pm$\,0.08  &           \nodata\\
   NGC\,2976   &$~\, 9   ~47   ~15.3   $&$+   67   ~55~~\, 0$  &      SAc  &3.55  &         SF  & 0.49\,$\pm$\,0.07  & 0.29\,$\pm$\,0.07\\
   NGC\,3049   &$~\, 9   ~54   ~49.6   $&$+~\, 9   ~16   ~17$  &     SBab  &19.2  &         SF  & 1.56\,$\pm$\,0.13  &           \nodata\\
   NGC\,3077   &$   10~~\, 3   ~19.1   $&$+   68   ~44~~\, 2$  &    I0pec  &3.83  &      SF(*)  & 7.55\,$\pm$\,0.40  &           \nodata\\
   NGC\,3190   &$   10   ~18~~\, 5.6   $&$+   21   ~49   ~54$  &     SAap  &19.3  &     AGN(*)  & 0.93\,$\pm$\,0.08  &           \nodata\\
   NGC\,3184   &$   10   ~18   ~16.7   $&$+   41   ~25   ~27$  &    SABcd  &11.7  &         SF  &         $<$\,0.18  &           \nodata\\
   NGC\,3198   &$   10   ~19   ~54.9   $&$+   45   ~32   ~58$  &      SBc  &14.1  &         SF  & 0.72\,$\pm$\,0.11  &           \nodata\\
    IC\,2574   &$   10   ~28   ~48.4   $&$+   68   ~28~~\, 1$  &     SABm  &3.79  &      SF(*)  & 0.62\,$\pm$\,0.05  &           \nodata\\
   NGC\,3265   &$   10   ~31~~\, 6.7   $&$+   28   ~47   ~48$  &        E  &19.6  &         SF  & 1.05\,$\pm$\,0.14  &           \nodata\\
   NGC\,3351   &$   10   ~43   ~57.8   $&$+   11   ~42   ~14$  &      SBb  &9.33  &         SF  & 4.86\,$\pm$\,0.27  &           \nodata\\
   NGC\,3521   &$   11~~\, 5   ~48.9   $&$-~\, 0~~\, 2~~\, 6$  &    SABbc  &11.2  &  SF/AGN(*)  & 1.45\,$\pm$\,0.10  &           \nodata\\
   NGC\,3627   &$   11   ~20   ~15.0   $&$+   12   ~59   ~30$  &     SABb  &9.38  &        AGN  & 2.70\,$\pm$\,0.16  &           \nodata\\
   NGC\,3773   &$   11   ~38   ~13.0   $&$+   12~~\, 6   ~45$  &      SA0  &12.4  &         SF  & 0.95\,$\pm$\,0.07  &           \nodata\\
   NGC\,3938   &$   11   ~52   ~49.5   $&$+   44~~\, 7   ~14$  &      SAc  &17.9  &      SF(*)  &         $<$\,0.19  &           \nodata\\
   NGC\,4254   &$   12   ~18   ~49.4   $&$+   14   ~24   ~59$  &      SAc  &14.4  &     SF/AGN  & 1.60\,$\pm$\,0.12  &           \nodata\\
   NGC\,4321   &$   12   ~22   ~54.9   $&$+   15   ~49   ~21$  &    SABbc  &14.3  &        AGN  & 4.34\,$\pm$\,0.24  &           \nodata\\
   NGC\,4536   &$   12   ~34   ~27.1   $&$+~\, 2   ~11   ~17$  &    SABbc  &14.5  &     SF/AGN  &17.49\,$\pm$\,0.88  &           \nodata\\
   NGC\,4559   &$   12   ~35   ~57.7   $&$+   27   ~57   ~35$  &    SABcd  &6.98  &         SF  & 0.70\,$\pm$\,0.05  &           \nodata\\
   NGC\,4569   &$   12   ~36   ~49.8   $&$+   13~~\, 9   ~46$  &    SABab  &9.86  &        AGN  & 1.81\,$\pm$\,0.13  &           \nodata\\
   NGC\,4579   &$   12   ~37   ~43.6   $&$+   11   ~49~~\, 1$  &     SABb  &16.4  &        AGN  &23.38\,$\pm$\,1.17  &           \nodata\\
   NGC\,4594   &$   12   ~39   ~59.4   $&$-   11   ~37   ~23$  &      SAa  &9.08  &        AGN  &69.30\,$\pm$\,3.47  &           \nodata\\
   NGC\,4625   &$   12   ~41   ~52.4   $&$+   41   ~16   ~23$  &    SABmp  & 9.3  &         SF  & 0.36\,$\pm$\,0.02  &           \nodata\\
   NGC\,4631   &$   12   ~42~~\, 5.9   $&$+   32   ~32   ~21$  &      SBd  &7.62  &      SF(*)  & 3.83\,$\pm$\,0.21  &           \nodata\\
   NGC\,4725   &$   12   ~50   ~26.6   $&$+   25   ~30~~\, 6$  &    SABab  &11.9  &        AGN  & 0.29\,$\pm$\,0.03  &           \nodata\\
   NGC\,4736   &$   12   ~50   ~53.0   $&$+   41~~\, 7   ~14$  &     SAab  &4.66  &     AGN(*)  & 2.86\,$\pm$\,0.16  &           \nodata\\
   NGC\,4826   &$   12   ~56   ~43.9   $&$+   21   ~40   ~59$  &     SAab  &5.27  &        AGN  & 5.88\,$\pm$\,0.31  &           \nodata\\
   NGC\,5055   &$   13   ~15   ~49.2   $&$+   42~~\, 1   ~49$  &     SAbc  &7.94  &        AGN  & 2.58\,$\pm$\,0.25  &           \nodata\\
   NGC\,5194   &$   13   ~29   ~52.7   $&$+   47   ~11   ~43$  &   SABbcp  &7.62  &        AGN  & 6.25\,$\pm$\,0.39  &           \nodata\\
   NGC\,5457   &$   14~~\, 3   ~12.6   $&$+   54   ~20   ~57$  &    SABcd  & 6.7  &      SF(*)  &         $<$\,0.73  &           \nodata\\
   NGC\,5474   &$   14~~\, 5~~\, 1.3   $&$+   53   ~39   ~43$  &     SAcd  & 6.8  &      SF(*)  & 0.17\,$\pm$\,0.05  & 0.16\,$\pm$\,0.05\\
   NGC\,5713   &$   14   ~40   ~11.3   $&$-~\, 0   ~17   ~26$  &   SABbcp  &21.4  &         SF  & 6.34\,$\pm$\,0.35  &           \nodata\\
   NGC\,5866   &$   15~~\, 6   ~29.5   $&$+   55   ~45   ~47$  &       S0  &15.3  &        AGN  & 6.21\,$\pm$\,0.32  &           \nodata\\
   NGC\,6946   &$   20   ~34   ~52.3   $&$+   60~~\, 9   ~14$  &    SABcd  & 6.8  &         SF  &15.84\,$\pm$\,0.80  &           \nodata\\
   NGC\,7331   &$   22   ~37~~\, 4.1   $&$+   34   ~24   ~56$  &      SAb  &14.5  &        AGN  & 1.42\,$\pm$\,0.11  &           \nodata  
\enddata
\tablenotetext{a}{Morphological type taken from the NASA Extragalactic Database (NED; http://nedwww.ipac.caltech.edu).}
\tablenotetext{b}{Redshift-independent distance taken from the list compiled by \citet{kf11}, except for the two non-KINGFISH galaxies NGC\,5194 \citep{rc02} and NGC\,2403 \citep{hkp01}.}
\tablenotetext{c}{Nuclear Type based optical spectroscopy: SF$\,=\,$ Star-Forming; AGN$\,=\,$ Non-thermal emission as given in Table~5 of \citet{jm10} or (*) Table~4 of \citet{hfs97}.}
\tablenotetext{d}{Measured 33\,GHz flux densities before applying a correction for the oversubtraction of emission due to reference nods falling on bright regions of the galaxy.  Corrections were only applied to cases where the estimated loss is $>$15\%. }
\end{deluxetable*}

%% file: tbl-2.tex
\begin{deluxetable*}{lcccc}
\tablecaption{Extranuclear Source Positions and 33\,GHz Photometry \label{tbl-2}}
\tabletypesize{\scriptsize}
\tablewidth{0pt}
\tablehead{
\colhead{ID}  & \colhead{R.A.} & \colhead{Decl.} & \colhead{$S_{\rm 33~GHz}$} & \colhead{$S_{\rm 33~GHz}$\tablenotemark{a}}\\
\colhead{} & \colhead{(J2000)} & \colhead{(J2000)} & \colhead{(mJy)} & \colhead{(mJy)}
}
   NGC\,628~Enuc.\,1   &$~\, 1   ~36   ~45.1   $&$+   15   ~47   ~51  $  & 0.54\,$\pm$\,0.17  &           \nodata\\
   NGC\,628~Enuc.\,2   &$~\, 1   ~36   ~37.5   $&$+   15   ~45   ~11  $  & 0.35\,$\pm$\,0.04  &           \nodata\\
   NGC\,628~Enuc.\,3   &$~\, 1   ~36   ~38.8   $&$+   15   ~44   ~25  $  & 0.54\,$\pm$\,0.04  & 0.46\,$\pm$\,0.04\\
   NGC\,628~Enuc.\,4   &$~\, 1   ~36   ~35.5   $&$+   15   ~50   ~11  $  & 0.23\,$\pm$\,0.05  &         $<$\,0.16\\
  NGC\,2403~Enuc.\,1   &$~\, 7   ~36   ~45.5   $&$+   65   ~37~~\, 0  $  & 1.68\,$\pm$\,0.14  &           \nodata\\
  NGC\,2403~Enuc.\,2   &$~\, 7   ~36   ~52.7   $&$+   65   ~36   ~46  $  & 1.33\,$\pm$\,0.13  &           \nodata\\
  NGC\,2403~Enuc.\,3   &$~\, 7   ~37~~\, 6.9   $&$+   65   ~36   ~38  $  & 3.04\,$\pm$\,0.19  &           \nodata\\
  NGC\,2403~Enuc.\,4   &$~\, 7   ~37   ~17.9   $&$+   65   ~33   ~46  $  & 1.37\,$\pm$\,0.08  &           \nodata\\
  NGC\,2403~Enuc.\,5   &$~\, 7   ~36   ~19.5   $&$+   65   ~37~~\, 4  $  & 1.28\,$\pm$\,0.21  &           \nodata\\
  NGC\,2403~Enuc.\,6   &$~\, 7   ~36   ~28.5   $&$+   65   ~33   ~50  $  & 0.73\,$\pm$\,0.06  &           \nodata\\
  NGC\,2976~Enuc.\,1   &$~\, 9   ~47~~\, 7.8   $&$+   67   ~55   ~51  $  & 3.62\,$\pm$\,0.21  &           \nodata\\
  NGC\,2976~Enuc.\,2   &$~\, 9   ~47   ~24.1   $&$+   67   ~53   ~56  $  & 2.12\,$\pm$\,0.15  & 1.92\,$\pm$\,0.15\\
  NGC\,3521~Enuc.\,1   &$   11~~\, 5   ~46.3   $&$-~\, 0~~\, 4~~\, 9  $  & 0.47\,$\pm$\,0.02  & 0.25\,$\pm$\,0.02\\
  NGC\,3521~Enuc.\,2   &$   11~~\, 5   ~49.9   $&$-~\, 0~~\, 3   ~38  $  & 0.90\,$\pm$\,0.09  &           \nodata\\
  NGC\,3521~Enuc.\,3   &$   11~~\, 5   ~47.6   $&$+~\, 0~~\, 0   ~33  $  & 0.37\,$\pm$\,0.03  &           \nodata\\
  NGC\,3627~Enuc.\,1   &$   11   ~20   ~16.2   $&$+   12   ~57   ~50  $  & 2.42\,$\pm$\,0.15  &           \nodata\\
  NGC\,3627~Enuc.\,2   &$   11   ~20   ~16.3   $&$+   12   ~58   ~44  $  & 6.16\,$\pm$\,0.32  &           \nodata\\
  NGC\,3627~Enuc.\,3   &$   11   ~20   ~16.0   $&$+   12   ~59   ~52  $  & 0.77\,$\pm$\,0.08  & 0.62\,$\pm$\,0.08\\
  NGC\,3938~Enuc.\,1   &$   11   ~52   ~46.4   $&$+   44~~\, 7~~\, 0  $  & 0.15\,$\pm$\,0.02  & 0.10\,$\pm$\,0.02\\
  NGC\,3938~Enuc.\,2   &$   11   ~53~~\, 0.0   $&$+   44~~\, 7   ~54  $  &         $<$\,0.66  &           \nodata\\
  NGC\,4254~Enuc.\,1   &$   12   ~18   ~49.1   $&$+   14   ~23   ~58  $  & 1.10\,$\pm$\,0.08  &           \nodata\\
  NGC\,4254~Enuc.\,2   &$   12   ~18   ~44.6   $&$+   14   ~24   ~24  $  & 0.19\,$\pm$\,0.04  &         $<$\,0.13\\
  NGC\,4321~Enuc.\,1   &$   12   ~22   ~58.9   $&$+   15   ~49   ~35  $  & 0.15\,$\pm$\,0.05  &         $<$\,0.14\\
  NGC\,4321~Enuc.\,2   &$   12   ~22   ~49.8   $&$+   15   ~50   ~29  $  & 0.16\,$\pm$\,0.04  &         $<$\,0.13\\
  NGC\,4631~Enuc.\,1   &$   12   ~41   ~40.8   $&$+   32   ~31   ~50  $  & 0.90\,$\pm$\,0.08  &           \nodata\\
  NGC\,4631~Enuc.\,2   &$   12   ~42   ~21.3   $&$+   32   ~33~~\, 6  $  & 0.67\,$\pm$\,0.06  &           \nodata\\
  NGC\,4736~Enuc.\,1   &$   12   ~50   ~56.2   $&$+   41~~\, 7   ~19  $  & 2.58\,$\pm$\,0.15  &           \nodata\\
  NGC\,5055~Enuc.\,1   &$   13   ~15   ~58.0   $&$+   42~~\, 0   ~25  $  & 0.43\,$\pm$\,0.12  &         $<$\,0.37\\
  NGC\,5194~Enuc.\,1   &$   13   ~29   ~53.1   $&$+   47   ~12   ~39  $  &         $<$\,0.87  &           \nodata\\
  NGC\,5194~Enuc.\,2   &$   13   ~29   ~44.1   $&$+   47   ~10   ~21  $  & 1.72\,$\pm$\,0.24  &           \nodata\\
  NGC\,5194~Enuc.\,3   &$   13   ~29   ~44.6   $&$+   47~~\, 9   ~55  $  & 1.00\,$\pm$\,0.23  &           \nodata\\
  NGC\,5194~Enuc.\,4   &$   13   ~29   ~56.2   $&$+   47   ~14~~\, 7  $  & 2.06\,$\pm$\,0.23  & 1.77\,$\pm$\,0.23\\
  NGC\,5194~Enuc.\,5   &$   13   ~29   ~59.6   $&$+   47   ~14~~\, 0  $  & 2.05\,$\pm$\,0.24  & 1.82\,$\pm$\,0.24\\
  NGC\,5194~Enuc.\,6   &$   13   ~29   ~39.5   $&$+   47~~\, 8   ~35  $  & 0.92\,$\pm$\,0.07  & 0.85\,$\pm$\,0.07\\
  NGC\,5194~Enuc.\,7   &$   13   ~30~~\, 2.5   $&$+   47~~\, 9   ~51  $  &         $<$\,0.64  &           \nodata\\
  NGC\,5194~Enuc.\,8   &$   13   ~30~~\, 1.6   $&$+   47   ~12   ~51  $  & 0.97\,$\pm$\,0.27  &           \nodata\\
  NGC\,5194~Enuc.\,9   &$   13   ~29   ~59.9   $&$+   47   ~11   ~12  $  & 1.03\,$\pm$\,0.22  & 0.78\,$\pm$\,0.22\\
 NGC\,5194~Enuc.\,10   &$   13   ~29   ~56.7   $&$+   47   ~10   ~45  $  & 0.85\,$\pm$\,0.22  &         $<$\,0.66\\
 NGC\,5194~Enuc.\,11   &$   13   ~29   ~49.7   $&$+   47   ~13   ~28  $  & 0.69\,$\pm$\,0.10  &         $<$\,0.30\\
  NGC\,5457~Enuc.\,1   &$   14~~\, 3   ~10.2   $&$+   54   ~20   ~57  $  & 0.39\,$\pm$\,0.12  &           \nodata\\
  NGC\,5457~Enuc.\,2   &$   14~~\, 2   ~55.0   $&$+   54   ~22   ~26  $  & 0.29\,$\pm$\,0.08  &           \nodata\\
  NGC\,5457~Enuc.\,3   &$   14~~\, 3   ~41.3   $&$+   54   ~19~~\, 4  $  & 8.27\,$\pm$\,0.52  &           \nodata\\
  NGC\,5457~Enuc.\,4   &$   14~~\, 3   ~53.1   $&$+   54   ~22~~\, 5  $  & 2.11\,$\pm$\,0.22  &           \nodata\\
  NGC\,5457~Enuc.\,5   &$   14~~\, 3~~\, 1.1   $&$+   54   ~14   ~27  $  & 2.49\,$\pm$\,0.27  &           \nodata\\
  NGC\,5457~Enuc.\,6   &$   14~~\, 2   ~28.1   $&$+   54   ~16   ~26  $  & 3.38\,$\pm$\,0.33  &           \nodata\\
  NGC\,5457~Enuc.\,7   &$   14~~\, 4   ~29.3   $&$+   54   ~23   ~45  $  & 4.69\,$\pm$\,0.35  &           \nodata\\
  NGC\,5713~Enuc.\,1   &$   14   ~40   ~12.1   $&$-~\, 0   ~17   ~47  $  & 1.18\,$\pm$\,0.14  &           \nodata\\
  NGC\,5713~Enuc.\,2   &$   14   ~40   ~10.5   $&$-~\, 0   ~17   ~47  $  & 1.48\,$\pm$\,0.15  &           \nodata\\
  NGC\,6946~Enuc.\,1   &$   20   ~35   ~16.7   $&$+   60   ~11~~\, 0  $  & 0.74\,$\pm$\,0.05  & 0.39\,$\pm$\,0.05\\
  NGC\,6946~Enuc.\,2   &$   20   ~35   ~25.5   $&$+   60   ~10~~\, 1  $  & 2.36\,$\pm$\,0.16  &           \nodata\\
  NGC\,6946~Enuc.\,3   &$   20   ~34   ~51.9   $&$+   60   ~12   ~45  $  & 1.01\,$\pm$\,0.07  &           \nodata\\
  NGC\,6946~Enuc.\,4   &$   20   ~34   ~19.2   $&$+   60   ~10~~\, 9  $  & 3.01\,$\pm$\,0.15  & 2.89\,$\pm$\,0.15\\
  NGC\,6946~Enuc.\,5   &$   20   ~34   ~39.3   $&$+   60~~\, 4   ~54  $  & 0.51\,$\pm$\,0.04  &           \nodata\\
  NGC\,6946~Enuc.\,6   &$   20   ~35~~\, 6.1   $&$+   60   ~11~~\, 0  $  & 2.86\,$\pm$\,0.18  &           \nodata\\
  NGC\,6946~Enuc.\,7   &$   20   ~35   ~11.2   $&$+   60~~\, 9~~\, 0  $  & 3.20\,$\pm$\,0.20  &           \nodata\\
  NGC\,6946~Enuc.\,8   &$   20   ~34   ~32.5   $&$+   60   ~10   ~22  $  & 1.80\,$\pm$\,0.14  &           \nodata\\
  NGC\,6946~Enuc.\,9   &$   20   ~35   ~12.7   $&$+   60~~\, 8   ~53  $  & 2.49\,$\pm$\,0.17  &           \nodata  
\enddata
\tablenotetext{a}{Measured 33\,GHz flux densities before applying a correction for the oversubtraction of emission due to reference nods falling on bright regions of the galaxy.  Corrections were only applied to cases where the estimated loss is $>$15\%. }
\end{deluxetable*}

%% file: tbl-3.tex
\begin{deluxetable*}{lcccccccc}
\tablecaption{Nuclear Photometry from Ancillary Data \label{tbl-3}}
\tabletypesize{\scriptsize}
\tablewidth{0pt}
\tablehead{
\colhead{Galaxy}  & \colhead{$f_{\nu} (24\,\micron)$}  & \colhead{$F_{\rm IR}/10^{-11}$} & \colhead{$S_{\rm 1.7GHz}$} & \colhead{$S_{\rm 1.4GHz}$} & \colhead{$f_{\nu} (1528$\,\AA)}  & \colhead{$f_{\nu} (2271$\,\AA)}  & \colhead{$f_{\rm H\alpha}/10^{-13}$}  & \colhead{$E(B-V)$\tablenotemark{a}}\\
\colhead{} & \colhead{(mJy)} & \colhead{(erg\,s$^{-1}$\,cm$^{-2}$)} & \colhead{(mJy)} & \colhead{(mJy)} &  \colhead{(mJy)} & \colhead{(mJy)} & \colhead{(erg\,s$^{-1}$\,cm$^{-2}$)} & \colhead{(mag)}
}
    NGC\,337  &   241.0\,$\pm$\,  42.7  &     23.6\,$\pm$\,  4.2  &             \nodata  &             \nodata  &    1.7\,$\pm$\, 0.4  &    3.8\,$\pm$\, 0.9  &       8.3\,$\pm$\, 2.2  & 0.112\\
    NGC\,628  &    57.4\,$\pm$\,  10.2  &      8.3\,$\pm$\,  1.6  &             \nodata  &             \nodata  &             \nodata  &             \nodata  &                \nodata  & 0.070\\
    NGC\,855  &    60.8\,$\pm$\,  10.8  &      7.6\,$\pm$\,  1.5  &             \nodata  &             \nodata  &    1.0\,$\pm$\, 0.2  &    1.6\,$\pm$\, 0.4  &                \nodata  & 0.071\\
    NGC\,925  &    46.2\,$\pm$\,   8.2  &      7.3\,$\pm$\,  1.3  &    1.8\,$\pm$\, 0.3  &   2.1\,$\pm$\,  0.4  &    2.2\,$\pm$\, 0.5  &    3.0\,$\pm$\, 0.7  &       3.6\,$\pm$\, 0.9  & 0.076\\
   NGC\,1266  &   811.6\,$\pm$\, 143.8  &     75.2\,$\pm$\, 27.7  &             \nodata  &             \nodata  &  0.02\,$\pm$\, 0.00  &  0.14\,$\pm$\, 0.03  &                \nodata  & 0.098\\
   NGC\,1377  &  1804.0\,$\pm$\, 319.7  &     64.0\,$\pm$\, 11.7  &             \nodata  &             \nodata  &             \nodata  &             \nodata  &                \nodata  & 0.028\\
     IC\,342  &  9561.4\,$\pm$\,1694.3  &    584.5\,$\pm$\,320.8  &             \nodata  &             \nodata  &   19.6\,$\pm$\, 4.4  &  119.4\,$\pm$\,27.1  &                \nodata  & 0.559\\
   NGC\,1482  &  1116.7\,$\pm$\, 197.9  &    152.8\,$\pm$\, 37.7  &             \nodata  &             \nodata  &  0.09\,$\pm$\, 0.02  &    0.3\,$\pm$\, 0.1  &                \nodata  & 0.040\\
   NGC\,2146  &  7414.2\,$\pm$\,1313.8  &    892.0\,$\pm$\,164.1  &  504.3\,$\pm$\,89.4  & 584.1\,$\pm$\,103.5  &  0.16\,$\pm$\, 0.04  &    0.5\,$\pm$\, 0.1  &                \nodata  & 0.096\\
   NGC\,2403  &    59.9\,$\pm$\,  10.6  &     11.3\,$\pm$\,  2.3  &    1.4\,$\pm$\, 0.2  &   1.8\,$\pm$\,  0.3  &    1.5\,$\pm$\, 0.3  &    2.5\,$\pm$\, 0.6  &       2.7\,$\pm$\, 0.7  & 0.040\\
Holmberg\,II  &    35.2\,$\pm$\,   6.2  &      1.8\,$\pm$\,  0.3  &    1.0\,$\pm$\, 0.2  &   1.2\,$\pm$\,  0.2  &             \nodata  &             \nodata  &                \nodata  & 0.032\\
   NGC\,2798  &   956.1\,$\pm$\, 169.4  &    120.7\,$\pm$\, 22.7  &             \nodata  &             \nodata  &             \nodata  &             \nodata  &                \nodata  & 0.020\\
   NGC\,2841  &    44.9\,$\pm$\,   8.0  &      4.4\,$\pm$\,  0.9  &    3.7\,$\pm$\, 0.7  &   4.4\,$\pm$\,  0.8  &    0.6\,$\pm$\, 0.1  &    1.0\,$\pm$\, 0.2  &       1.0\,$\pm$\, 0.3  & 0.015\\
   NGC\,2976  &    93.6\,$\pm$\,  16.6  &     10.1\,$\pm$\,  2.0  &    1.9\,$\pm$\, 0.3  &   2.1\,$\pm$\,  0.4  &    1.2\,$\pm$\, 0.3  &    2.0\,$\pm$\, 0.5  &       4.4\,$\pm$\, 1.2  & 0.072\\
   NGC\,3049  &   387.3\,$\pm$\,  68.6  &     20.6\,$\pm$\,  3.9  &             \nodata  &             \nodata  &    1.3\,$\pm$\, 0.3  &    2.2\,$\pm$\, 0.5  &       3.9\,$\pm$\, 1.0  & 0.038\\
   NGC\,3077  &   868.6\,$\pm$\, 153.9  &     72.5\,$\pm$\, 12.5  &             \nodata  &             \nodata  &             \nodata  &             \nodata  &      21.8\,$\pm$\, 5.7  & 0.067\\
   NGC\,3190  &   108.5\,$\pm$\,  19.2  &     22.7\,$\pm$\,  4.0  &             \nodata  &             \nodata  &  0.07\,$\pm$\, 0.01  &    0.3\,$\pm$\, 0.1  &       0.4\,$\pm$\, 0.1  & 0.025\\
   NGC\,3184  &   129.7\,$\pm$\,  23.0  &     11.6\,$\pm$\,  2.0  &    2.1\,$\pm$\, 0.4  &   2.3\,$\pm$\,  0.4  &    0.4\,$\pm$\, 0.1  &    0.9\,$\pm$\, 0.2  &       0.8\,$\pm$\, 0.2  & 0.017\\
   NGC\,3198  &   364.6\,$\pm$\,  64.6  &     20.8\,$\pm$\,  3.6  &    2.9\,$\pm$\, 0.5  &   3.2\,$\pm$\,  0.6  &    0.2\,$\pm$\, 0.1  &    0.4\,$\pm$\, 0.1  &       0.7\,$\pm$\, 0.2  & 0.012\\
    IC\,2574  &    41.2\,$\pm$\,   7.3  &      3.0\,$\pm$\,  0.5  &    0.7\,$\pm$\, 0.1  &   0.9\,$\pm$\,  0.2  &    1.0\,$\pm$\, 0.2  &    1.2\,$\pm$\, 0.3  &       4.3\,$\pm$\, 1.1  & 0.037\\
   NGC\,3265  &   278.8\,$\pm$\,  49.4  &     18.4\,$\pm$\,  3.2  &             \nodata  &             \nodata  &    0.4\,$\pm$\, 0.1  &    0.8\,$\pm$\, 0.2  &                \nodata  & 0.024\\
   NGC\,3351  &  1435.0\,$\pm$\, 254.3  &    103.7\,$\pm$\, 23.1  &             \nodata  &             \nodata  &    2.0\,$\pm$\, 0.4  &    5.3\,$\pm$\, 1.2  &      11.8\,$\pm$\, 3.1  & 0.028\\
   NGC\,3521  &   287.3\,$\pm$\,  50.9  &     37.9\,$\pm$\,  9.0  &             \nodata  &             \nodata  &    0.5\,$\pm$\, 0.1  &    2.0\,$\pm$\, 0.4  &       5.7\,$\pm$\, 1.5  & 0.057\\
   NGC\,3627  &   633.2\,$\pm$\, 112.2  &     90.4\,$\pm$\, 16.2  &             \nodata  &             \nodata  &    0.5\,$\pm$\, 0.1  &    1.7\,$\pm$\, 0.4  &       4.9\,$\pm$\, 1.3  & 0.033\\
   NGC\,3773  &   128.0\,$\pm$\,  22.7  &      9.4\,$\pm$\,  1.8  &             \nodata  &             \nodata  &    3.2\,$\pm$\, 0.7  &    4.0\,$\pm$\, 0.9  &                \nodata  & 0.027\\
   NGC\,3938  &    68.2\,$\pm$\,  12.1  &     11.3\,$\pm$\,  2.1  &    1.3\,$\pm$\, 0.2  &   1.5\,$\pm$\,  0.3  &    0.7\,$\pm$\, 0.2  &    1.4\,$\pm$\, 0.3  &       1.2\,$\pm$\, 0.3  & 0.021\\
   NGC\,4254  &   436.0\,$\pm$\,  77.3  &     54.8\,$\pm$\, 16.2  &             \nodata  &             \nodata  &    1.1\,$\pm$\, 0.2  &    3.1\,$\pm$\, 0.7  &       6.4\,$\pm$\, 1.7  & 0.039\\
   NGC\,4321  &   723.2\,$\pm$\, 128.2  &     76.6\,$\pm$\, 14.3  &             \nodata  &             \nodata  &    2.4\,$\pm$\, 0.5  &    5.6\,$\pm$\, 1.3  &       6.3\,$\pm$\, 1.7  & 0.026\\
   NGC\,4536  &  2474.1\,$\pm$\, 438.4  &    193.3\,$\pm$\, 33.2  &             \nodata  &             \nodata  &    0.3\,$\pm$\, 0.1  &    0.7\,$\pm$\, 0.2  &       7.0\,$\pm$\, 1.8  & 0.018\\
   NGC\,4559  &    84.2\,$\pm$\,  14.9  &     12.0\,$\pm$\,  2.0  &    2.5\,$\pm$\, 0.4  &   3.0\,$\pm$\,  0.5  &    1.2\,$\pm$\, 0.3  &    1.9\,$\pm$\, 0.4  &       4.0\,$\pm$\, 1.0  & 0.018\\
   NGC\,4569  &   701.4\,$\pm$\, 124.3  &     47.3\,$\pm$\,  9.5  &             \nodata  &             \nodata  &    0.9\,$\pm$\, 0.2  &    3.3\,$\pm$\, 0.8  &       8.9\,$\pm$\, 2.3  & 0.047\\
   NGC\,4579  &   170.7\,$\pm$\,  30.2  &     15.5\,$\pm$\,  2.7  &             \nodata  &             \nodata  &    0.3\,$\pm$\, 0.1  &    0.8\,$\pm$\, 0.2  &       5.3\,$\pm$\, 1.4  & 0.041\\
   NGC\,4594  &   110.9\,$\pm$\,  19.6  &      7.9\,$\pm$\,  1.6  &             \nodata  &             \nodata  &    0.6\,$\pm$\, 0.1  &    1.5\,$\pm$\, 0.3  &                \nodata  & 0.051\\
   NGC\,4625  &    45.4\,$\pm$\,   8.1  &      6.4\,$\pm$\,  1.2  &             \nodata  &             \nodata  &    1.1\,$\pm$\, 0.2  &    1.6\,$\pm$\, 0.4  &       1.7\,$\pm$\, 0.5  & 0.018\\
   NGC\,4631  &   566.5\,$\pm$\, 100.4  &     79.2\,$\pm$\, 13.9  &   39.9\,$\pm$\, 7.1  &  45.8\,$\pm$\,  8.1  &    1.6\,$\pm$\, 0.4  &    2.4\,$\pm$\, 0.6  &       4.8\,$\pm$\, 1.3  & 0.017\\
   NGC\,4725  &    54.2\,$\pm$\,   9.6  &      4.4\,$\pm$\,  0.8  &    0.3\,$\pm$\, 0.0  &             \nodata  &  0.17\,$\pm$\, 0.04  &    0.4\,$\pm$\, 0.1  &       0.3\,$\pm$\, 0.1  & 0.012\\
   NGC\,4736  &  1094.8\,$\pm$\, 194.0  &    150.4\,$\pm$\, 33.1  &   23.8\,$\pm$\, 4.2  &  24.8\,$\pm$\,  4.4  &    1.7\,$\pm$\, 0.4  &    5.8\,$\pm$\, 1.3  &       4.9\,$\pm$\, 1.3  & 0.018\\
   NGC\,4826  &  1088.9\,$\pm$\, 193.0  &    167.6\,$\pm$\, 30.5  &   33.9\,$\pm$\, 6.0  &             \nodata  &    1.0\,$\pm$\, 0.2  &    2.7\,$\pm$\, 0.6  &      17.1\,$\pm$\, 4.5  & 0.041\\
   NGC\,5055  &   353.0\,$\pm$\,  62.6  &     76.8\,$\pm$\, 13.6  &   18.1\,$\pm$\, 3.2  &  20.6\,$\pm$\,  3.7  &    0.6\,$\pm$\, 0.1  &    1.9\,$\pm$\, 0.4  &       5.8\,$\pm$\, 1.5  & 0.018\\
   NGC\,5194  &   608.9\,$\pm$\, 107.9  &     95.2\,$\pm$\, 26.0  &   52.2\,$\pm$\, 9.3  &  58.5\,$\pm$\, 10.4  &    2.2\,$\pm$\, 0.5  &    6.1\,$\pm$\, 1.4  &      11.1\,$\pm$\, 2.9  & 0.035\\
   NGC\,5457  &   177.7\,$\pm$\,  31.5  &     20.6\,$\pm$\,  3.7  &             \nodata  &             \nodata  &    0.9\,$\pm$\, 0.2  &    1.9\,$\pm$\, 0.4  &       2.0\,$\pm$\, 0.5  & 0.009\\
   NGC\,5474  &    14.4\,$\pm$\,   2.6  &      2.5\,$\pm$\,  0.5  &             \nodata  &             \nodata  &    1.2\,$\pm$\, 0.3  &    1.5\,$\pm$\, 0.3  &       1.7\,$\pm$\, 0.5  & 0.010\\
   NGC\,5713  &  1137.6\,$\pm$\, 201.6  &     85.6\,$\pm$\, 16.6  &             \nodata  &             \nodata  &    1.7\,$\pm$\, 0.4  &    3.6\,$\pm$\, 0.8  &       5.6\,$\pm$\, 1.5  & 0.039\\
   NGC\,5866  &    95.5\,$\pm$\,  16.9  &     33.8\,$\pm$\,  5.9  &             \nodata  &             \nodata  &  0.12\,$\pm$\, 0.03  &    0.6\,$\pm$\, 0.1  &                \nodata  & 0.013\\
   NGC\,6946  &  4355.3\,$\pm$\, 771.8  &    301.0\,$\pm$\, 60.0  &   83.8\,$\pm$\,14.8  &  91.9\,$\pm$\, 16.3  &    0.4\,$\pm$\, 0.1  &    2.0\,$\pm$\, 0.5  &      18.1\,$\pm$\, 4.8  & 0.343\\
   NGC\,7331  &   301.9\,$\pm$\,  53.5  &     49.0\,$\pm$\,  8.6  &   17.2\,$\pm$\, 3.0  &  20.1\,$\pm$\,  3.6  &    0.3\,$\pm$\, 0.1  &    1.6\,$\pm$\, 0.4  &       2.3\,$\pm$\, 0.6  & 0.091  
\enddata
\tablenotetext{a}{Galactic extinction taken from \citep{ds98} used to correct the H$\alpha$ and GALEX FUV and NUV flux densities assuming $A_{V}/E(B-V)=3.1$ and the modeled extinction curves of \citet{wd01,bd03}.}
\end{deluxetable*}

%% file: tbl-4.tex
\begin{deluxetable*}{lcccccccc}
\tablecaption{Extranuclear Photometry from Ancillary Data \label{tbl-4}}
\tabletypesize{\scriptsize}
\tablewidth{0pt}
\tablehead{
\colhead{Galaxy}  & \colhead{$f_{\nu} (24\,\micron)$} & \colhead{$F_{\rm IR}/10^{-11}$} & \colhead{$S_{\rm 1.7GHz}$} & \colhead{$S_{\rm 1.4GHz}$} &\colhead{$f_{\nu} (1528\,$\AA)}  & \colhead{$f_{\nu} (2271\,$\AA)}  & \colhead{$f_{\rm H\alpha}/10^{-13}$}  & \colhead{$E(B-V)$\tablenotemark{a}}\\
\colhead{} & \colhead{(mJy)} & \colhead{(erg\,s$^{-1}$\,cm$^{-2}$)} & \colhead{(mJy)} & \colhead{(mJy)} & \colhead{(mJy)} & \colhead{(mJy)} & \colhead{(erg\,s$^{-1}$\,cm$^{-2}$)} &\colhead{(mag)}
}
   NGC\,628~Enuc.\,1  &    167.5\,$\pm$\, 29.7  &     10.0\,$\pm$\,  1.8  &             \nodata  &             \nodata  &    0.6\,$\pm$\, 0.1  &    0.9\,$\pm$\, 0.2  &       2.2\,$\pm$\, 0.6  & 0.070\\
   NGC\,628~Enuc.\,2  &     87.9\,$\pm$\, 15.6  &      5.5\,$\pm$\,  1.0  &             \nodata  &             \nodata  &    0.5\,$\pm$\, 0.1  &    0.7\,$\pm$\, 0.2  &       1.3\,$\pm$\, 0.3  & 0.070\\
   NGC\,628~Enuc.\,3  &     86.2\,$\pm$\, 15.3  &      6.4\,$\pm$\,  1.1  &             \nodata  &             \nodata  &    1.5\,$\pm$\, 0.3  &    2.0\,$\pm$\, 0.4  &       2.4\,$\pm$\, 0.6  & 0.070\\
   NGC\,628~Enuc.\,4  &     25.8\,$\pm$\,  4.6  &      2.1\,$\pm$\,  0.4  &             \nodata  &             \nodata  &    0.4\,$\pm$\, 0.1  &    0.5\,$\pm$\, 0.1  &       1.2\,$\pm$\, 0.3  & 0.072\\
  NGC\,2403~Enuc.\,1  &    247.5\,$\pm$\, 43.9  &     20.5\,$\pm$\,  3.8  &    4.3\,$\pm$\, 0.8  &    4.7\,$\pm$\, 0.8  &    6.4\,$\pm$\, 1.5  &    8.2\,$\pm$\, 1.9  &      13.3\,$\pm$\, 3.5  & 0.040\\
  NGC\,2403~Enuc.\,2  &    189.3\,$\pm$\, 33.5  &     18.3\,$\pm$\,  5.0  &    3.7\,$\pm$\, 0.6  &    4.0\,$\pm$\, 0.7  &    3.7\,$\pm$\, 0.8  &    4.9\,$\pm$\, 1.1  &      10.0\,$\pm$\, 2.6  & 0.040\\
  NGC\,2403~Enuc.\,3  &    808.9\,$\pm$\,143.3  &     45.8\,$\pm$\, 10.5  &    9.6\,$\pm$\, 1.7  &    9.9\,$\pm$\, 1.8  &    5.9\,$\pm$\, 1.3  &    7.0\,$\pm$\, 1.6  &      21.6\,$\pm$\, 5.7  & 0.040\\
  NGC\,2403~Enuc.\,4  &     68.8\,$\pm$\, 12.2  &      7.9\,$\pm$\,  1.4  &    2.2\,$\pm$\, 0.4  &    2.5\,$\pm$\, 0.4  &    1.1\,$\pm$\, 0.3  &    1.4\,$\pm$\, 0.3  &       3.4\,$\pm$\, 0.9  & 0.040\\
  NGC\,2403~Enuc.\,5  &    171.3\,$\pm$\, 30.4  &     14.0\,$\pm$\,  2.5  &    3.6\,$\pm$\, 0.6  &    3.9\,$\pm$\, 0.7  &    4.4\,$\pm$\, 1.0  &    4.8\,$\pm$\, 1.1  &      10.3\,$\pm$\, 2.7  & 0.040\\
  NGC\,2403~Enuc.\,6  &     27.7\,$\pm$\,  4.9  &      3.3\,$\pm$\,  0.6  &    1.5\,$\pm$\, 0.3  &    1.8\,$\pm$\, 0.3  &    1.2\,$\pm$\, 0.3  &    1.4\,$\pm$\, 0.3  &       3.8\,$\pm$\, 1.0  & 0.040\\
  NGC\,2976~Enuc.\,1  &    411.4\,$\pm$\, 72.9  &     32.8\,$\pm$\,  7.4  &    5.8\,$\pm$\, 1.0  &    6.0\,$\pm$\, 1.1  &    1.3\,$\pm$\, 0.3  &    2.1\,$\pm$\, 0.5  &      15.1\,$\pm$\, 4.0  & 0.070\\
  NGC\,2976~Enuc.\,2  &    186.5\,$\pm$\, 33.1  &     18.8\,$\pm$\,  3.5  &    3.2\,$\pm$\, 0.6  &    3.3\,$\pm$\, 0.6  &    1.2\,$\pm$\, 0.3  &    1.6\,$\pm$\, 0.4  &       8.8\,$\pm$\, 2.3  & 0.072\\
  NGC\,3521~Enuc.\,1  &     23.8\,$\pm$\,  4.2  &      2.2\,$\pm$\,  0.4  &             \nodata  &             \nodata  &  0.17\,$\pm$\, 0.04  &    0.2\,$\pm$\, 0.1  &       0.8\,$\pm$\, 0.2  & 0.057\\
  NGC\,3521~Enuc.\,2  &    138.9\,$\pm$\, 24.6  &     19.7\,$\pm$\,  3.6  &             \nodata  &             \nodata  &    0.2\,$\pm$\, 0.1  &    0.5\,$\pm$\, 0.1  &       2.0\,$\pm$\, 0.5  & 0.057\\
  NGC\,3521~Enuc.\,3  &     30.3\,$\pm$\,  5.4  &      3.7\,$\pm$\,  0.7  &             \nodata  &             \nodata  &  0.20\,$\pm$\, 0.05  &    0.3\,$\pm$\, 0.1  &       0.7\,$\pm$\, 0.2  & 0.059\\
  NGC\,3627~Enuc.\,1  &    405.9\,$\pm$\, 71.9  &     32.6\,$\pm$\,  5.8  &             \nodata  &             \nodata  &    0.3\,$\pm$\, 0.1  &    0.5\,$\pm$\, 0.1  &       2.5\,$\pm$\, 0.7  & 0.035\\
  NGC\,3627~Enuc.\,2  &   1389.2\,$\pm$\,246.2  &    111.7\,$\pm$\, 19.2  &             \nodata  &             \nodata  &    1.1\,$\pm$\, 0.3  &    2.5\,$\pm$\, 0.6  &       7.3\,$\pm$\, 1.9  & 0.033\\
  NGC\,3627~Enuc.\,3  &    139.3\,$\pm$\, 24.7  &     19.6\,$\pm$\,  3.4  &             \nodata  &             \nodata  &    1.8\,$\pm$\, 0.4  &    3.4\,$\pm$\, 0.8  &       3.0\,$\pm$\, 0.8  & 0.033\\
  NGC\,3938~Enuc.\,1  &     46.8\,$\pm$\,  8.3  &      6.2\,$\pm$\,  1.1  &    1.4\,$\pm$\, 0.2  &    1.7\,$\pm$\, 0.3  &    0.7\,$\pm$\, 0.2  &    1.0\,$\pm$\, 0.2  &       1.1\,$\pm$\, 0.3  & 0.021\\
  NGC\,3938~Enuc.\,2  &     57.4\,$\pm$\, 10.2  &      3.4\,$\pm$\,  0.6  &    0.9\,$\pm$\, 0.2  &    0.9\,$\pm$\, 0.2  &    0.5\,$\pm$\, 0.1  &    0.6\,$\pm$\, 0.1  &       1.4\,$\pm$\, 0.4  & 0.021\\
  NGC\,4254~Enuc.\,1  &    114.4\,$\pm$\, 20.3  &     10.0\,$\pm$\,  2.4  &             \nodata  &             \nodata  &    0.7\,$\pm$\, 0.2  &    1.2\,$\pm$\, 0.3  &       2.4\,$\pm$\, 0.6  & 0.039\\
  NGC\,4254~Enuc.\,2  &     49.7\,$\pm$\,  8.8  &      6.5\,$\pm$\,  1.3  &             \nodata  &             \nodata  &    0.5\,$\pm$\, 0.1  &    0.8\,$\pm$\, 0.2  &       1.3\,$\pm$\, 0.4  & 0.039\\
  NGC\,4321~Enuc.\,1  &     54.9\,$\pm$\,  9.7  &      7.1\,$\pm$\,  1.3  &             \nodata  &             \nodata  &    0.4\,$\pm$\, 0.1  &    0.8\,$\pm$\, 0.2  &       1.0\,$\pm$\, 0.3  & 0.026\\
  NGC\,4321~Enuc.\,2  &     47.8\,$\pm$\,  8.5  &      6.5\,$\pm$\,  1.2  &             \nodata  &             \nodata  &    0.7\,$\pm$\, 0.2  &    1.1\,$\pm$\, 0.3  &       1.1\,$\pm$\, 0.3  & 0.026\\
  NGC\,4631~Enuc.\,1  &     46.8\,$\pm$\,  8.3  &      6.3\,$\pm$\,  1.1  &    4.4\,$\pm$\, 0.8  &    4.8\,$\pm$\, 0.8  &    2.0\,$\pm$\, 0.5  &    2.6\,$\pm$\, 0.6  &       4.5\,$\pm$\, 1.2  & 0.018\\
  NGC\,4631~Enuc.\,2  &     84.7\,$\pm$\, 15.0  &      8.8\,$\pm$\,  1.6  &    6.0\,$\pm$\, 1.1  &    6.8\,$\pm$\, 1.2  &    1.3\,$\pm$\, 0.3  &    1.6\,$\pm$\, 0.4  &       3.9\,$\pm$\, 1.0  & 0.017\\
  NGC\,4736~Enuc.\,1  &    506.4\,$\pm$\, 89.7  &     54.4\,$\pm$\,  9.8  &   14.0\,$\pm$\, 2.5  &   14.2\,$\pm$\, 2.5  &    4.1\,$\pm$\, 0.9  &    5.7\,$\pm$\, 1.3  &       7.0\,$\pm$\, 1.8  & 0.018\\
  NGC\,5055~Enuc.\,1  &     78.3\,$\pm$\, 13.9  &      8.0\,$\pm$\,  1.4  &    2.4\,$\pm$\, 0.4  &    2.9\,$\pm$\, 0.5  &    0.4\,$\pm$\, 0.1  &    0.6\,$\pm$\, 0.1  &       2.2\,$\pm$\, 0.6  & 0.018\\
  NGC\,5194~Enuc.\,1  &    250.9\,$\pm$\, 44.5  &     29.1\,$\pm$\,  8.8  &   12.1\,$\pm$\, 2.2  &   14.0\,$\pm$\, 2.5  &    0.8\,$\pm$\, 0.2  &    1.7\,$\pm$\, 0.4  &       5.0\,$\pm$\, 1.3  & 0.035\\
  NGC\,5194~Enuc.\,2  &    299.9\,$\pm$\, 53.1  &     37.4\,$\pm$\, 11.4  &    7.4\,$\pm$\, 1.3  &    8.4\,$\pm$\, 1.5  &    1.7\,$\pm$\, 0.4  &    2.9\,$\pm$\, 0.6  &       5.6\,$\pm$\, 1.5  & 0.036\\
  NGC\,5194~Enuc.\,3  &    227.1\,$\pm$\, 40.2  &     22.8\,$\pm$\,  6.2  &    7.6\,$\pm$\, 1.3  &    8.4\,$\pm$\, 1.5  &    2.1\,$\pm$\, 0.5  &    3.4\,$\pm$\, 0.8  &       4.8\,$\pm$\, 1.3  & 0.036\\
  NGC\,5194~Enuc.\,4  &    168.5\,$\pm$\, 29.9  &     21.1\,$\pm$\,  4.1  &    9.6\,$\pm$\, 1.7  &   11.0\,$\pm$\, 1.9  &    1.1\,$\pm$\, 0.2  &    1.7\,$\pm$\, 0.4  &       0.3\,$\pm$\, 0.1  & 0.035\\
  NGC\,5194~Enuc.\,5  &    196.5\,$\pm$\, 34.8  &     23.2\,$\pm$\,  4.5  &    9.2\,$\pm$\, 1.6  &   10.5\,$\pm$\, 1.9  &    2.1\,$\pm$\, 0.5  &    3.4\,$\pm$\, 0.8  &       0.4\,$\pm$\, 0.1  & 0.035\\
  NGC\,5194~Enuc.\,6  &     66.9\,$\pm$\, 11.9  &      8.1\,$\pm$\,  1.6  &    2.7\,$\pm$\, 0.5  &    3.1\,$\pm$\, 0.6  &    0.3\,$\pm$\, 0.1  &    0.4\,$\pm$\, 0.1  &       1.9\,$\pm$\, 0.5  & 0.038\\
  NGC\,5194~Enuc.\,7  &    180.8\,$\pm$\, 32.0  &     17.0\,$\pm$\,  3.3  &    4.7\,$\pm$\, 0.8  &    5.3\,$\pm$\, 0.9  &    1.2\,$\pm$\, 0.3  &    1.8\,$\pm$\, 0.4  &       5.0\,$\pm$\, 1.3  & 0.036\\
  NGC\,5194~Enuc.\,8  &    327.5\,$\pm$\, 58.0  &     31.2\,$\pm$\,  9.0  &   10.7\,$\pm$\, 1.9  &   11.8\,$\pm$\, 2.1  &    2.3\,$\pm$\, 0.5  &    3.5\,$\pm$\, 0.8  &       4.8\,$\pm$\, 1.2  & 0.035\\
  NGC\,5194~Enuc.\,9  &    146.1\,$\pm$\, 25.9  &     17.3\,$\pm$\,  3.2  &    5.5\,$\pm$\, 1.0  &    6.0\,$\pm$\, 1.1  &    1.3\,$\pm$\, 0.3  &    2.3\,$\pm$\, 0.5  &       2.2\,$\pm$\, 0.6  & 0.035\\
 NGC\,5194~Enuc.\,10  &    175.7\,$\pm$\, 31.1  &     24.4\,$\pm$\,  5.9  &    9.1\,$\pm$\, 1.6  &   10.4\,$\pm$\, 1.8  &    1.0\,$\pm$\, 0.2  &    1.8\,$\pm$\, 0.4  &       2.5\,$\pm$\, 0.7  & 0.036\\
 NGC\,5194~Enuc.\,11  &     78.4\,$\pm$\, 13.9  &      9.8\,$\pm$\,  2.3  &    6.1\,$\pm$\, 1.1  &    7.3\,$\pm$\, 1.3  &    0.7\,$\pm$\, 0.2  &    1.2\,$\pm$\, 0.3  &       1.3\,$\pm$\, 0.4  & 0.035\\
  NGC\,5457~Enuc.\,1  &     97.8\,$\pm$\, 17.3  &     12.3\,$\pm$\,  2.3  &             \nodata  &             \nodata  &    0.6\,$\pm$\, 0.1  &    1.1\,$\pm$\, 0.2  &       1.3\,$\pm$\, 0.3  & 0.009\\
  NGC\,5457~Enuc.\,2  &     60.8\,$\pm$\, 10.8  &      5.4\,$\pm$\,  0.9  &             \nodata  &             \nodata  &    0.8\,$\pm$\, 0.2  &    1.0\,$\pm$\, 0.2  &       2.3\,$\pm$\, 0.6  & 0.009\\
  NGC\,5457~Enuc.\,3  &   1279.4\,$\pm$\,226.7  &     50.2\,$\pm$\,  9.9  &             \nodata  &             \nodata  &    2.9\,$\pm$\, 0.6  &    3.5\,$\pm$\, 0.8  &      19.6\,$\pm$\, 5.1  & 0.009\\
  NGC\,5457~Enuc.\,4  &    143.3\,$\pm$\, 25.4  &     10.9\,$\pm$\,  1.9  &             \nodata  &             \nodata  &    3.8\,$\pm$\, 0.9  &    4.2\,$\pm$\, 1.0  &       8.0\,$\pm$\, 2.1  & 0.009\\
  NGC\,5457~Enuc.\,5  &    186.6\,$\pm$\, 33.1  &     10.9\,$\pm$\,  2.2  &             \nodata  &             \nodata  &    2.9\,$\pm$\, 0.7  &    3.3\,$\pm$\, 0.8  &      10.0\,$\pm$\, 2.6  & 0.009\\
  NGC\,5457~Enuc.\,6  &    248.0\,$\pm$\, 43.9  &     17.8\,$\pm$\,  3.3  &             \nodata  &             \nodata  &    3.9\,$\pm$\, 0.9  &    4.3\,$\pm$\, 1.0  &      10.5\,$\pm$\, 2.8  & 0.009\\
  NGC\,5457~Enuc.\,7  &    162.7\,$\pm$\, 28.8  &      8.1\,$\pm$\,  1.9  &             \nodata  &             \nodata  &             \nodata  &             \nodata  &                \nodata  & 0.010\\
  NGC\,5713~Enuc.\,1  &     97.8\,$\pm$\, 17.3  &     14.2\,$\pm$\,  2.7  &             \nodata  &             \nodata  &    0.4\,$\pm$\, 0.1  &    1.0\,$\pm$\, 0.2  &       1.1\,$\pm$\, 0.3  & 0.039\\
  NGC\,5713~Enuc.\,2  &    248.6\,$\pm$\, 44.1  &     20.3\,$\pm$\,  3.7  &             \nodata  &             \nodata  &    0.6\,$\pm$\, 0.1  &    1.2\,$\pm$\, 0.3  &       2.2\,$\pm$\, 0.6  & 0.039\\
  NGC\,6946~Enuc.\,1  &    129.4\,$\pm$\, 22.9  &     11.5\,$\pm$\,  2.0  &    3.7\,$\pm$\, 0.7  &    4.3\,$\pm$\, 0.8  &    2.2\,$\pm$\, 0.5  &    4.7\,$\pm$\, 1.1  &      10.5\,$\pm$\, 2.8  & 0.342\\
  NGC\,6946~Enuc.\,2  &    211.9\,$\pm$\, 37.5  &     16.3\,$\pm$\,  2.9  &    6.5\,$\pm$\, 1.2  &    7.4\,$\pm$\, 1.3  &    6.2\,$\pm$\, 1.4  &   11.4\,$\pm$\, 2.6  &      31.4\,$\pm$\, 8.2  & 0.344\\
  NGC\,6946~Enuc.\,3  &     92.6\,$\pm$\, 16.4  &     10.1\,$\pm$\,  1.8  &    3.4\,$\pm$\, 0.6  &    3.8\,$\pm$\, 0.7  &    2.4\,$\pm$\, 0.6  &    4.7\,$\pm$\, 1.1  &      13.7\,$\pm$\, 3.6  & 0.342\\
  NGC\,6946~Enuc.\,4  &    102.2\,$\pm$\, 18.1  &     10.3\,$\pm$\,  1.8  &    3.3\,$\pm$\, 0.6  &    3.8\,$\pm$\, 0.7  &    1.9\,$\pm$\, 0.4  &    4.1\,$\pm$\, 0.9  &                \nodata  & 0.343\\
  NGC\,6946~Enuc.\,5  &     36.4\,$\pm$\,  6.5  &      4.7\,$\pm$\,  0.8  &    2.2\,$\pm$\, 0.4  &    2.7\,$\pm$\, 0.5  &    1.4\,$\pm$\, 0.3  &    3.0\,$\pm$\, 0.7  &       8.9\,$\pm$\, 2.3  & 0.338\\
  NGC\,6946~Enuc.\,6  &    445.3\,$\pm$\, 78.9  &     41.8\,$\pm$\,  7.4  &   11.8\,$\pm$\, 2.1  &   13.0\,$\pm$\, 2.3  &    2.9\,$\pm$\, 0.7  &    6.6\,$\pm$\, 1.5  &      23.2\,$\pm$\, 6.1  & 0.342\\
  NGC\,6946~Enuc.\,7  &    454.0\,$\pm$\, 80.4  &     43.5\,$\pm$\,  7.6  &    9.8\,$\pm$\, 1.7  &   10.9\,$\pm$\, 1.9  &    2.5\,$\pm$\, 0.6  &    5.9\,$\pm$\, 1.3  &      22.3\,$\pm$\, 5.9  & 0.342\\
  NGC\,6946~Enuc.\,8  &    233.4\,$\pm$\, 41.4  &     24.1\,$\pm$\,  4.3  &    6.5\,$\pm$\, 1.1  &    7.5\,$\pm$\, 1.3  &             \nodata  &             \nodata  &                \nodata  & 0.343\\
  NGC\,6946~Enuc.\,9  &    367.6\,$\pm$\, 65.1  &     36.8\,$\pm$\,  6.7  &    8.5\,$\pm$\, 1.5  &    9.6\,$\pm$\, 1.7  &    1.7\,$\pm$\, 0.4  &    4.1\,$\pm$\, 0.9  &      14.2\,$\pm$\, 3.7  & 0.342  
\enddata
\tablenotetext{a}{Galactic extinction taken from \citep{ds98} used to correct the H$\alpha$ and GALEX FUV and NUV flux densities assuming $A_{V}/E(B-V)=3.1$ and the modeled extinction curves of \citet{wd01,bd03}.}
\end{deluxetable*}

%% file: tbl-5.tex
\begin{deluxetable*}{lcccc}
\tablecaption{Average Infrared to 33\,GHz Star Formation Rate Ratios \label{tbl-5}}
\tablewidth{0pt}
\tablehead{
\colhead{Sources} & \colhead{$<\log[{\rm SFR_{33\,GHz}}/\nu L_{\nu}(24\,\mu{\rm m})]>$} & \colhead{$\sigma$} & \colhead{$<\log({\rm SFR_{33\,GHz}}/L_{\rm IR})>$} & \colhead{$\sigma$} \\
\colhead{} & \colhead{(dex)} & \colhead{(dex)} & \colhead{(dex)} & \colhead{(dex)}
}
\startdata
      Nuc.&  -42.61&  0.27&  -43.55&  0.28\\
     Enuc.&  -42.61&  0.26&  -43.49&  0.29\\
       All&  -42.61&  0.26&  -43.50&  0.29
\enddata
\tablecomments{Median ratios are given. Only non-AGN detected sources included.}
\end{deluxetable*}

%% file: tbl-A1.tex
\begin{deluxetable*}{lccccccc}
\tablecaption{Nuclear Photometry for Individual CCB Ports \label{tbl-A1}}
\tablewidth{0pt}
\tablehead{
\colhead{Galaxy}  & \colhead{$S_{\rm 22.75\,GHz}$}  & \colhead{$S_{\rm 31.25\,GHz}$}  & \colhead{$S_{\rm 34.75\,GHz}$}  & \colhead{$S_{\rm 38.25\,GHz}$} &
\colhead{$\nu_{\rm eff}$}  &\colhead{$N_{\rm nods}$}  & \colhead{$t_{\rm on}$}\\
\colhead{} &  \colhead{(mJy)} & \colhead{(mJy)} & \colhead{(mJy)} & \colhead{(mJy)} &
 \colhead{(GHz)} & \colhead{used/taken}  & \colhead{(min)}
}
    NGC\,337  & 3.5\,$\pm$\,0.39  & 2.7\,$\pm$\,0.26  & 2.5\,$\pm$\,0.34  & 2.1\,$\pm$\,0.34  &33.11  & 8/ 8  & 5.3\\
    NGC\,628  & 0.1\,$\pm$\,0.09  &-0.0\,$\pm$\,0.06  &-0.1\,$\pm$\,0.05  &-0.0\,$\pm$\,0.08  &33.18  &43/60  &28.7\\
    NGC\,855  & 0.9\,$\pm$\,0.08  & 0.7\,$\pm$\,0.05  & 0.6\,$\pm$\,0.05  & 0.6\,$\pm$\,0.08  &33.11  &48/60  &32.0\\
    NGC\,925  & 0.7\,$\pm$\,0.06  & 0.6\,$\pm$\,0.04  & 0.4\,$\pm$\,0.05  & 0.4\,$\pm$\,0.08  &32.83  &60/77  &40.0\\
   NGC\,1266  & 9.8\,$\pm$\,0.35  & 9.6\,$\pm$\,0.25  & 8.2\,$\pm$\,0.34  & 7.6\,$\pm$\,0.33  &33.24  & 8/ 8  & 5.3\\
   NGC\,1377  &-0.2\,$\pm$\,0.35  &-0.2\,$\pm$\,0.25  &-0.2\,$\pm$\,0.33  &-0.5\,$\pm$\,0.32  &32.48  & 8/ 8  & 5.3\\
     IC\,342  &29.7\,$\pm$\,0.27  &27.6\,$\pm$\,0.17  &25.0\,$\pm$\,0.18  &23.3\,$\pm$\,0.16  &33.29  & 8/ 8  & 5.3\\
   NGC\,1482  &12.8\,$\pm$\,0.34  &13.1\,$\pm$\,0.24  &10.6\,$\pm$\,0.33  & 9.1\,$\pm$\,0.32  &33.58  & 8/ 8  & 5.3\\
   NGC\,2146  &99.8\,$\pm$\,0.21  &85.5\,$\pm$\,0.19  &73.7\,$\pm$\,0.14  &64.9\,$\pm$\,0.24  &33.80  & 4/ 4  & 2.7\\
   NGC\,2403  & 0.3\,$\pm$\,0.08  & 0.1\,$\pm$\,0.06  & 0.1\,$\pm$\,0.05  & 0.1\,$\pm$\,0.09  &32.88  &23/23  &15.3\\
Holmberg\,II  & 0.9\,$\pm$\,0.04  & 0.7\,$\pm$\,0.03  & 0.7\,$\pm$\,0.03  & 0.6\,$\pm$\,0.04  &33.36  &70/70  &46.7\\
   NGC\,2798  &10.0\,$\pm$\,1.10  & 8.9\,$\pm$\,0.94  & 7.8\,$\pm$\,0.35  & 6.6\,$\pm$\,0.60  &34.15  & 1/ 2  & 0.7\\
   NGC\,2841  & 1.7\,$\pm$\,0.06  & 1.6\,$\pm$\,0.06  & 1.6\,$\pm$\,0.04  & 1.5\,$\pm$\,0.06  &32.84  &50/50  &33.3\\
   NGC\,2976  & 0.3\,$\pm$\,0.25  & 0.1\,$\pm$\,0.16  & 0.4\,$\pm$\,0.09  & 0.2\,$\pm$\,0.20  &33.64  &10/10  & 6.7\\
   NGC\,3049  & 2.2\,$\pm$\,0.30  & 1.8\,$\pm$\,0.21  & 1.5\,$\pm$\,0.15  & 1.2\,$\pm$\,0.21  &33.78  & 4/ 4  & 2.7\\
   NGC\,3077  & 9.1\,$\pm$\,0.32  & 8.4\,$\pm$\,0.24  & 7.4\,$\pm$\,0.14  & 6.6\,$\pm$\,0.26  &33.46  & 4/ 4  & 2.7\\
   NGC\,3190  & 1.3\,$\pm$\,0.17  & 1.0\,$\pm$\,0.13  & 0.9\,$\pm$\,0.09  & 0.8\,$\pm$\,0.13  &33.59  &10/10  & 6.7\\
   NGC\,3184  & 0.2\,$\pm$\,0.18  & 0.2\,$\pm$\,0.13  & 0.1\,$\pm$\,0.09  & 0.2\,$\pm$\,0.13  &33.63  &10/10  & 6.7\\
   NGC\,3198  & 0.8\,$\pm$\,0.30  & 0.8\,$\pm$\,0.27  & 0.7\,$\pm$\,0.13  & 0.7\,$\pm$\,0.28  &33.57  & 4/ 4  & 2.7\\
    IC\,2574  & 0.7\,$\pm$\,0.12  & 0.7\,$\pm$\,0.07  & 0.6\,$\pm$\,0.09  & 0.6\,$\pm$\,0.07  &33.46  &54/54  &36.0\\
   NGC\,3265  & 1.2\,$\pm$\,0.37  & 1.3\,$\pm$\,0.30  & 1.2\,$\pm$\,0.20  & 0.7\,$\pm$\,0.24  &33.94  & 2/ 2  & 1.3\\
   NGC\,3351  & 6.3\,$\pm$\,0.36  & 5.4\,$\pm$\,0.29  & 4.9\,$\pm$\,0.19  & 4.1\,$\pm$\,0.22  &33.85  & 2/ 2  & 1.3\\
   NGC\,3521  & 2.2\,$\pm$\,0.18  & 1.8\,$\pm$\,0.17  & 1.2\,$\pm$\,0.10  & 1.3\,$\pm$\,0.18  &33.63  & 4/ 4  & 2.7\\
   NGC\,3627  & 3.2\,$\pm$\,0.19  & 2.7\,$\pm$\,0.18  & 2.8\,$\pm$\,0.11  & 2.4\,$\pm$\,0.20  &33.05  & 4/ 4  & 2.7\\
   NGC\,3773  & 1.1\,$\pm$\,0.09  & 1.0\,$\pm$\,0.10  & 0.9\,$\pm$\,0.12  & 0.8\,$\pm$\,0.20  &31.71  & 8/ 8  & 5.3\\
   NGC\,3938  & 0.2\,$\pm$\,0.11  & 0.2\,$\pm$\,0.11  & 0.1\,$\pm$\,0.17  & 0.0\,$\pm$\,0.17  &31.17  &42/44  &28.0\\
   NGC\,4254  & 2.1\,$\pm$\,0.27  & 1.8\,$\pm$\,0.22  & 1.6\,$\pm$\,0.11  & 1.2\,$\pm$\,0.21  &33.79  & 4/ 4  & 2.7\\
   NGC\,4321  & 5.2\,$\pm$\,0.26  & 4.5\,$\pm$\,0.21  & 4.3\,$\pm$\,0.13  & 4.0\,$\pm$\,0.20  &33.33  & 4/ 4  & 2.7\\
   NGC\,4536  &20.2\,$\pm$\,0.26  &18.5\,$\pm$\,0.21  &17.2\,$\pm$\,0.11  &16.5\,$\pm$\,0.20  &33.15  & 4/ 4  & 2.7\\
   NGC\,4559  & 0.9\,$\pm$\,0.10  & 0.7\,$\pm$\,0.08  & 0.7\,$\pm$\,0.06  & 0.6\,$\pm$\,0.09  &33.31  &21/30  &14.0\\
   NGC\,4569  & 2.3\,$\pm$\,0.25  & 2.0\,$\pm$\,0.21  & 1.8\,$\pm$\,0.12  & 1.5\,$\pm$\,0.20  &33.58  & 4/ 4  & 2.7\\
   NGC\,4579  &26.1\,$\pm$\,0.24  &24.7\,$\pm$\,0.20  &23.3\,$\pm$\,0.12  &22.6\,$\pm$\,0.22  &32.98  & 4/ 4  & 2.7\\
   NGC\,4594  &75.3\,$\pm$\,0.08  &74.5\,$\pm$\,0.06  &71.8\,$\pm$\,0.07  &65.4\,$\pm$\,0.14  &32.95  &14/14  & 9.3\\
   NGC\,4625  & 0.5\,$\pm$\,0.04  & 0.4\,$\pm$\,0.03  & 0.3\,$\pm$\,0.03  & 0.3\,$\pm$\,0.04  &33.21  &50/50  &33.3\\
   NGC\,4631  & 6.4\,$\pm$\,0.14  & 4.7\,$\pm$\,0.15  & 3.4\,$\pm$\,0.11  & 3.0\,$\pm$\,0.20  &34.27  & 4/ 4  & 2.7\\
   NGC\,4725  & 0.3\,$\pm$\,0.05  & 0.3\,$\pm$\,0.04  & 0.3\,$\pm$\,0.06  & 0.2\,$\pm$\,0.07  &32.01  &42/74  &28.0\\
   NGC\,4736  & 3.7\,$\pm$\,0.15  & 3.2\,$\pm$\,0.15  & 2.6\,$\pm$\,0.11  & 2.5\,$\pm$\,0.21  &33.26  & 4/ 4  & 2.7\\
   NGC\,4826  & 7.7\,$\pm$\,0.18  & 6.4\,$\pm$\,0.16  & 5.7\,$\pm$\,0.12  & 5.1\,$\pm$\,0.20  &33.61  & 4/ 4  & 2.7\\
   NGC\,5055  & 3.2\,$\pm$\,0.47  & 2.9\,$\pm$\,0.37  & 2.4\,$\pm$\,0.33  & 2.0\,$\pm$\,0.76  &32.37  & 4/ 4  & 2.7\\
   NGC\,5194  & 7.7\,$\pm$\,0.45  & 7.6\,$\pm$\,0.39  & 5.6\,$\pm$\,0.31  & 5.4\,$\pm$\,0.72  &32.91  & 4/ 4  & 2.7\\
   NGC\,5457  &-0.0\,$\pm$\,0.44  &-2.2\,$\pm$\,0.39  &-1.6\,$\pm$\,0.46  &-0.9\,$\pm$\,1.11  &31.05  & 4/ 4  & 2.7\\
   NGC\,5474  & 0.3\,$\pm$\,0.12  & 0.1\,$\pm$\,0.08  & 0.1\,$\pm$\,0.08  & 0.2\,$\pm$\,0.11  &32.99  &53/57  &35.3\\
   NGC\,5713  & 8.5\,$\pm$\,0.43  & 7.2\,$\pm$\,0.37  & 6.1\,$\pm$\,0.14  & 5.3\,$\pm$\,0.28  &33.93  & 4/ 4  & 2.7\\
   NGC\,5866  & 6.7\,$\pm$\,0.23  & 6.3\,$\pm$\,0.17  & 6.4\,$\pm$\,0.08  & 6.3\,$\pm$\,0.16  &32.77  &12/12  & 8.0\\
   NGC\,6946  &19.9\,$\pm$\,0.16  &17.4\,$\pm$\,0.23  &15.3\,$\pm$\,0.18  &13.9\,$\pm$\,0.25  &33.55  & 4/ 4  & 2.7\\
   NGC\,7331  & 2.2\,$\pm$\,0.22  & 1.6\,$\pm$\,0.30  & 1.3\,$\pm$\,0.19  & 1.3\,$\pm$\,0.22  &34.48  & 3/ 4  & 2.0  
\enddata
\end{deluxetable*}

%% file: tbl-A2.tex
\begin{deluxetable*}{lccccccc}
\tablecaption{Extranuclear Photometry for Individual CCB ports\label{tbl-A2}}
\tablewidth{0pt}
\tablehead{
\colhead{ID}  & \colhead{$S_{\rm 22.75\,GHz}$}  & \colhead{$S_{\rm 31.25\,GHz}$}  & \colhead{$S_{\rm 34.75\,GHz}$}  & \colhead{$S_{\rm 38.25\,GHz}$} &
\colhead{$\nu_{\rm eff}$}  & \colhead{$N_{\rm nods}$}  & \colhead{$t_{\rm on}$}\\
\colhead{} & \colhead{(mJy)} & \colhead{(mJy)} & \colhead{(mJy)} & \colhead{(mJy)} &
\colhead{(GHz)}  &\colhead{used/taken}  & \colhead{(min)}
}
   NGC\,628~Enuc.\,1  & 0.6\,$\pm$\,0.39  & 0.6\,$\pm$\,0.29  & 0.6\,$\pm$\,0.37  & 0.3\,$\pm$\,0.38  &32.48  & 7/ 7  & 4.7\\
   NGC\,628~Enuc.\,2  & 0.4\,$\pm$\,0.11  & 0.3\,$\pm$\,0.08  & 0.4\,$\pm$\,0.05  & 0.3\,$\pm$\,0.09  &33.51  &40/40  &26.7\\
   NGC\,628~Enuc.\,3  & 0.6\,$\pm$\,0.12  & 0.5\,$\pm$\,0.08  & 0.5\,$\pm$\,0.05  & 0.4\,$\pm$\,0.09  &33.49  &37/40  &24.7\\
   NGC\,628~Enuc.\,4  & 0.1\,$\pm$\,0.15  & 0.1\,$\pm$\,0.08  & 0.1\,$\pm$\,0.10  & 0.2\,$\pm$\,0.12  &32.95  &37/60  &24.7\\
  NGC\,2403~Enuc.\,1  & 1.9\,$\pm$\,0.36  & 1.9\,$\pm$\,0.27  & 1.6\,$\pm$\,0.14  & 1.5\,$\pm$\,0.30  &33.50  & 4/ 4  & 2.7\\
  NGC\,2403~Enuc.\,2  & 1.7\,$\pm$\,0.33  & 1.2\,$\pm$\,0.24  & 1.5\,$\pm$\,0.14  & 1.1\,$\pm$\,0.29  &33.31  & 4/ 4  & 2.7\\
  NGC\,2403~Enuc.\,3  & 4.2\,$\pm$\,0.33  & 3.6\,$\pm$\,0.23  & 3.0\,$\pm$\,0.13  & 2.3\,$\pm$\,0.28  &33.73  & 4/ 4  & 2.7\\
  NGC\,2403~Enuc.\,4  & 1.7\,$\pm$\,0.12  & 1.5\,$\pm$\,0.08  & 1.3\,$\pm$\,0.05  & 1.2\,$\pm$\,0.11  &33.37  &24/24  &16.0\\
  NGC\,2403~Enuc.\,5  & 1.6\,$\pm$\,0.82  & 1.7\,$\pm$\,0.70  & 1.3\,$\pm$\,0.24  & 0.9\,$\pm$\,0.42  &34.62  & 2/ 2  & 1.3\\
  NGC\,2403~Enuc.\,6  & 0.9\,$\pm$\,0.15  & 0.7\,$\pm$\,0.08  & 0.8\,$\pm$\,0.09  & 0.7\,$\pm$\,0.11  &33.08  &59/59  &39.3\\
  NGC\,2976~Enuc.\,1  & 4.6\,$\pm$\,0.31  & 4.0\,$\pm$\,0.24  & 3.4\,$\pm$\,0.13  & 3.2\,$\pm$\,0.26  &33.46  & 4/ 4  & 2.7\\
  NGC\,2976~Enuc.\,2  & 2.0\,$\pm$\,0.31  & 2.0\,$\pm$\,0.25  & 2.1\,$\pm$\,0.14  & 1.7\,$\pm$\,0.27  &33.21  & 4/ 4  & 2.7\\
  NGC\,3521~Enuc.\,1  & 0.3\,$\pm$\,0.05  & 0.2\,$\pm$\,0.04  & 0.3\,$\pm$\,0.04  & 0.2\,$\pm$\,0.05  &32.39  &65/66  &43.3\\
  NGC\,3521~Enuc.\,2  & 1.2\,$\pm$\,0.18  & 1.3\,$\pm$\,0.21  & 0.9\,$\pm$\,0.10  & 0.5\,$\pm$\,0.18  &33.37  & 4/ 4  & 2.7\\
  NGC\,3521~Enuc.\,3  & 0.5\,$\pm$\,0.05  & 0.4\,$\pm$\,0.05  & 0.3\,$\pm$\,0.05  & 0.3\,$\pm$\,0.06  &32.77  &59/60  &39.3\\
  NGC\,3627~Enuc.\,1  & 3.0\,$\pm$\,0.19  & 2.5\,$\pm$\,0.18  & 2.3\,$\pm$\,0.11  & 2.2\,$\pm$\,0.20  &33.17  & 4/ 4  & 2.7\\
  NGC\,3627~Enuc.\,2  & 7.4\,$\pm$\,0.18  & 6.7\,$\pm$\,0.18  & 5.9\,$\pm$\,0.11  & 5.6\,$\pm$\,0.19  &33.31  & 4/ 4  & 2.7\\
  NGC\,3627~Enuc.\,3  & 0.9\,$\pm$\,0.18  & 0.8\,$\pm$\,0.19  & 0.5\,$\pm$\,0.11  & 0.5\,$\pm$\,0.20  &33.17  & 4/ 4  & 2.7\\
  NGC\,3938~Enuc.\,1  & 0.2\,$\pm$\,0.05  & 0.1\,$\pm$\,0.04  & 0.1\,$\pm$\,0.03  & 0.1\,$\pm$\,0.05  &33.01  &49/50  &32.7\\
  NGC\,3938~Enuc.\,2  & 0.4\,$\pm$\,0.38  & 0.4\,$\pm$\,0.44  & 0.3\,$\pm$\,0.53  & 0.3\,$\pm$\,0.62  &31.23  &44/44  &29.3\\
  NGC\,4254~Enuc.\,1  & 1.4\,$\pm$\,0.10  & 1.2\,$\pm$\,0.10  & 1.0\,$\pm$\,0.13  & 0.8\,$\pm$\,0.22  &31.92  & 8/ 8  & 5.3\\
  NGC\,4254~Enuc.\,2  & 0.0\,$\pm$\,0.12  &-0.0\,$\pm$\,0.07  &-0.1\,$\pm$\,0.08  & 0.0\,$\pm$\,0.10  &32.94  &50/50  &33.3\\
  NGC\,4321~Enuc.\,1  & 0.1\,$\pm$\,0.10  & 0.1\,$\pm$\,0.09  & 0.1\,$\pm$\,0.08  &-0.0\,$\pm$\,0.11  &32.52  &43/51  &28.7\\
  NGC\,4321~Enuc.\,2  &-0.4\,$\pm$\,0.11  &-0.3\,$\pm$\,0.07  &-0.2\,$\pm$\,0.07  &-0.2\,$\pm$\,0.09  &33.39  &49/49  &32.7\\
  NGC\,4631~Enuc.\,1  & 1.1\,$\pm$\,0.16  & 1.0\,$\pm$\,0.10  & 1.0\,$\pm$\,0.15  & 0.7\,$\pm$\,0.17  &32.43  &25/50  &16.7\\
  NGC\,4631~Enuc.\,2  & 0.9\,$\pm$\,0.12  & 0.8\,$\pm$\,0.09  & 0.7\,$\pm$\,0.07  & 0.5\,$\pm$\,0.09  &33.62  &24/26  &16.0\\
  NGC\,4736~Enuc.\,1  & 3.3\,$\pm$\,0.15  & 2.8\,$\pm$\,0.15  & 2.6\,$\pm$\,0.12  & 2.1\,$\pm$\,0.21  &33.15  & 4/ 4  & 2.7\\
  NGC\,5055~Enuc.\,1  & 0.4\,$\pm$\,0.21  & 0.1\,$\pm$\,0.38  & 0.1\,$\pm$\,0.26  & 0.2\,$\pm$\,0.24  &32.28  &17/17  &11.3\\
  NGC\,5194~Enuc.\,1  & 0.8\,$\pm$\,0.53  &-4.3\,$\pm$\,0.52  &-3.3\,$\pm$\,0.38  &-0.3\,$\pm$\,0.84  &31.89  & 3/ 4  & 2.0\\
  NGC\,5194~Enuc.\,2  & 2.9\,$\pm$\,0.49  & 3.5\,$\pm$\,0.43  & 0.9\,$\pm$\,0.34  &-0.1\,$\pm$\,0.82  &32.71  & 3/ 3  & 2.0\\
  NGC\,5194~Enuc.\,3  & 1.2\,$\pm$\,0.46  & 1.1\,$\pm$\,0.47  & 1.1\,$\pm$\,0.35  & 0.5\,$\pm$\,0.75  &32.15  & 4/ 4  & 2.7\\
  NGC\,5194~Enuc.\,4  & 1.7\,$\pm$\,0.45  & 2.7\,$\pm$\,0.45  & 1.3\,$\pm$\,0.34  & 2.5\,$\pm$\,0.73  &32.22  & 4/ 4  & 2.7\\
  NGC\,5194~Enuc.\,5  & 1.9\,$\pm$\,0.46  & 2.3\,$\pm$\,0.51  & 1.9\,$\pm$\,0.32  & 1.2\,$\pm$\,0.69  &32.39  & 4/ 4  & 2.7\\
  NGC\,5194~Enuc.\,6  & 1.0\,$\pm$\,0.14  & 0.9\,$\pm$\,0.12  & 0.8\,$\pm$\,0.07  & 0.9\,$\pm$\,0.12  &33.30  &23/23  &15.3\\
  NGC\,5194~Enuc.\,7  & 0.8\,$\pm$\,0.45  & 0.2\,$\pm$\,0.51  & 0.6\,$\pm$\,0.32  & 0.7\,$\pm$\,0.69  &32.41  & 4/ 4  & 2.7\\
  NGC\,5194~Enuc.\,8  & 1.1\,$\pm$\,0.51  &-1.5\,$\pm$\,0.60  & 2.1\,$\pm$\,0.40  & 1.8\,$\pm$\,0.81  &32.05  & 3/ 4  & 2.0\\
  NGC\,5194~Enuc.\,9  & 1.2\,$\pm$\,0.44  & 0.8\,$\pm$\,0.48  & 0.6\,$\pm$\,0.35  & 0.5\,$\pm$\,0.73  &32.19  & 4/ 4  & 2.7\\
 NGC\,5194~Enuc.\,10  & 0.8\,$\pm$\,0.45  & 1.2\,$\pm$\,0.47  & 0.3\,$\pm$\,0.34  &-0.8\,$\pm$\,0.72  &32.23  & 4/ 4  & 2.7\\
 NGC\,5194~Enuc.\,11  &-0.6\,$\pm$\,0.20  &-0.5\,$\pm$\,0.23  &-0.3\,$\pm$\,0.16  &-0.3\,$\pm$\,0.24  &32.65  &14/14  & 9.3\\
  NGC\,5457~Enuc.\,1  & 0.3\,$\pm$\,0.17  & 0.7\,$\pm$\,0.39  & 0.6\,$\pm$\,0.28  & 0.5\,$\pm$\,0.25  &31.26  &17/17  &11.3\\
  NGC\,5457~Enuc.\,2  & 0.3\,$\pm$\,0.15  & 0.4\,$\pm$\,0.15  & 0.2\,$\pm$\,0.18  & 0.2\,$\pm$\,0.19  &31.93  &32/40  &21.3\\
  NGC\,5457~Enuc.\,3  & 8.5\,$\pm$\,0.54  & 9.1\,$\pm$\,0.45  & 9.2\,$\pm$\,0.55  & 7.4\,$\pm$\,1.14  &31.84  & 4/ 4  & 2.7\\
  NGC\,5457~Enuc.\,4  & 2.4\,$\pm$\,0.33  & 3.3\,$\pm$\,0.26  & 1.3\,$\pm$\,0.37  & 1.1\,$\pm$\,0.86  &31.48  & 7/ 7  & 4.7\\
  NGC\,5457~Enuc.\,5  & 2.3\,$\pm$\,0.47  & 2.8\,$\pm$\,0.36  & 2.8\,$\pm$\,0.49  & 2.8\,$\pm$\,1.12  &31.20  & 4/ 4  & 2.7\\
  NGC\,5457~Enuc.\,6  & 3.1\,$\pm$\,0.52  & 4.1\,$\pm$\,0.38  & 3.7\,$\pm$\,0.64  & 2.4\,$\pm$\,1.37  &31.02  & 3/ 3  & 2.0\\
  NGC\,5457~Enuc.\,7  & 5.0\,$\pm$\,0.45  & 5.1\,$\pm$\,0.35  & 4.8\,$\pm$\,0.54  & 4.5\,$\pm$\,1.21  &31.33  & 4/ 4  & 2.7\\
  NGC\,5713~Enuc.\,1  & 1.8\,$\pm$\,0.42  & 1.8\,$\pm$\,0.42  & 1.0\,$\pm$\,0.16  & 1.1\,$\pm$\,0.28  &34.28  & 4/ 4  & 2.7\\
  NGC\,5713~Enuc.\,2  & 2.4\,$\pm$\,0.42  & 1.7\,$\pm$\,0.38  & 1.3\,$\pm$\,0.16  & 1.3\,$\pm$\,0.29  &34.08  & 4/ 4  & 2.7\\
  NGC\,6946~Enuc.\,1  & 0.3\,$\pm$\,0.09  & 0.5\,$\pm$\,0.08  & 0.5\,$\pm$\,0.10  & 0.5\,$\pm$\,0.12  &31.70  &16/16  &10.7\\
  NGC\,6946~Enuc.\,2  & 2.8\,$\pm$\,0.17  & 2.6\,$\pm$\,0.22  & 2.2\,$\pm$\,0.21  & 2.1\,$\pm$\,0.25  &32.64  & 4/ 4  & 2.7\\
  NGC\,6946~Enuc.\,3  & 1.2\,$\pm$\,0.10  & 1.1\,$\pm$\,0.09  & 0.9\,$\pm$\,0.12  & 0.9\,$\pm$\,0.13  &32.52  &15/16  &10.0\\
  NGC\,6946~Enuc.\,4  & 3.4\,$\pm$\,0.10  & 3.2\,$\pm$\,0.10  & 2.8\,$\pm$\,0.12  & 2.7\,$\pm$\,0.13  &33.07  &15/16  &10.0\\
  NGC\,6946~Enuc.\,5  & 0.6\,$\pm$\,0.07  & 0.6\,$\pm$\,0.07  & 0.4\,$\pm$\,0.09  & 0.5\,$\pm$\,0.08  &32.37  &37/37  &24.7\\
  NGC\,6946~Enuc.\,6  & 3.6\,$\pm$\,0.16  & 3.1\,$\pm$\,0.22  & 3.0\,$\pm$\,0.21  & 2.2\,$\pm$\,0.26  &32.98  & 4/ 4  & 2.7\\
  NGC\,6946~Enuc.\,7  & 3.6\,$\pm$\,0.21  & 3.5\,$\pm$\,0.24  & 3.0\,$\pm$\,0.31  & 3.1\,$\pm$\,0.29  &32.48  & 3/ 4  & 2.0\\
  NGC\,6946~Enuc.\,8  & 2.1\,$\pm$\,0.19  & 1.9\,$\pm$\,0.20  & 1.7\,$\pm$\,0.26  & 1.7\,$\pm$\,0.24  &32.41  & 4/ 4  & 2.7\\
  NGC\,6946~Enuc.\,9  & 3.1\,$\pm$\,0.20  & 2.6\,$\pm$\,0.20  & 2.4\,$\pm$\,0.28  & 2.2\,$\pm$\,0.23  &32.81  & 4/ 4  & 2.7  
\enddata
\end{deluxetable*}

%% file: tbl-A3.tex
\begin{deluxetable*}{lll}
\tablecaption{Summary of Star Formation Rate Conversions \label{tbl-A3}}
\tablehead{
\colhead{Observable} & \colhead{Equation} & \colhead{Fading Timescale\tablenotemark{a}}
}
\startdata
\cutinhead{Theoretical Relations\tablenotemark{b}}
Extinction corrected FUV flux & \({\rm SFR} = 4.42\times10^{-44}L_{\rm FUV}\) & $\sim 5, 100$\,Myr\\\\
Extinction corrected NUV flux & \({\rm SFR} = 7.15\times10^{-44}L_{\rm NUV}\) & $\sim 5, 150$\,Myr\\\\
Extinction corrected H$\alpha$ recombination line flux  & \({\rm SFR} = 5.37\times10^{-42}L_{\rm H\alpha}\) & $\sim 2, 5$\,Myr\\\\
Free-free (radio) flux density & \({\rm SFR} = 4.6\times10^{-28} T_{\rm e, 4}^{-0.45} \nu_{\rm GHz}^{0.1} L_{\nu}^{\rm T} \) & $\sim 2, 5$\,Myr\\\\
Non-thermal (radio) flux density & \({\rm SFR} = 6.6\times10^{-29}  \nu_{\rm GHz}^{\alpha_{\rm NT}} L_{\nu}^{\rm NT} \) & $\sim 40, 100$\,Myr\\\\
Total (radio) flux density & \({\rm SFR} =10^{-27}/(2.18 T_{\rm e, 4}^{0.45} \nu_{\rm GHz}^{-0.1} + 15.1 \nu_{\rm GHz}^{-\alpha_{\rm NT}}) L_{\nu} \) & $\sim 2-40, 5-100$\,Myr\\
\cutinhead{Empirical Relations\tablenotemark{c}}
24\,$\mu$m flux density & \({\rm SFR} = 2.45\times10^{-43} \nu L_{\nu}(24\,\micron)\) &$\sim 5, 100$\,Myr \\\\
Total infrared (IR; $8-1000\,\mu$m) flux  & \({\rm SFR} = 3.15\times10^{-44} L_{\rm IR}\) & $\sim 5, 100$\,Myr \\\\
24\,$\mu$m flux density and H$\alpha$ line flux & \({\rm SFR} = 5.37\times10^{-42} [L_{\rm H\alpha} + 0.018 \nu L_{\nu}(24\,\micron)]\) & $\sim 2-5, 5-100$\,Myr\\\\
Total IR and FUV fluxes & \({\rm SFR} = 4.42\times10^{-44}(L_{\rm FUV} + 0.50 L_{\rm IR})\) & $\sim 5, 100$\,Myr
\enddata
\tablenotetext{a}{
The timescales given are those for the emission to drop to 50 and 5\% of its peak value after having had 100\,Myr of continuous star formation come to a halt.  
These values, which are sensitive to the star formation history, were calculated using Starburst99 \citep[][]{cl99} following the same assumptions that went into deriving the corresponding theoretical relations.  
The non-thermal radio continuum timescale sums the fading timescale of supernovae activity and appropriate fractions of the radiating lifetimes for 1.4\,GHz emitting CR electrons in the ISM assuming a magnetic field strength of 5\,$\mu$G, equipartition between the magnetic and radiation field energy densities, and an ISM density of $n_{\rm ISM} = 0.1\,$cm$^{-3}$ \citep[for details on the CR cooling timescales see][]{ejm09c}.  
}
\tablenotetext{b}{ 
Each theoretical calibration was calculated using  using Starburst99 \citep[][]{cl99} and assumes solar metallicity, continuous star formation, and a \citet{pk01} IMF, having a slope of $-1.3$ for stellar masses between $0.1-0.5~M_{\sun}$ and $-2.3$ for stellar masses ranging between $0.5-100\,M_{\sun}$.  
Star formation rates are in units of $M_{\sun}\,{\rm yr}^{-1}$ and observables are in cgs units.  
The UV calibrations are calculated using the {\it GALEX} FUV and NUV basspands, which are centered at 1528 and 2271\,\AA, respecively.  
For radio-based star formation rates, $\nu_{\rm GHz}$ and $T_{\rm e, 4}$ are frequency and thermal electron temperature in units of GHz and $10^{4}$\,K, respectively.  
Additionally, $\alpha^{\rm NT}$ is the non-thermal radio spectral index, which is assumed to be $\sim0.85$ in the present analysis.  
}
\tablenotetext{c}{
The empirical relations each have a scatter that is roughly a factor of $\la$2.  
Consequently, the empirical calibration given here for the total IR luminosity is conistent with the theoretical relation given in \citet[][\({\rm SFR} = 3.88\times10^{-44} L_{\rm IR}\) ]{ejm11b}, which, using starburst99 with the same assumptions above, assumed that the entire Balmer continuum was absorbed and reradiated in the infrared.
See \S\ref{sec-sfr}, \ref{sec-sfr24}, and \ref{sec-sfrha24} for more details. 
} 
\end{deluxetable*}